\documentclass[12pt]{elsart}                                              
\usepackage{array}
\usepackage{epsfig}
\usepackage{latexsym}
\usepackage{rotating}
\usepackage{supertabular}
\newcommand{\PreserveBackslash}[1]{\let\temp=\\#1\let\\=\temp}

\newcommand{\GeV}{\ensuremath{\mathrm{GeV}}}
\newcommand{\MeV}{\ensuremath{\mathrm{MeV}}}
\newcommand{\keV}{\ensuremath{\mathrm{keV}}}
\newcommand{\mm}{\ensuremath{\mathrm{mm}}}
\newcommand{\cm}{\ensuremath{\mathrm{cm}}}
\newcommand{\meter}{\ensuremath{\mathrm{m}}}
\newcommand{\MHz}{\ensuremath{\mathrm{MHz}}}
\newcommand{\ns}{\ensuremath{\mathrm{ns}}}
\newcommand{\nA}{\ensuremath{\mathrm{nA}}}
\newcommand{\Dy}{\displaystyle}
\begin{document}
\newcommand{\tdr}{\tau_{\rm dr}}
\newcommand{\tcal}{\tau_{\rm cal}}
%
%auto-ignore
%
% cover material
%
% - title
% - author list
% - abstract
% - keywords
\begin{center}
%\today\ Draft Version 2.2  
\end{center}

\begin{frontmatter}
\title{Hadronic Calibration of the ATLAS Liquid Argon End-Cap
    Calorimeter in the Pseudorapidity Region $1.6<|\eta|<1.8$ in Beam
    Tests}
\collab{ATLAS Liquid Argon EMEC/HEC Collaboration}

\author{C. Cojocaru},
\author{J. Pinfold},
\author{J. Soukup},
\author{M. Vincter}
\address{University of Alberta, Edmonton, Canada}
\author{V. Datskov},
\author{A. Fedorov},
\author{S. Golubykh},
\author{N. Javadov\thanksref{a-Baku}},
\author{V. Kalinnikov},
\author{S. Kakurin},
\author{M. Kazarinov},
\author{V. Kukhtin},
\author{E. Ladygin},
\author{A. Lazarev},
\author{A. Neganov},
\author{I. Pisarev},
\author{N. Rousakovitch},
\author{E. Serochkin},
\author{S. Shilov},
\author{A. Shalyugin},
\author{Yu. Usov}
\address{Joint Institute for Nuclear Research, Dubna, Russia}
% ...

\begin{abstract}
 A full azimuthal $\phi$-wedge of the ATLAS liquid argon end-cap
 calorimeter has been exposed to beams of electrons, muons and pions
 in the energy range $6\,\GeV\le E\le 200\,\GeV$ at the CERN SPS.  The
 angular region studied corresponds to the ATLAS impact position
 around the pseudorapidity interval $1.6<|\eta|<1.8$. The beam test
 set-up is described.  A detailed study of the performance is given as
 well as the related intercalibration constants obtained. Following
 the ATLAS hadronic calibration proposal, a first study of the hadron
 calibration using a weighting ansatz is presented. The results are
 compared to predictions from Monte Carlo simulations, based on
 GEANT~3 and GEANT~4 models.
\end{abstract}
\end{frontmatter}

\section{Introduction}
  %auto-ignore
%
% Introduction
%
The ATLAS calorimeter has to provide an accurate measurement of the
energy and position of electrons and photons, of the energy and
direction of jets and of the missing transverse energy in a given
event and to provide information on particle identification. Previous
beam runs with individual set-ups of the
electromagnetic~\cite{r-emec,r-embar}, hadronic~\cite{r-hec} and
forward calorimeters respectively provided important information on
the stand-alone calibration. They also contributed substantially to
the assessment of the production quality of the calorimeters or
modules.  Only combined runs with all calorimeter types, in a set-up
as close as possible to the final ATLAS detector, can yield
calibration constants for single pions in ATLAS. This can be
transferred to ATLAS via detailed comparison between Monte Carlo
(MC) simulations and jets in ATLAS.  This beam
test studied the forward region corresponding to the pseudorapidity
interval $1.6<|\eta|<1.8$ in ATLAS and the analysis
\begin{itemize}
\item	obtained intercalibration constants for electrons and pions in the 
       energy range $6\,\GeV<E<200\,\GeV$;
\item   did a detailed comparison with simulation to allow extrapolation 
       to jets;
\item	tested methods and algorithms for optimal hadronic energy 
       reconstruction in ATLAS.
\end{itemize}
Details of the set-up, data analysis and simulations can be found
in~\cite{hendrik}.  It should be stressed that the set-up did not
reproduce the exact ATLAS projective geometry. The beam was incident
perpendicular to the face of the calorimeters, rather than tilted at
an angle corresponding to the ATLAS impact region
$1.6<|\eta|<1.8$. This tilt angle has no major impact on the hadronic
response. For electrons the difference is more relevant. Therefore the
results given in~\cite{r-emec} for the performance of electrons are
those which can be directly transfered to ATLAS.

\section{General Set-up, Read-out and Calibration}

\subsection{Description of the Electromagnetic End-Cap and Hadronic End-Cap 
Calorimeter} 
  %auto-ignore
The electromagnetic end-cap calorimeter (EMEC) \cite{r-emec} is a
liquid argon sampling calorimeter with lead as absorber material. One
end-cap wheel is structured in eight azimuthal wedge-shaped modules,
with an inner and outer section.  The absorber plates are mounted in a
radial arrangement like spokes of a wheel, with the accordion waves
running in depth parallel to the front and back edges. The liquid
argon gap increases with the radius and the accordion wave amplitude
and the related folding angle varies at each radius. In the outer
section there are 96 absorbers, segmented into three longitudinal
sections. In total there are 3888 read-out cells per module.

The presampler is placed in front of the EMEC module and it consists
of two $2\,\mm$ thick active liquid argon layers, formed by three
electrodes parallel to the front face of the EMEC calorimeter.

The hadronic end-cap calorimeter (HEC) \cite{r-hec} is a liquid argon
sampling calorimeter with flat copper absorber plates. The thickness
of the absorber plates is $25\,\mm$ for the front wheel (HEC1) and
$50\,\mm$ for the rear wheel (HEC2).  Each wheel is made out of 32
modules. In total 24 gaps ($8.5\,\mm$) for HEC1 and 16 gaps for HEC2
are instrumented with a read-out structure. Longitudinally the
read-out is segmented in 8 and 16 gaps for HEC1 and 8 and 8 gaps for
HEC2. The total number of read-out channels for a $\phi$-wedge
consisting of one HEC1 and one HEC2 module is 88.
 
 \subsection{General Beam Test Set-up} 
\begin{figure}[htb]
  \begin{center}
%    \psdraft   
    \mbox{\epsfig{file=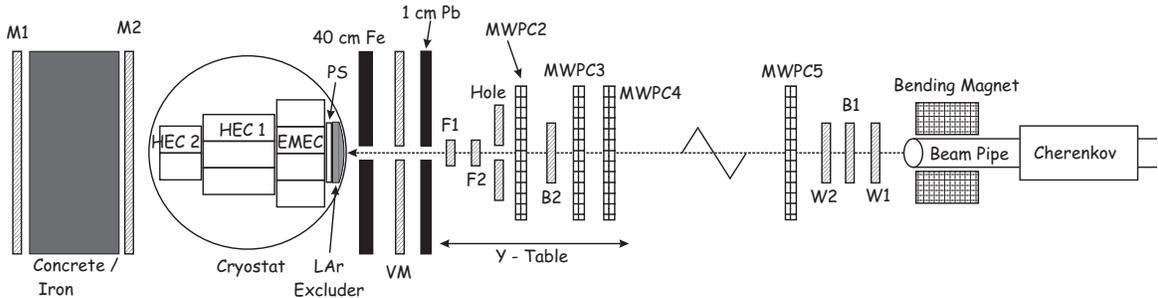,width=1.1\textwidth}}
%    \psfull
  \end{center}
  \caption{Schematic view of the general beam test set-up. Shown are
   the modules in the cryostat as well as the beam instrumentation
   used.}
  \label{nim_2_general_set-up}
\end{figure}

  %auto-ignore
The beam tests have been carried out in the H6 beamline at the CERN
SPS providing hadrons, electrons or muons in the energy range
$6\,\GeV\le E\le 200\,\GeV$. The general set-up is shown in
Fig.~\ref{nim_2_general_set-up}.

\begin{figure}[htb] 
  \begin{center}
%    \psdraft   
    \mbox{\epsfig{file=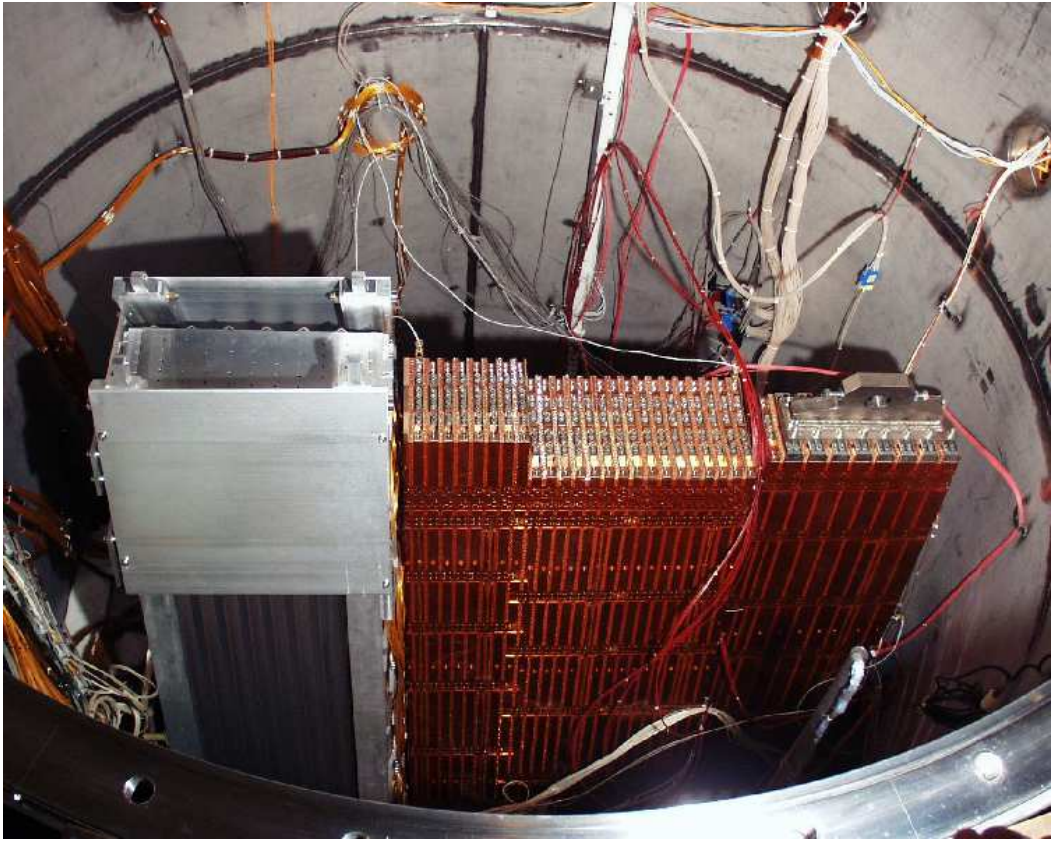,angle=-90,width=1.0\textwidth}}
%    \psfull
  \end{center}
  \caption{Top view of the cryostat, showing (from left to right) the
  EMEC module with the presampler, the three HEC1 modules and the two
  HEC2 modules of reduced longitudinal size.}
  \label{nim_2_photo}
\end{figure}

The load in the liquid argon (LAr) cryostat consists of the outer
section of one EMEC module (1/8 of the full EMEC wheel), three HEC1
modules (3/32 of the full HEC1 wheel) and two HEC2
modules. Constrained by the cryostat dimensions the depth of the HEC2
modules was half of the ATLAS modules.  The impact angle of beam
particles was $90^\circ$ with respect to the front plane, yielding a
non-pointing geometry of the set-up in $\eta$ (vertically) unlike the
ATLAS situation.  Fig.~\ref{nim_2_general_set-up} shows the location
of the trigger and veto counters. The trigger is derived from the
scintillation counter B1, three scintillator walls VM, M1, M2, and two
scintillation counters F1, F2 (pretrigger, for details
see~\cite{r-hec}) for fast timing. Up to $80\,\GeV$ the Cherenkov
counter was used for the event trigger as well.  The impact position
and angle of particles were derived from four multiwire proportional
chambers (MWPC) with vertical and horizontal planes per chamber having
$1\,\mm$ (MWPC 2, MWPC3, MWPC4) and $2\,\mm$ (MWPC5) wire spacing. The
veto wall VM rejects beam halo particles; muons are tagged using the
coincidence of M1 and M2 signals.  The M1 and M2 scintillator walls
are separated by an iron wall.  The cryostat has an inner diameter of
$2.50\,\meter$, it can be filled with LAr up to a height of
$2.20\,\meter$. It can be moved horizontally by $\pm30\,\cm$. The
beamline vertical bending magnet allows the beam to be deflected in a
range of $\pm25\,\cm$ at the front face of the cryostat. The circular
cryostat beam window has a reduced wall thickness ($5.5\,\mm$
stainless steel) and a diameter of $60\,\cm$. Thus an area of
$60\times50\,\cm^2$ is available for horizontal and vertical scans.
The calorimeter modules in the cryostat are shown in
Fig.~\ref{nim_2_photo}.

The top view of the cryostat shows (from left to right) the EMEC
module including the presampler, the three HEC1 modules and the two
HEC2 modules of reduced longitudinal size.  Simulation studies show
that the typical leakage for pions is at the level of $4$-$6\,$\% (see
section Monte Carlo simulation). Monitoring of the LAr purity and temperature
are included in the cryostat; details may be found in~\cite{r-hec}.

  \subsection{Read-out Electronics and Data Acquisition}
  %auto-ignore
The HEC cold front-end electronics was identical to the one used in
the previous HEC stand-alone tests~\cite{r-hec}. The output signals of
the cold HEC summing amplifiers as well as the raw signals from the
EMEC were carried to the front-end boards (FEB) outside the
cryostat. Here the amplification of the EMEC signals and signal
shaping of all signals was performed. The crate with the FEB's was
directly located on the two related feedthroughs, thus extending the
Faraday cage of the cryostat.  The signals were sampled and stored in
the switched capacity array of the FEB at a rate of $40\,\MHz$. Upon
arrival of the trigger the FEB's stopped the sampling, performed the
digitization and sent the data to the read-out driver (ROD) via a
serial electrical link. The two types of FEB's used for the EMEC and
HEC channels are described in refs.~\cite{r-emec,r-embar,r-hec}. The
read-out of the EMEC data was performed by nine MINI-ROD modules,
exploited previously for the EMEC stand-alone tests~\cite{r-emec}. The
prototype of the ATLAS ROD module (ROD-demo~\cite{r-rod}), designed to
validate the final ATLAS-LAr calorimetry read-out system, was used for
the HEC read-out.  The ROD-demo was a 9U VME motherboard with four
mezzanine processing unit (PU) cards. Each PU processed the data from
one half FEB, so in total six PU's on two ROD boards were used to read
out the three HEC FEB's. 

\begin{figure}[htb] \begin{center}
%    \psdraft   
    \mbox{\epsfig{file=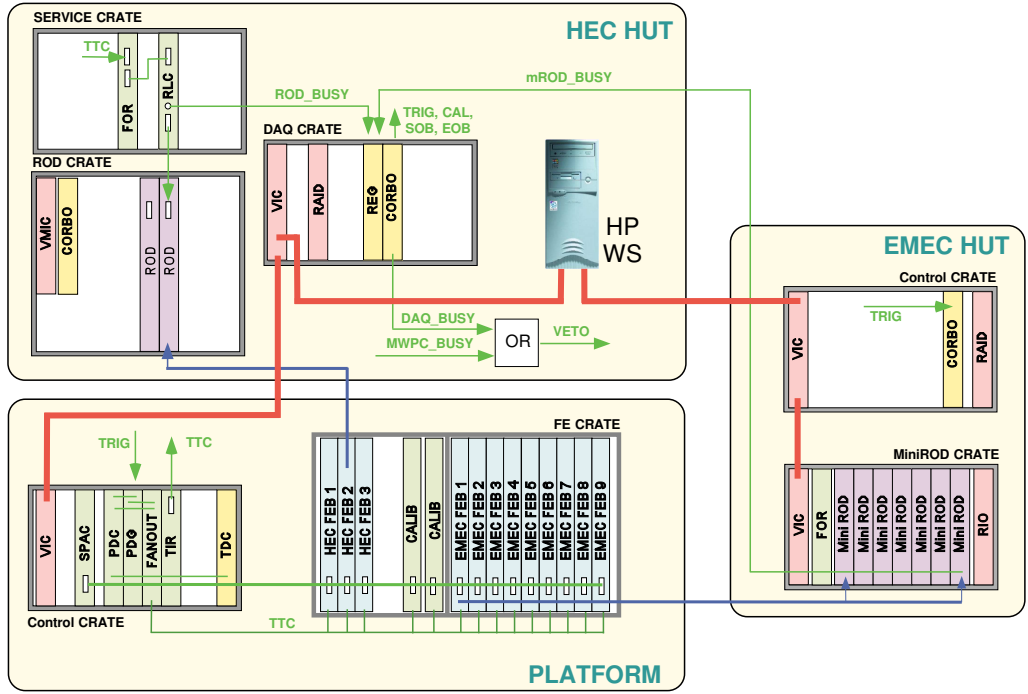,angle=0,width=1.0\textwidth}}
%    \psfull
  \end{center}
  \caption{Block diagram of the front-end, read-out and trigger electronics.}
  \label{nim_2_readout_diagram}
\end{figure}

The triggering and the synchronization of the nine EMEC FEB's and the
three HEC FEB's as well as the nine MINI-ROD's, two ROD-demo modules
and two calibration boards was done using the TTC-0 system as employed
in the EMEC stand-alone tests~\cite{r-emec}. In addition, a simple
level converter adapted the TTC-0 signals to the ROD-demo
standard. The relative timing of the pre-trigger (see~\cite{r-hec} for
details) with respect to the $40\,\MHz$ sampling clock was logged by a
TDC module. The beam trigger scintillation counters and the related
fast logic as well as the multiwire proportional chambers used for
tracking the beam particles were identical to those previously used in
the HEC stand-alone test~\cite{r-hec}.  The block diagram of the
front-end, read-out and trigger electronics is shown in
Fig.~\ref{nim_2_readout_diagram}.

The data acquisition system included a 6U control crate with the TTC-0
system and the TDC module, located near the front-end crate on the
cryostat platform. The 6U crate with the MINI-ROD modules and the 6U
service crate were placed in the EMEC control room. All other
components of the system, the 9U crate with the ROD-demo boards, the
CAMAC crate with the MWPC read-out, the 6U crate with the registers
for the scintillation counters and the event builder (CES RAID module
with a RISC CPU, running the real time system CDC EP/LX) were in the
HEC control room.

  \subsection{Calibration and Signal Reconstruction}
  %auto-ignore

The signal was sampled every $25\,\ns$. Five samplings were typically
used to reconstruct the full amplitude. For the EMEC typically 7
samplings have been read-out. For the HEC, where due to the larger
read-out cell capacities in comparison to the EMEC a larger noise is
expected, usually 16 samplings have been used. Here the first
samplings were preceding the pulse, so the noise could be
reconstructed the same way as the signal amplitude.  But for signal
shape studies even a larger number of samplings has been used.

The hardware calibration system used was the same as in the previous
HEC stand-alone beam test runs~\cite{r-hec}.  For the HEC the
calibration pulse is injected directly at the pad, for the EMEC at the
motherboards. Therefore the procedure used for the signal
reconstruction and the prediction of the calibration parameters was
different for the HEC and EMEC and is described briefly here.

Because the optimal filtering method for amplitude
reconstruction~\cite{clel1} will be used in ATLAS, we followed the
same procedure in the beam test calibration.  A detailed knowledge of
both the amplitude response and the waveform dependence on the
amplitude is needed.  The goal of the optimal filtering method is to
reconstruct the amplitude and time of a signal with a known signal
shape from discrete measurements of the signal. Thus the method
minimizes the noise contribution to the amplitude reconstruction.

To unfold the particle signal shape from the calibration signal shape
an inverse Laplace transformation has been used for the HEC and a
Fourier Transformation for the EMEC. This choice is mostly driven by
the difference in the injection of the calibration pulse. For each
method we stay within the traditional set of variables,
e.g. $s=j\omega$, with $\omega$ being the angular frequency.

\subsubsection{HEC calibration}
For the HEC signal reconstruction a new procedure, compared to
previous beam test runs~\cite{r-hec,r-calhec}, 
was applied, using the detailed
knowledge of the electronic chain.  The full response function in the
beam test set-up can be written in the frequency domain
(multiplicative constants not included) as ($\tau_x$ are various time
constants and $\alpha$ is the parameter of the calibration current
shape $I^{\rm cal}(t) = \alpha + (1-\alpha)e^{-t/\tcal}$):
\begin{eqnarray}
H(s) &=&  \frac{\alpha + s\tau_{\rm cal}}{s\cdot(1 + s\tau_{\rm cal})}\cdot
 \frac{(1+s\tau_{\rm zc})(1+s\tau_{\rm sl})}
   {(1+s\tau_{\rm pc})(1+s\tau_{\rm 0c})}\times \nonumber \\
 & &\frac{1}{(1+s\tau_{\rm a})(1+s\tau_{\rm d})}\cdot
 \frac{(1+s\tau_{\rm zs})}{(1+s\tau_{\rm ps})(1+s\tau_{\rm 0s})}\times \nonumber \\ 
 & & \frac{(1+s\tau_{\rm pz})}{(1+s\tau_{\rm i})(1+s\tau_{0})}\cdot
\frac{s\tau_{\rm s}}{(1+s\tau_{\rm s})^3\cdot(1+s\tau_{\rm fd})\cdot(1+s\tau_{\rm ac})}, \nonumber
\end{eqnarray}
where the first line corresponds to the calibration generator chain
($H^{\rm c}(s)$), the second line to the cold electronics and cables,
and the third line to the warm electronics part ($H^{\rm e}(s)$).
This function cannot be generated in time domain by directly
performing an inverse Laplace transformation (ILT). Instead we made
use of the well known method of ILT for rational functions, known as
the expansion to poles, based on the fact that any rational function
can be expanded as ($n>m$, all poles are different):
$$
\frac{(s+z_1)\cdot(s+z_2)\ldots (s+z_m)}
     {(s+p_1)\cdot(s+p_2)\ldots(s+p_n)} \equiv
\sum_{k=1}^{n}\frac{d_k}{s+p_k}.
$$
The values $d_k$ are determined by zeros and poles, and can be
calculated numerically.  Then the ILT becomes,
$$
\sum \frac{1}{s+p_k} \Rightarrow \sum{\rm e}^{-p_k\cdot t}.
$$
This method can be applied separately to the calibration signal
$H^{\rm c}(s)$ with $N_{\rm c}$ poles, and to the electronic chain response
function $H^{\rm e}(s)$ with $N_{\rm e}$ poles, so that
$$H^{\rm c}(s) = \sum_{i=1}^{N_{\rm c}} \frac{C_i}{1 + s \cdot S^{\rm c}_i} ;
 ~~~~~H^{\rm e}(s) =  \sum_{i=1}^{N_{\rm e}} \frac{E_i}{1 + s \cdot S^{\rm e}_i},$$
where $S^{\rm c}_i$ are all poles of $H^{\rm c}(s)$ and $S^{\rm e}_i$
are all poles of $H^{\rm e}(s)$.  Using the property of convolution of
two exponential functions
$$
%\int_0^t {\rm e}^{-s_1\cdot x}\cdot{\rm e}^{-s_2\cdot x} {\rm dx} \equiv
 {\rm e}^{-s_1\cdot t}\otimes{\rm e}^{-s_2\cdot t}  \equiv
\frac{{\rm e}^{-s_1\cdot t} - {\rm e}^{-s_2\cdot t}}{s_2 - s_1},
$$
the calibration signal is the combination of exponential functions:
$$
U^{\rm c}(t) = \sum_{i=1}^{N_{\rm c}} \sum_{j=1}^{N_{\rm e}}
E_{j}\cdot C_{i}\cdot \frac{{\rm e}^{-S^{\rm c}_{i}\cdot t} - 
                                {\rm e}^{-S^{\rm e}_{j}\cdot t}}
                           {S^{\rm e}_{j} - S^{\rm c}_{i}}.
$$

\begin{figure}[hbt]
\begin{center}
\includegraphics[width=0.48\textwidth,height=0.35\textheight]{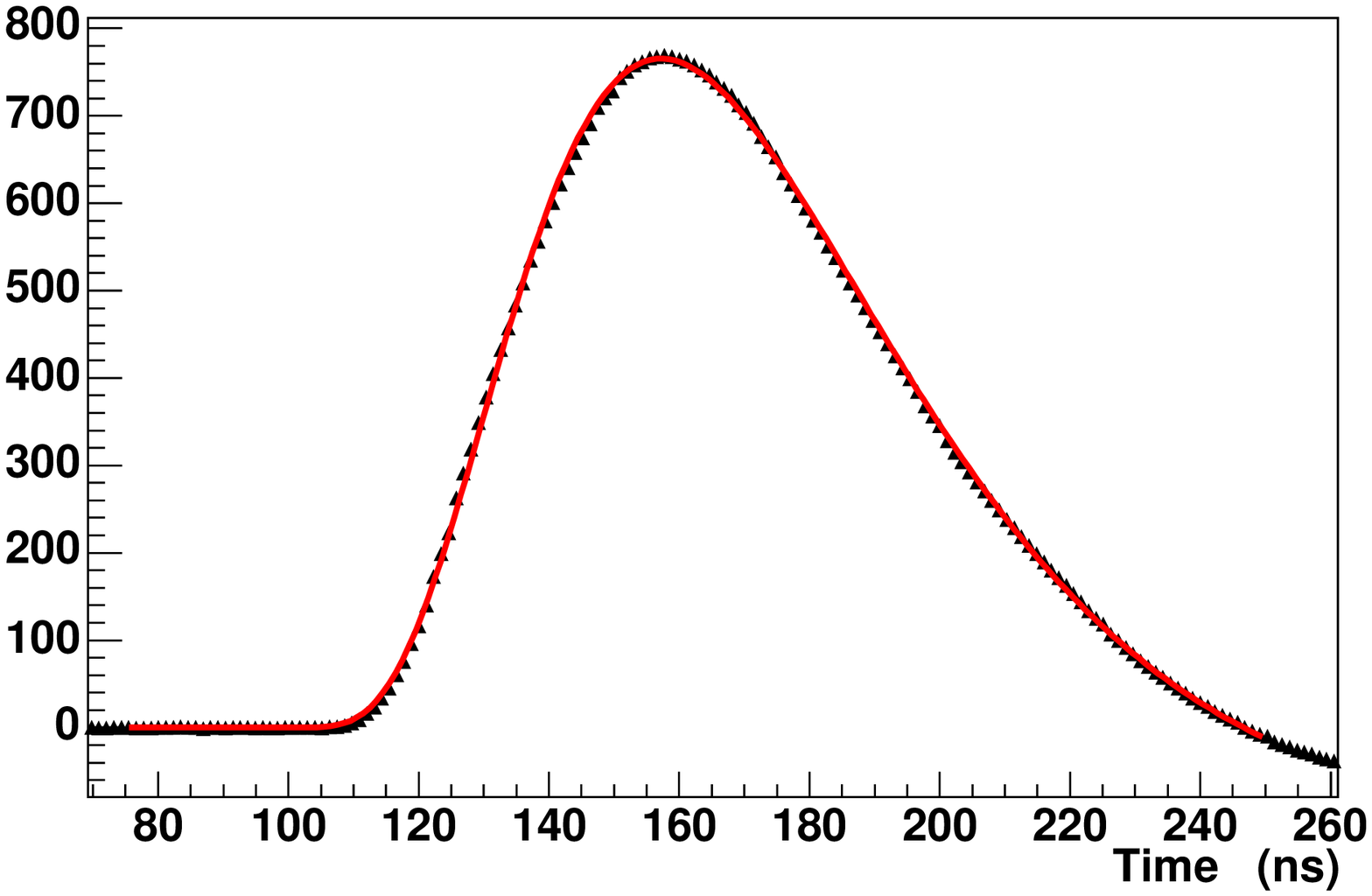}
\includegraphics[width=0.48\textwidth,height=0.35\textheight]{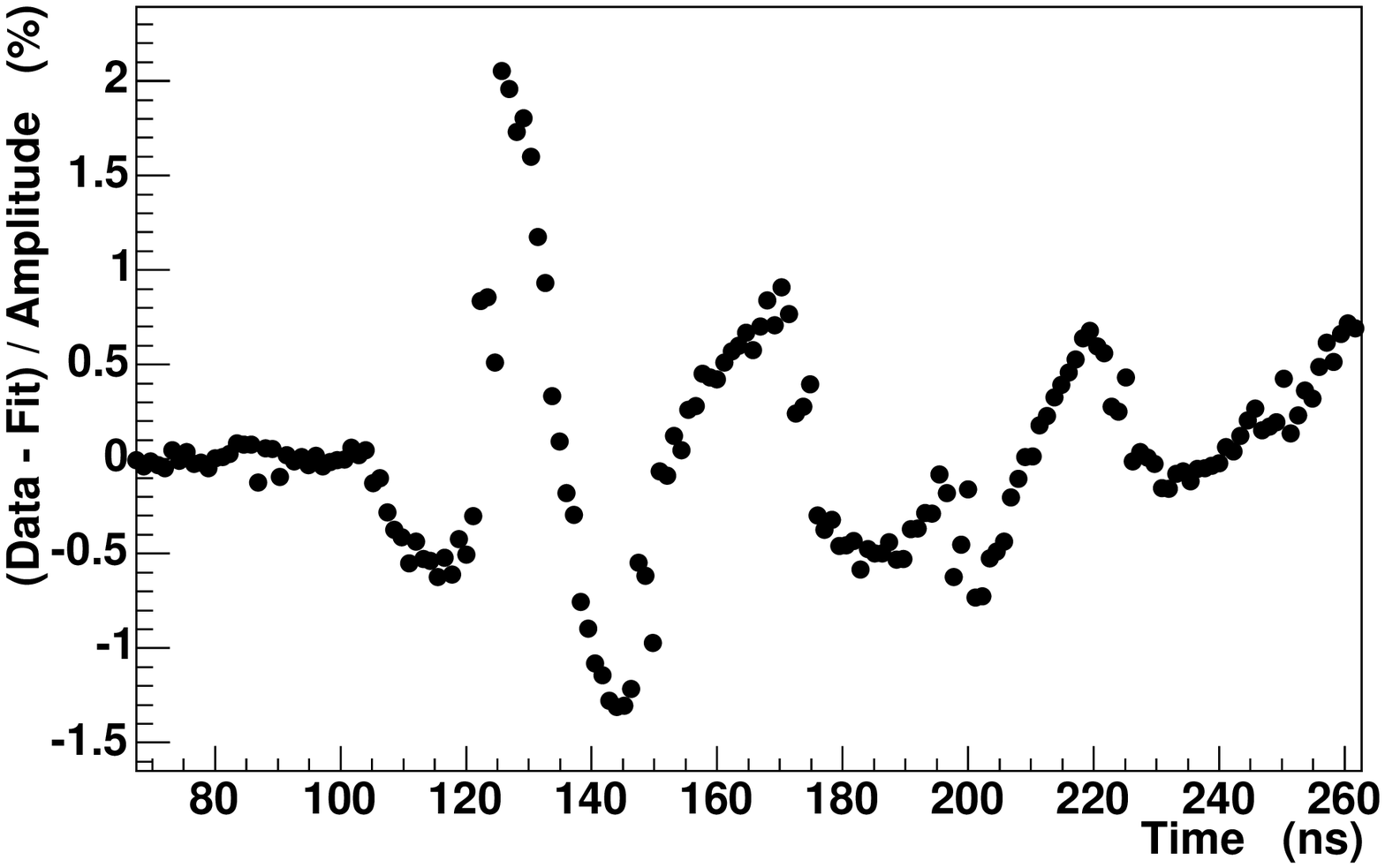}
%\mbox{\epsfig{file=nim_2_calibration1_fit.eps,width=0.40\textwidth,height=0.35\textheight},
%     \epsfig{file=nim_2_calibration1_res.eps,width=0.40\textwidth,height=0.35\textheight}}

\caption{For a typical HEC read-out channel:
Left: the measured calibration signal (points) fitted by the full 
electronics function (line). The amplitude is shown in units of ADC counts.
 Right: the residua of the fit.}
\label{nim_2_calibration1}
\end{center}
\end{figure}

The measured calibration signal shape was fitted with two free
parameters, namely the $\tau_{\rm s}$-shaper pole and $\tau_{\rm pz}$-preshaper
pole.  Other parameters are known from laboratory measurements of the
electronic chain.  We computed $C_{i}$ analytically (2 zeroes and 3
poles), and $E_{j}$ during numerical fitting resolving the system of
$n$ linear equations. To be able to use the method we approximated by
substituting the triple pole for the shaper by three single poles with
very close time constants ($\pm0.1\,\ns$). The result of the fit for
one HEC channel is shown in Fig.~\ref{nim_2_calibration1}.
The left figure shows the calibration signal (points) fitted by the
full electronics function (red line). The right figure shows the
corresponding residua of the fit. The residua are well within
$\pm1.5\,$\%, except for the signal rise, where the influence of small
distortions of the calibration pulse shape is not taken into account.
 
\begin{figure}[hbt]
\begin{center}
\includegraphics[width=0.95\textwidth,height=0.45\textheight]{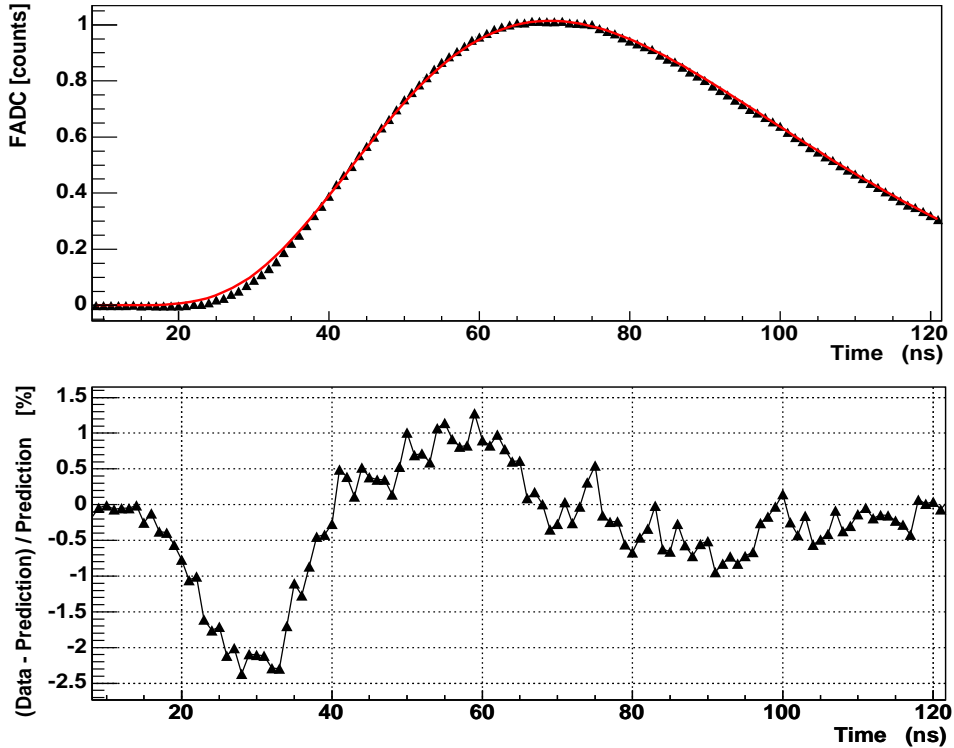}
\caption{Upper figure: comparison of the measured normalized particle signal
(points) with the prediction (line) using the parameters as obtained from the
fit of the calibration signal. Lower figure: corresponding residua.}
\label{nim_2_calibration2}
\end{center}
\end{figure}

The fitted parameters of the response function were used to predict
the ionization current by convolution of $H^{\rm e}(s)$ with the
triangle current where the only free parameter is the drift time. The
predicted function was used to compute the optimal filter
weights~\cite{clel1}.  The comparison of the pion data with the
predicted particle shape for the same HEC channel is shown in
Fig.~\ref{nim_2_calibration2}.
  
The upper figure shows the comparison of the measured normalized
particle signal (points) with the prediction (line) using the
parameters as obtained from the fit of the calibration signal. The
lower figure shows the corresponding residua.  Apart from near the
signal start, where the influence of calibration pulse imperfections
is seen, the shape is predicted again rather well (residua within
$\pm1.5\,$\%).  In summary, the precision of the resulting signal
reconstruction of real particles, is at the level of $\pm1\,$\%.

Finally the transformation of the measured amplitude from voltage to
current was done using a 3$^{\rm rd}$ order polynomial or a linear
function. The coefficients of the polynomial were computed in two
steps:
\begin{itemize}
\item
 the calibration signals with different currents were transformed to
 triangular signals corresponding to the same currents;
\item
 the resulting amplitudes were fitted by the calibration polynomial to
 obtain the coefficients $p_i$ for the transformation of the amplitude
 $A$ to current $I$: $ I=\sum_{i=0}^{n}p_i\cdot A^i $, with $n=3$
 typically.
\end{itemize}

\subsubsection{EMEC calibration}
In comparison to the HEC, for the EMEC the electronic paths of the
calibration signal and of the particle-induced signal are more
complicated, and therefore a different approach was used to calibrate
the EMEC.  In addition to the main difference between the calibration
pulse shape (exponential function) and the particle shape (triangle),
there is also a substantial inductive and capacitive difference
between the two paths.

The particle pulse shape can be predicted from the calibration pulse
shape by applying two transformations, one for the current signal shape and
the other for the injection point:
\begin{equation}\label{eq:Hcur}
          H^{\rm cur}(s) = \frac{I^{\rm phy}(s)}{I^{\rm cal}(s)} = 
          \frac{(st_{\rm dr} +e^{-st_{\rm dr}}-1)(1+s\tcal)}{st_{\rm dr}(\alpha + 
s\tcal)},
\end{equation}
\begin{equation}\label{eq:Hlc}
          H^{\rm lc}(s) = \frac{H^{\rm phy}(s)}{H^{\rm cal}(s)} =
          \frac{1}{1+sRC+s^2LC},
\end{equation}
where $t_{\rm dr}$ is the drift time, $\tcal$ and $\alpha$ are
parameters of the calibration current shape, and $I^{\rm cal}(t) = \alpha
+ (1-\alpha)e^{-t/\tcal}$.

\begin{figure}[h]
\begin{center}
\includegraphics[width=.8\textwidth]{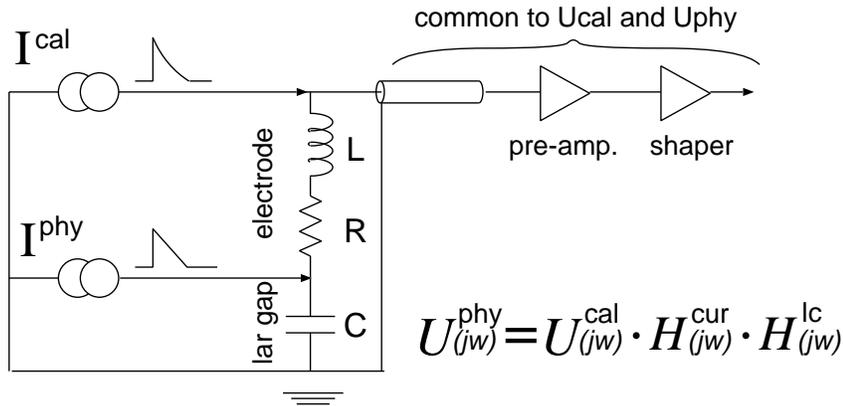}
\caption{Schematic EMEC read-out chain and calibration network.
The variables $U^{\rm phy} (j\omega)$ and $U^{\rm cal} (j\omega)$
are the  pulse shapes in the frequency domain for the physics and calibration
pulses, respectively.}
\label{nim_2_calibration4}
\end{center}
\end{figure}

In this calculation, we assume that the calorimeter read-out pad
behaves as a simple capacitor $C$, and that the strip-line on the
detector between the pad and the summing board (where the calibration
signal is injected) can be modeled with a series inductor $L$, and
series resistor $R$. The model circuit used in the method is shown in
Fig.~\ref{nim_2_calibration4}.

The method is based on a technique developed previously for the ATLAS
electromagnetic calorimeters~\cite{lappmethod,delmastro}.  The
particle-induced signal pulse shape is predicted from the calibration
pulse shapes using the Fast Fourier Transform\,(FFT) in the following
steps :
\begin{itemize}
\item Signal pulse shapes are obtained by averaging many pulses in each channel
in high energy pion and electron runs.  Since the sampling time in
each channel is $25\,\ns$ but $1\,\ns$ bins are used in the pulse
shapes, there is a set of bin-to-bin statistical normalization
constants, which are initially assumed to be unity;
\item Extrapolate the calibration pulse (800 time samples) smoothly by using 
an exponential function to avoid the discontinuity at the edges of the
time window, expanding the time window to 2048 time samples;
\item Apply a FTT to this expanded calibration pulse, to transform it 
into the frequency domain;
\item Apply the two transformations to the calibration pulse shape in 
the frequency domain, and then convert it to the time domain with the
FFT.  This is the ``predicted physics signal'' in the time domain;
\item Calculate the free parameters ($LC$, $RC$ and the drift-time) using
$\chi$-square minimization.
\end{itemize}
The above calculation was repeated for different signal pulse shape
bin-to-bin normalizations until the fit converged.

This procedure provides a pulse shape for each channel from the
calibration signal that is suitable for use in providing optimal
filtering weights for analyzing physics pulses in the channel.  More
importantly, it also gives predictions for $LC$ and $RC$, which are
needed to compute the absolute normalization of the signal pulses in
units of ionization current using the calibration system.  All cells
were not scanned during the beam test runs, and therefore there are
some cells without valid signal pulse shapes.  Signal pulse shapes
from cells with similar characteristics are used for cells without a
signal pulse shape.

\begin{figure}[h]
\begin{center}
\includegraphics[width=.95\textwidth]{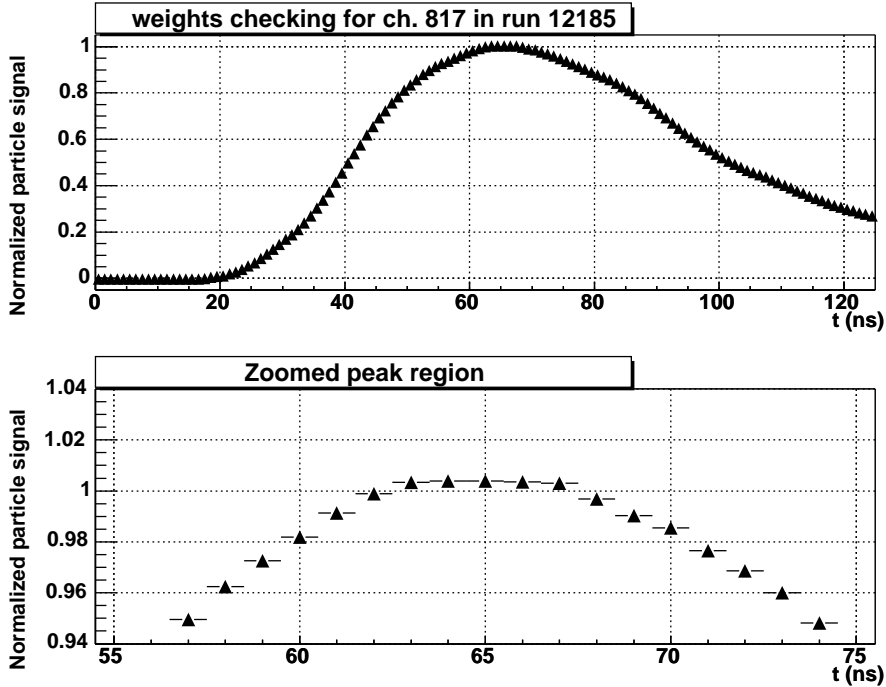}
\caption{Upper figure: normalized particle signal, using optimal 
filter weights computed from the calibration signal. Lower figure: close up
of the peak region.}
\label{nim_2_calibration5}
\end{center}
\end{figure}

 The optimal filtering coefficients are applied to 5 pulse samples
with $25\,\ns$ spacing. Because the beam particles arrive at the
calorimeter asynchronously with respect to the $40\,\MHz$ clock, only
a fraction of the pulses are actually sampled near the peak. The
expected reconstructed pulse height is defined to be that of an ideal
continuous pulse passing through these samples. The quality of the
signal reconstruction is assessed by taking a large sample of events
with hits in a channel, and plotting the average value of the sample
in a given time bin normalized event-by-event to the reconstructed
pulse height.  Because this average pulse is reconstructed for many
pulses with different sampling times, the complete average pulse can
be reconstructed with fine time bins, and is shown for one example
channel in Fig.~\ref{nim_2_calibration5}.  If the optimal filtering
weights are correct, the observed height of this normalized plot
should be unity.  The check verifies that the pulse height is
reconstructed to better than $1\,$\% accuracy.

%% circuit
%% check

  \section{Data Analysis}
  \subsection{Data}
\begin{figure}[htb]
  \begin{center}
%    \psdraft   
    \mbox{\epsfig{file=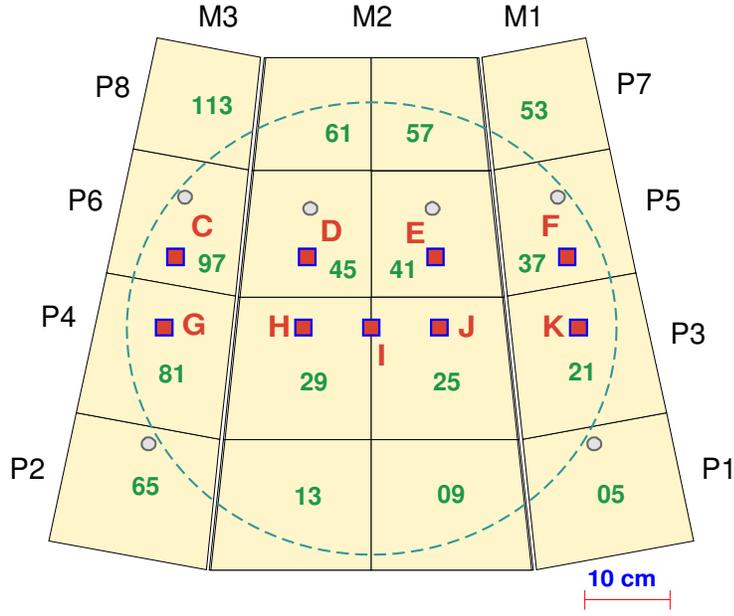,width=0.8\textwidth}}
%    \psfull
  \end{center}
  \caption{Position of the beam impact points (C-K) on the front face of HEC1.
   The large circle indicates the beam window; the small circles the position
    of the tie rods}
  \label{nim_3_data}
\end{figure}

  %auto-ignore
In total 743 runs have been taken with electrons, pions or muons in
the energy range $6\,\GeV\le E\le 200\,\GeV$ with about 25 million
triggers in total.  The data for the energy scans have been taken
typically at 9 beam impact points. They are shown in
Fig.~\ref{nim_3_data}, projected onto the front face of the three HEC1
modules.  Shown is also the related pad structure, the projection of
the beam window (circle) and the position of the calorimeter tie rods
(open circles).
The lower row (G-K) corresponds roughly to an $\eta$-range of $1.51$
(PS) - $1.65$ (layer 3) for the EMEC and to $1.67$ (layer 1) - $1.87$
(layer 3) for the HEC, the upper row (C-F) to $1.55$-$1.67$ (EMEC) and
to $1.72$-$1.92$ (HEC) respectively.  Horizontal and vertical scans
have been performed for all particle types and at various beam
energies. In the region of the cryostat window the amount of dead
material in front of the calorimeter is minimal, about $0.6\,X_0$ in
total.  To study the variation of the response in the presampler, EMEC
and HEC modules with dead material in front, dedicated scans have been
done introducing varying amounts of additional dead material in front
of the cryostat.

  \subsection{Electronic Noise}
\begin{figure}[htb]
  \begin{center}
%    \psdraft   
    \mbox{\epsfig{file=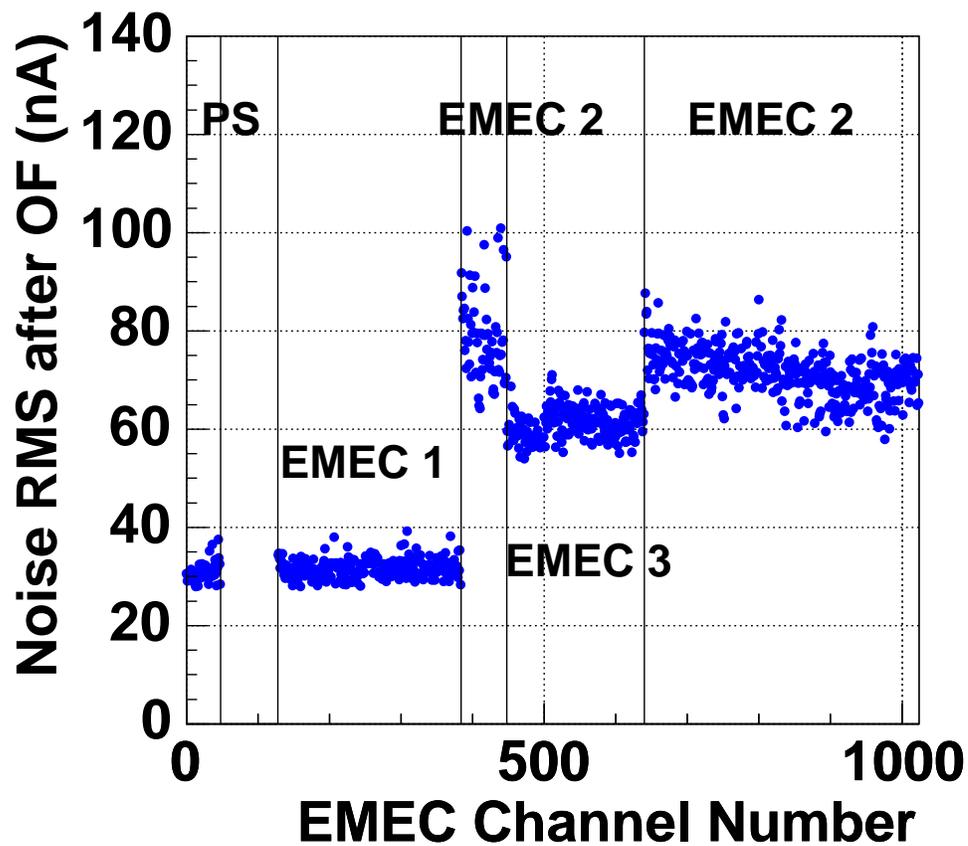,width=1.0\textwidth}}
%    \psfull
  \end{center}
  \caption{Electronic noise in the EMEC: shown is the RMS noise of the pedestal
   (after optimal filtering) for the individual read-out channels}
  \label{nim_3_noise_emec}
\end{figure}

\begin{figure}[htb]
  \begin{center}
%    \psdraft   
    \mbox{\epsfig{file=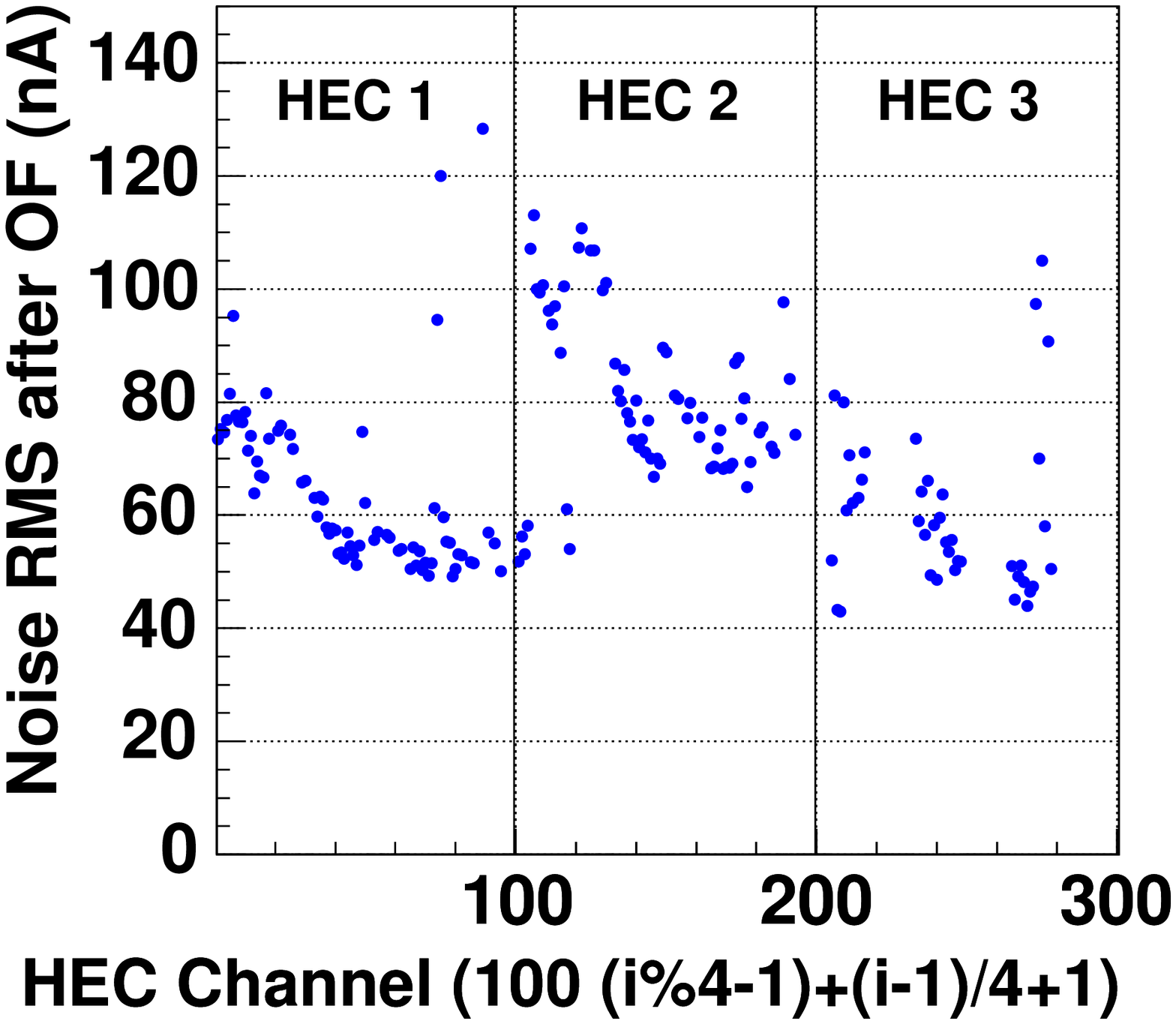,width=1.0\textwidth}}
%    \psfull
  \end{center}
  \caption{Electronic noise in the HEC: shown is the RMS noise of the pedestal
   (after optimal filtering) for the individual channels.
   A special channel numbering has been used to map the
   read-out channels $\rm {i}$ to the layer structure of the HEC.}
  \label{nim_3_noise_hec}
\end{figure}

\begin{figure}[htb]
  \begin{center}
%    \psdraft   
    \mbox{\epsfig{file=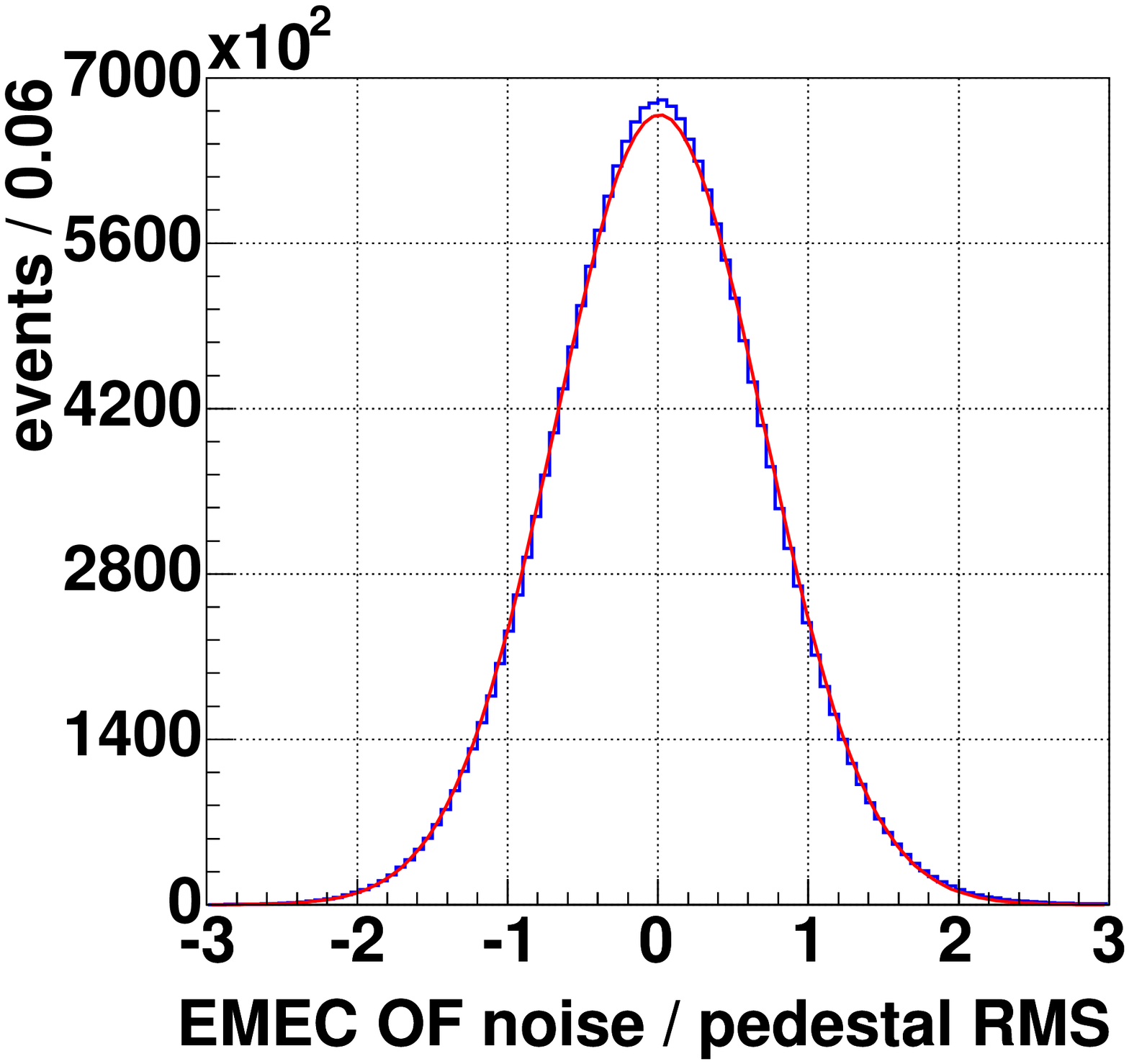,width=0.6\textwidth}}
%    \psfull
  \end{center}
  \caption{Pedestal subtracted signal of individual EMEC channels
  with no energy deposition as obtained from
  OF reconstruction. This noise is shown in units of the
  pedestal RMS. The histogram shows the data; the line is a gaussian
  fit to the distribution.}
  \label{nim_3_noise_emec_OF}
\end{figure}

  %auto-ignore
The optimal filtering (OF) method minimizes the noise-to-signal ratio
for the signal reconstruction using the known particle signal shape
and the resulting noise autocorrelation
weights. Figs.~\ref{nim_3_noise_emec} and~\ref{nim_3_noise_hec} show
the RMS noise of the pedestal (using optimal filtering) for the
individual read-out channels of the EMEC and HEC. For the HEC a special channel
numbering has been used (given in the Figure) to map the read-out channels
$\rm {i}$ to the layer structure of the HEC.  The variation of the
noise is as expected from the related channel capacitance. For the
EMEC the groups of channels in layer 0 (presampler) and layer 1, 2 and
3 are clearly visible. For the HEC the capacitance variation that
depends on $\eta$ can be seen. It is structured into individual bands
related to the corresponding longitudinal segment. For more details
see~\cite{Manuella}.
For the EMEC data only the very first time sample is preceding the
signal pulse.  Therefore the impact of the OF method on the noise has
been studied using the muon data.  Here only the channels hit directly
by the muon had to be excluded.  Fig.~\ref{nim_3_noise_emec_OF} shows
the corresponding result for the EMEC channels. The
noise is normalized to the related pedestal RMS of the corresponding
channel. From the fit a $\sigma$ of about $69\,$\% of the pedestal RMS
was obtained, showing that the OF result is somewhat worse than
theoretically expected (see HEC below). The quality of the prediction of the
signal pulse shape as well as residual coherent noise seem to set some
limitation.
\begin{figure}[htb]
  \begin{center}
%    \psdraft   
    \mbox{\epsfig{file=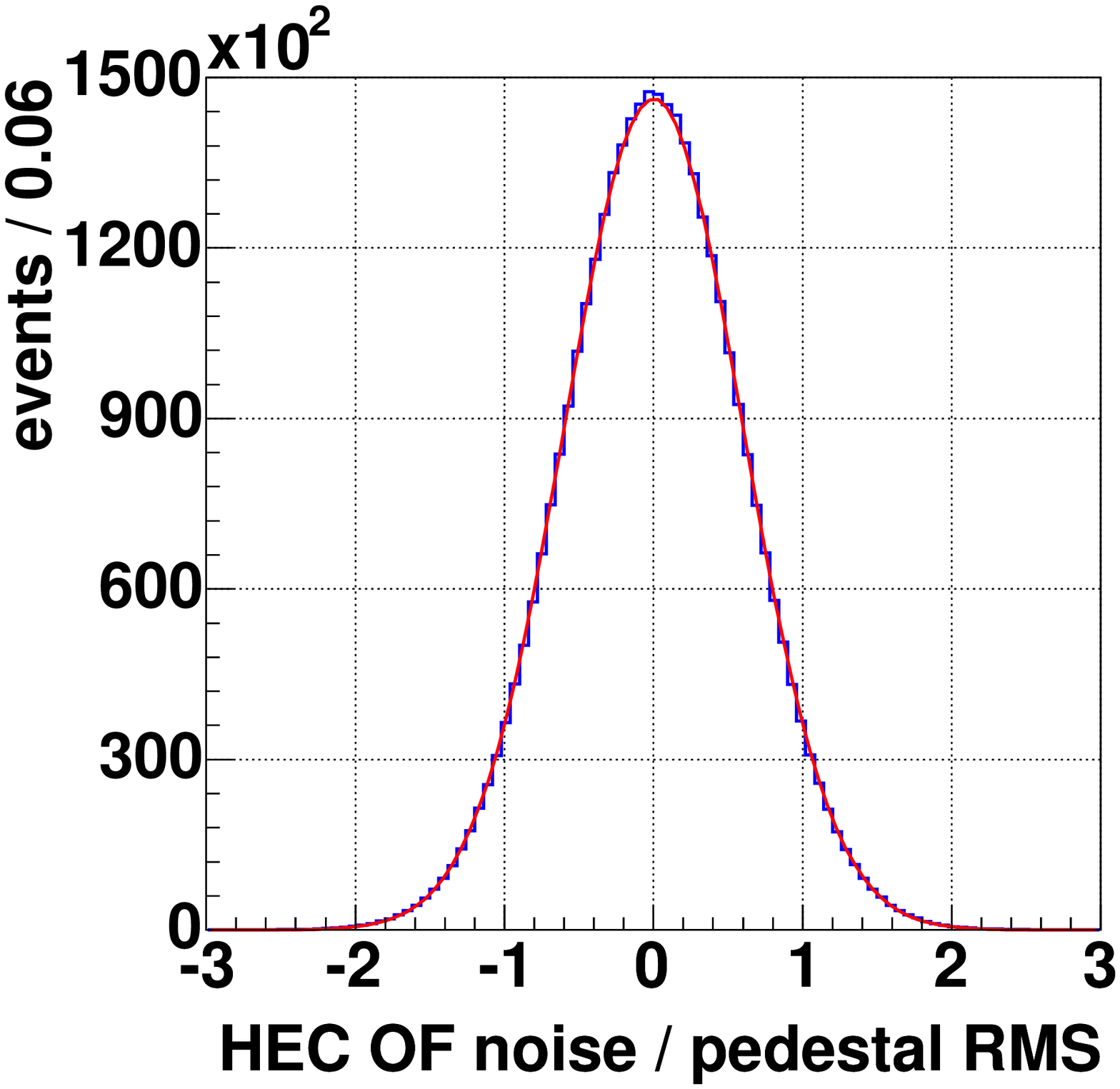,width=0.6\textwidth}}
%    \psfull
  \end{center}
  \caption{Pedestal subtracted signal of individual HEC channels
  with no energy deposition as obtained from
   OF reconstruction. This noise is shown in units of the
   pedestal RMS. The histogram shows the data; the line is a gaussian
   fit to the distribution.}
  \label{nim_3_noise_hec_OF}
\end{figure}  
For the HEC data, where the first time samples are preceding the
signal pulse, the first five time samples have been used to obtain the
noise on an event by event basis after applying the OF
weights. Fig.~\ref{nim_3_noise_hec_OF} shows the pedestal noise of
individual HEC channels as obtained from the OF reconstruction.  The
noise obtained is given in units of the pedestal RMS. The histogram
shows the data, the line is a gaussian fit to the distribution. From
the fit a $\sigma$ of about $60\,$\% of the pedestal RMS can be
obtained, demonstrating a noise reduction by the OF method of
$\sim1.7$. This is exactly the value theoretically expected for five
samplings used in the amplitude reconstruction. It also shows that any
coherent contribution to the noise is very low, and that the signal
shape reconstruction is well modeled.

  \clearpage
  
  \subsection{Online Monitoring and Offline Software}
  %auto-ignore
For this EMEC/HEC combined run the analysis and monitoring software
were integrated into the ATLAS object-oriented C++ ATHENA framework in
contrast to the previously used software~\cite{r-hecadc}.  This
software package includes tools for decoding the online data, building
data 'objects', storing them in the ATHENA transient data store, and
finally developing all analysis tools within the ATHENA framework.
This new software package has many benefits.  In particular, the
software being developed for beam test calibration, monitoring and
analysis will be directly available for analysing eventual ATLAS data,
encouraging the transfer of expertise from the detector groups leading
the beam test effort into the offline software and analysis. The
implementation of the beam test analysis into ATHENA, its first
application to real data, is currently driving a number of significant
modifications to the ATLAS software and analysis framework. These
modifications include many features allowing ATHENA to cope with real
conditions such as imperfect time varying data with changing
calibration constants and detailed event data content.  The EMEC/HEC
combined beam test was the first use of ATHENA for beam test analysis,
and the expertise gained in this effort is being carried forward into
upcoming beam tests and offline analysis.

 \subsection{Global Event Timing} 
  %auto-ignore
In contrast to the ATLAS situation at the LHC, the trigger in the beam
test is asynchronous with respect to the $40\,\MHz$ clock. The
relative timing between the clock and the trigger can be defined using
either the TDC information or it can be derived from global event
timing. From a 3$^{\rm rd}$ order polynomial fit to the signal shape
of an individual channel $i$ with a significant pulse height the
`cubic time' $t_i$ can be derived. Mainly due to differences in
cabling the timing of individual channels varies by up to $\pm10\,\ns$
for the EMEC and $\pm3\,\ns$ for the HEC. These channel offsets have
been defined from a global fit minimizing the channel-to-channel time
differences for all EMEC and HEC channels over all electron and pion
runs. Knowing these offsets $\Delta t_i$ finally a 'global cubic time'
for each event can be determined with:

$$t_{\rm global} =  \frac{\Dy\sum_{i}{\left[ t_i  -  \Delta t_i\right]/\sigma_{t,i}^2 }}{\Dy\sum_{i}{1/\sigma_{t,i}^2}},\qquad {\rm with}\quad
\sigma_{t,i} = \frac{24.1\,\ns}{{\rm signal}_i/\sigma_{{\rm noise},i}} \oplus 0.55\,\ns.$$

\begin{figure}[htb] 
  \begin{center}
%    \psdraft   
    \mbox{\epsfig{file=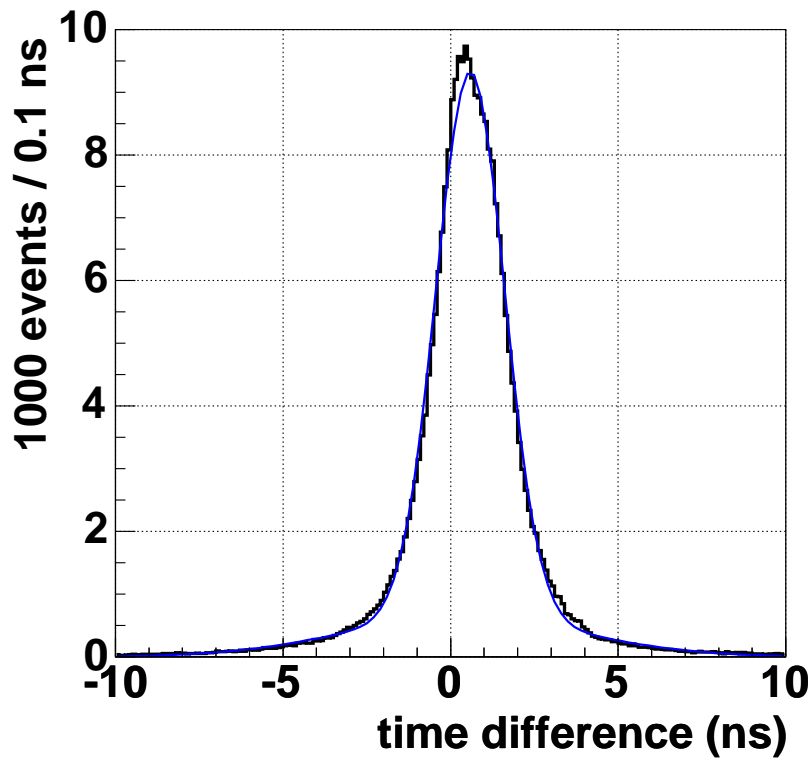,width=0.8\textwidth}}
%    \psfull
  \end{center} \caption{Time difference: comparison of the global time
  $t_{\rm global}$ with the cubic time of the largest signal channel.}
  \label{nim_3_timing}
\end{figure}

Fig.~\ref{nim_3_timing} shows the comparison of the global time
$t_{\rm global}$ thus obtained but without using the channel with the
largest signal with the cubic time of the channel with the largest
signal over all runs. The precision of the global time $t_{\rm
global}$ is better than $1.5\,\ns$. The slight offset from zero might
be partially related to the time evolution of the shower development,
which is for the hottest channel the shower core and for other
channels more in the shower tail.

  \subsection{Cluster Algorithm} 
\begin{figure}[htb]
  \begin{center}
%    \psdraft   
    \mbox{\epsfig{file=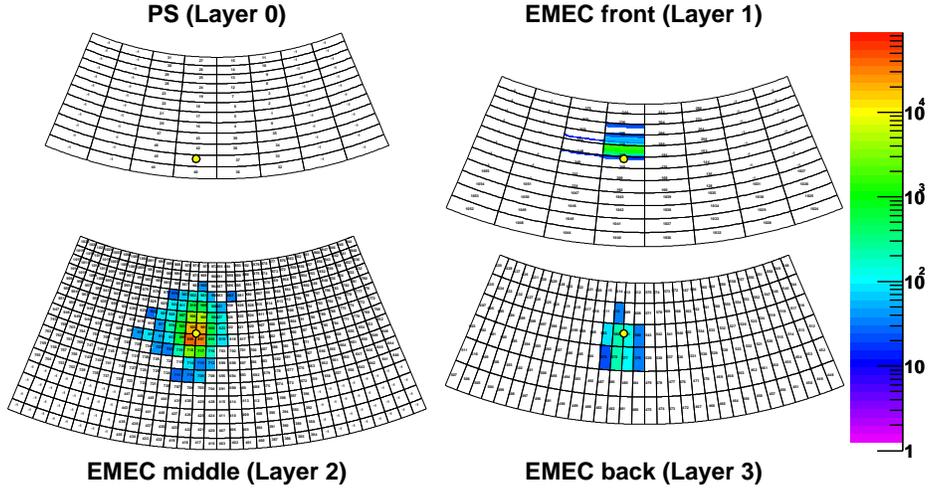,width=1.0\textwidth}}
%    \psfull
  \end{center}
  \caption{Reconstructed EMEC cluster for an electron of
  $148\,\GeV$. The yellow dot shows the beam impact point; the scale
  is given in nA.}
  \label{nim_3_cluster_elec}
\end{figure}

\begin{figure}[htb]
  \begin{center}
%    \psdraft   
    \mbox{\epsfig{file=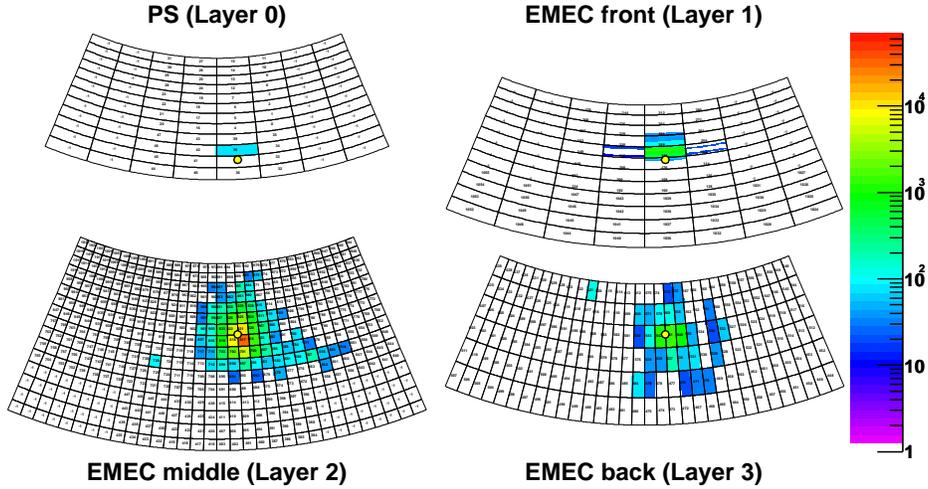,width=1.0\textwidth}}         
%    \psfull
  \end{center}
 \caption{Reconstructed EMEC clusters for a pion of $180\,\GeV$. The
  yellow dot shows the beam impact point; the scale is given in nA.}
  \label{nim_3_cluster_pion_EMEC}
\end{figure}

\begin{figure}[htb]
  \begin{center}
%    \psdraft   
    \mbox{\epsfig{file=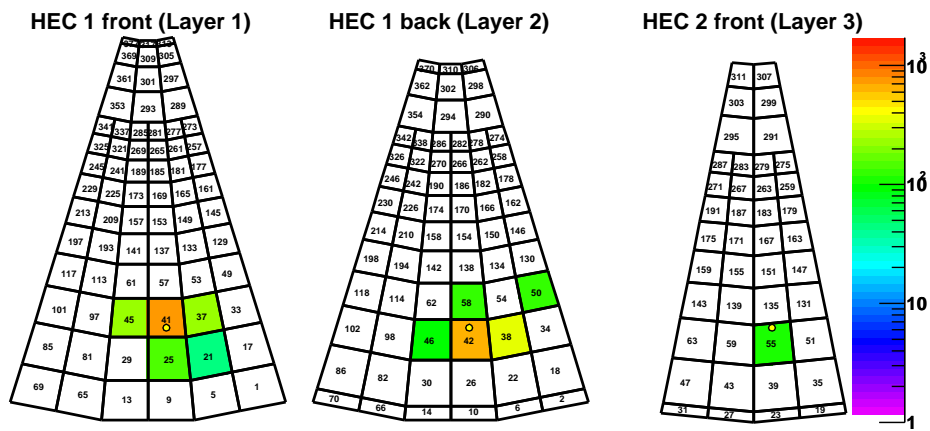,width=1.0\textwidth}}         
%    \psfull
  \end{center} 
  \caption{Reconstructed HEC clusters for a pion of
  $180\,\GeV$. The yellow dot shows the beam impact point; the scale
  is given in nA.}  
  \label{nim_3_cluster_pion_HEC}
\end{figure}

  %auto-ignore
A cluster algorithm has been developed to reconstruct the energy of
the beam particle. In each layer the topological neighbours of a given
cell are considered: they have to share at least one common corner. A
seed is chosen if the energy is significantly above the noise: $E_{\rm
seed}>4\sigma_{\rm noise}$. The general cut-off at the cell level is
$|E_{\rm cell}|>2\sigma_{\rm noise}$, the cluster is expanded with
neighbours of all cells satisfying $|E_{\rm neighbour}|>3\sigma_{\rm
noise}$. Symmetric cuts have been chosen in order to avoid major
energy offsets due to noise pick-up.  Fig.~\ref{nim_3_cluster_elec}
shows the read-out cells of a typical $148\,\GeV$ electron event: in
total there are four EMEC clusters (one per layer) and no HEC clusters
with 83 channels in total. Due to the non pointing geometry and the
different cluster algorithm used, the cluster size obtained for
electrons is bigger than the corresponding one used in the EMEC
modules 0 analysis.
For a typical pion of $180\,\GeV$ the corresponding distribution of
read-out cells selected is shown in
Figs.~\ref{nim_3_cluster_pion_EMEC} and~\ref{nim_3_cluster_pion_HEC}:
in total there are six clusters in the EMEC with 128 channels and
three HEC clusters with 11 channels.
 
The colour code indicates the related signal height of the individual
read-out cells.  The cluster algorithm proves to be very efficient for
the reconstruction of the particle energy deposited and avoids
including too much noise in the particle signal.

  \clearpage
  
  \subsection{Alignment} 
\begin{figure}[htb]
  \begin{center}
%    \psdraft   
    \mbox{\epsfig{file=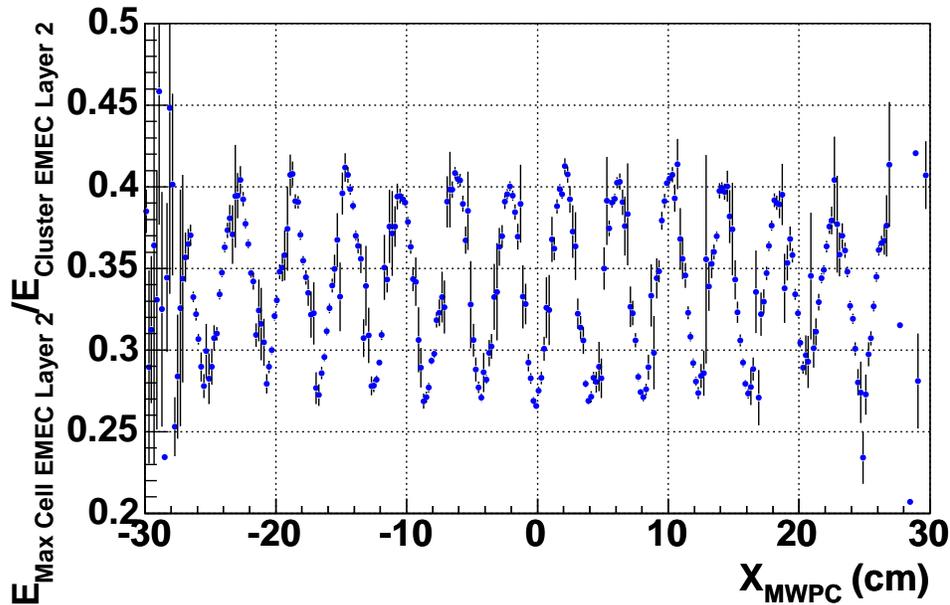,width=1.0\textwidth}}
%    \psfull
  \end{center}
  \caption{Signal response of pions of $200\,\GeV$ in the second EMEC
  layer from a horizontal scan as function of the beam impact
  position. Shown is the maximum signal in this layer with respect to
  the corresponding cluster energy.}
  \label{nim_3_alignment1}
\end{figure}

\begin{figure}[htb]
  \begin{center}
%    \psdraft   
    \mbox{\epsfig{file=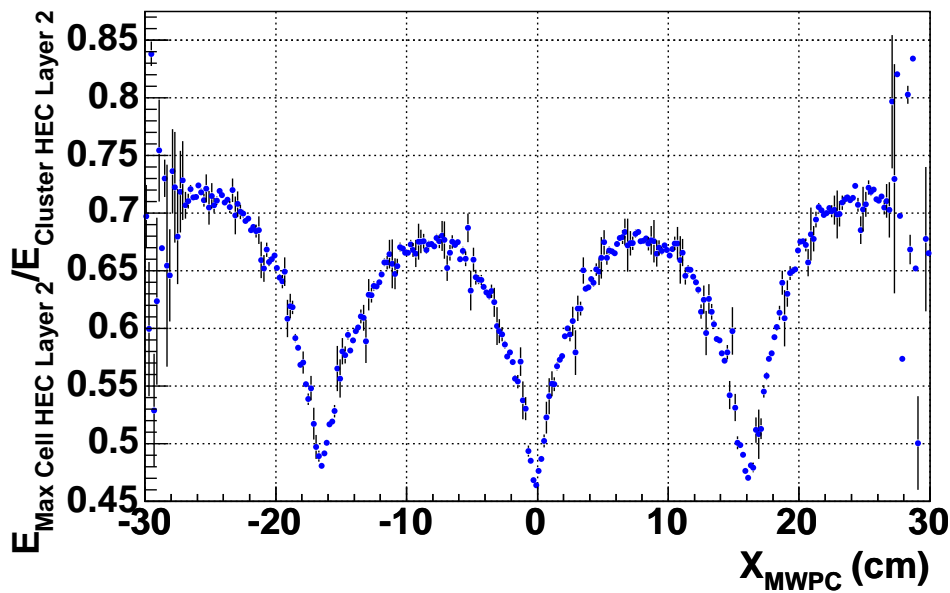,width=1.0\textwidth}}
%    \psfull
  \end{center}
  \caption{Signal response of pions of $200\,\GeV$ in the second HEC
  layer from a horizontal scan as function of beam impact
  position. Shown is the maximum signal in this layer with respect to
  the corresponding cluster energy.}
  \label{nim_3_alignment2}
\end{figure}

  %auto-ignore
In ATLAS the three sub-detectors EMEC, HEC1 and HEC2 form three
wheels, which are placed in one common cryostat. In the beam test
set-up the relative positions of the three sub-detectors follow the
ATLAS dimensions within tolerance (typically tolerances are
$<3\,\mm$). Nevertheless, to be able to simulate details of the shower
formation and cluster reconstruction, a precise horizontal and
vertical alignment of the sub-detectors is required.  Using the track
reconstruction based on the MWPC data, the signal response in the
EMEC, HEC1 and HEC2 can be closely followed in horizontal and vertical
scans. Module and pad boundaries as well as tie-rod positions can be
used to extract the required information with high precision. As an
example, Figs.~\ref{nim_3_alignment1} and~\ref{nim_3_alignment2} show
the relative response for pions of $200\,\GeV$ in the second EMEC
layer and in the second HEC1 layer when running a horizontal
scan. Plotted is in each case the ratio between the maximum signal
cell response and the corresponding cluster energy.  The pad
boundaries are clearly visible. Using the combined information of all
these data, the precision of the relative alignment is typically at
the level of $\pm1\,\mm$. In Fig.~\ref{nim_3_alignment2} the lateral
energy leakage is visible when the beam impact approaches the outer
modules: the related cluster energy is reduced.

  \subsection{High Voltage Corrections for Central HEC1 Module}
\begin{figure}[htb]
  \begin{center}
%    \psdraft   
    \mbox{\epsfig{file=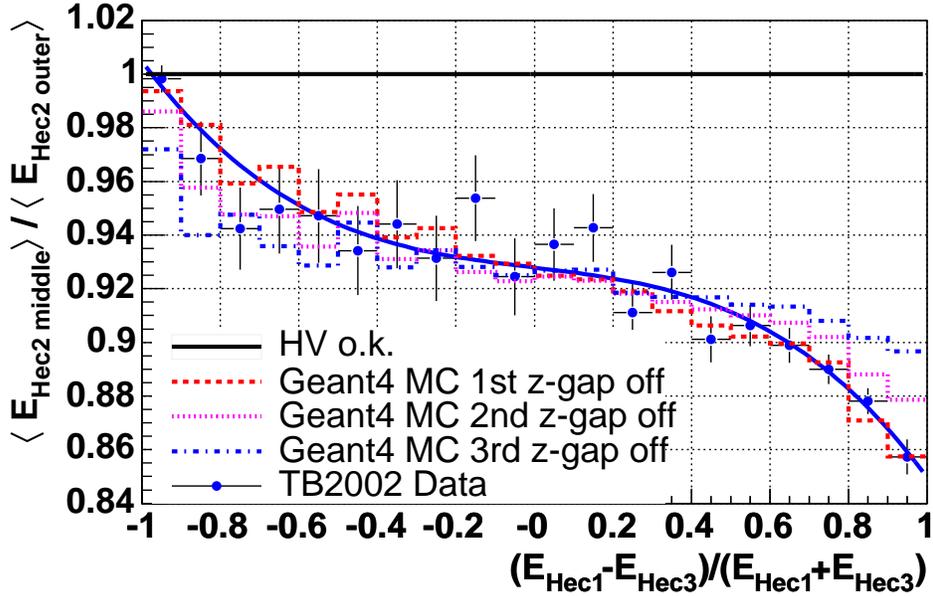,width=1.0\textwidth}}
%    \psfull
  \end{center}
  \caption{Ratio of the energy deposited for pions of $200\,\GeV$ in
           the 2$^{\rm nd}$ longitudinal segment of the central HEC1
           module with respect to the outer HEC1 modules as function
           of the asymmetry of the total energy depositions in the
           1$^{\rm st}$ and 3$^{\rm rd}$ (last) longitudinal segments}
  \label{nim_6_hvcorrections1}
\end{figure}

\begin{figure}[htb]
  \begin{center}
%    \psdraft   
    \mbox{\epsfig{file=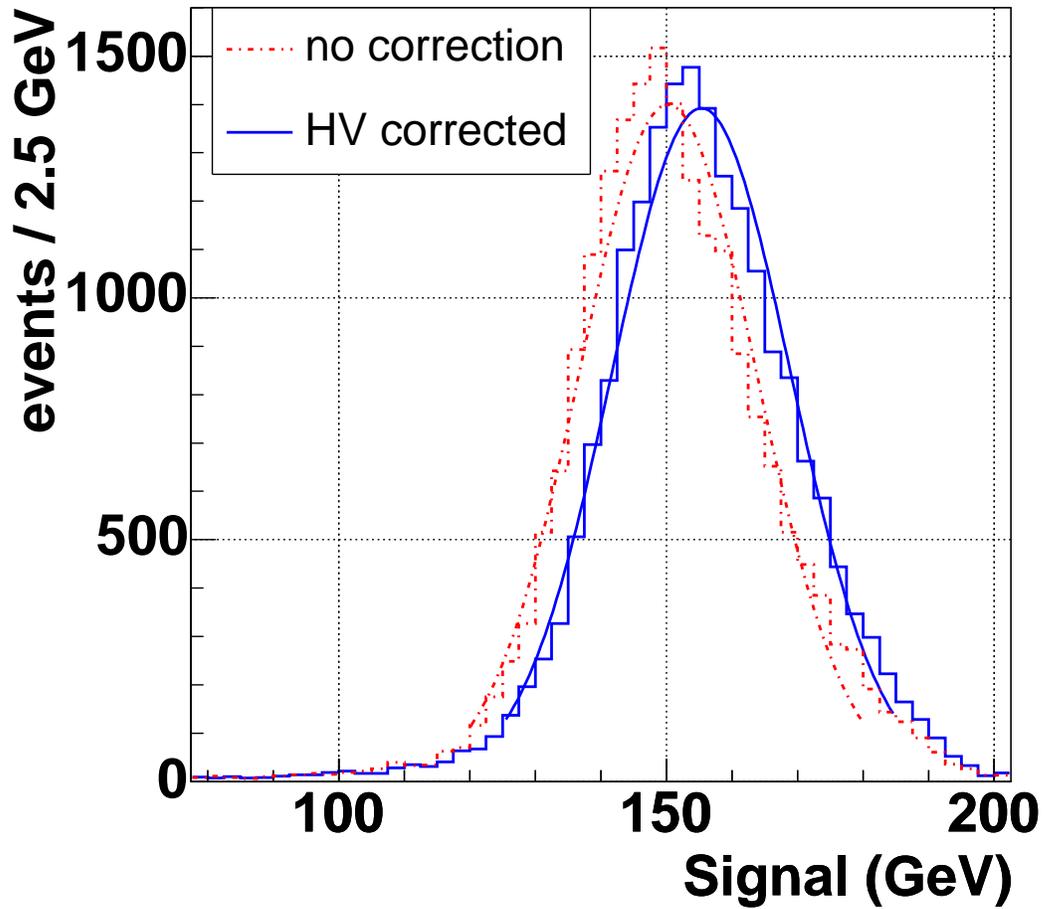,width=1.0\textwidth}}
%    \psfull
  \end{center}
  \caption{Energy deposition and resolution for pions of $200\,\GeV$
           at the impact point of $x=-10\,\cm$ before and after the HV
           correction for the central HEC1 module}
  \label{nim_6_hvcorrections2}
\end{figure}

  %auto-ignore
In the second longitudinal segment of the central HEC1 module during
the run one of the 4 HV lines, which are feeding the four sub-gaps of
this section, had to be disconnected because of a short. The related
correction 4/3 has been applied to those data. As the signal is still
measured correctly in three out of four sub-gaps, all fluctuations of
the hadronic shower are measured correctly as well. The main
consequence due to this HV short is a reduced signal to noise ratio
only.  The horizontal scans with pions revealed a particular problem
in the second longitudinal segment of the central HEC1 module: in
particular for pions that started early showering, the response was
weaker than normally. Fig.~\ref{nim_6_hvcorrections1} shows for
$200\,\GeV$ pions from a horizontal scan the ratio of the average energy
deposited in the 2$^{\rm nd}$ longitudinal segment of the central HEC1
module over the average energy deposited in the same longitudinal segment
but in the outer HEC1 modules and for different beam impact points, as
function of the
asymmetry of the total energy depositions in the 1$^{\rm st}$ and
3$^{\rm rd}$ (last) longitudinal segments (the last longitudinal
segment being HEC2).
The effect is clearly visible and well reproduced by the MC
simulation.  As seen in the MC distributions, even the position of the
disconnected gap can be inferred within limits, even though it is not
really required. Affected were gaps at the very beginning of the
2$^{\rm nd}$ longitudinal segment. After the run this problem could be
partially traced back to bad ground connections of the EST boards in
this region of the central HEC1 module.  To correct the data, the
observed asymmetry has been determined from a fit (line). The result
of this fit has been used for corrections.  Thus the data have been
used directly to correct for the effect observed without any further
MC assumptions. Repeating this analysis for $30\,\GeV$ pions shows
that this correction is rather energy independent. In consequence,
this correction on an event-by-event basis takes care to a large
extent of fluctuations in the energy deposition and thus also improves
the energy resolution.  This shows clearly
Fig.~\ref{nim_6_hvcorrections2}: the energy resolution $\sigma/E$
for $200\,\GeV$ pions (electromagnetic scale) improves typically from
$\sigma/E=(8.97\pm0.12)\,$\% to $\sigma/E=(8.61\pm0.12)\,$\%.

  \clearpage
  
  \section{Monte Carlo Simulation}
  %auto-ignore
In ATLAS the final hadronic calibration is driven by jets rather than
single particles. Following a weighting approach for the hadronic
calibration~\cite{r-hadcal} in ATLAS, weighting parameters are defined
relative to the electromagnetic scale, which can be well defined in
ATLAS as well as in the beam test set-up. These parameters have to be
derived from MC studies. Therefore it is important that the MC
describes the data well and the comparison of data with MC in the beam
test is of utmost importance. Thus special software packages have been
developed.

The first package uses the GEANT~3 program~\cite{r-GEANT3} (version
3.21) to simulate the response of various beam particles in the EMEC
and HEC.  The geometry description is very detailed: cryostat, liquid
argon excluder, all beam elements (MWPC's, scintillation counters
etc.) as well as all details of the calorimeter modules are
included. This code is based on the development done for the HEC
stand-alone tests~\cite{r-kiryunin}. For the hadronic shower
simulation the GCALOR code~\cite{r-GCALOR} is used. For particle
tracking the threshold has been set to $100\,\keV$ and for secondary
production of photons and electrons to $1\,\MeV$.

The second package uses the GEANT~4 code~\cite{r-GEANT4}, developed by
a worldwide collaboration, and since the first release in 1999
maintained by the international GEANT~4 collaboration. The
implementation of the ATLAS detector including the calorimeters is
still ongoing. Within this process the validation of the physics
models in GEANT~4 is one of the most important tasks. From the
proposed physics lists~\cite{r-wellisch} for hadronic shower
simulations in GEANT~4 two models have been selected for the
comparison of the data with MC: LHEP and QGSP. The LHEP physics list
uses the low and high energy pion parameterisation models for
inelastic scattering.  The QGSP physics list is based on theory driven
models: it uses the quark-gluon string model for interactions and a
pre-equilibrium decay model for the fragmentation. The geometrical
description of the set-up is at the same level of detail as in
GEANT~3. The range threshold for production has been set to
$700\,\mu\meter$ in general, irrespective of the material. This
corresponds to cut-off energies of $1.03\,\MeV$, $1.00\,\MeV$ and
$0.27\,\MeV$ for copper, lead and liquid argon respectively.

\begin{figure}[htb]
  \begin{center}
%    \psdraft   
    \mbox{\epsfig{file=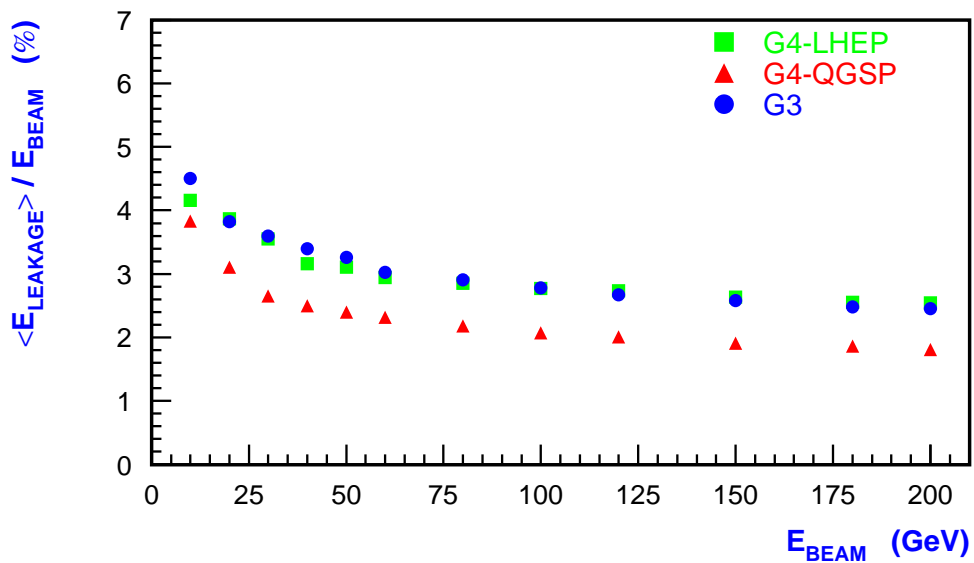,width=1.0\textwidth}}
%    \psfull
  \end{center}
  \caption{Energy dependence for $\pi^-$ of the lateral leakage for
  the beam test as obtained from different MC models.}
  \label{nim_4_simulation1}
\end{figure}

\begin{figure}[htb]
  \begin{center}
%    \psdraft   
    \mbox{\epsfig{file=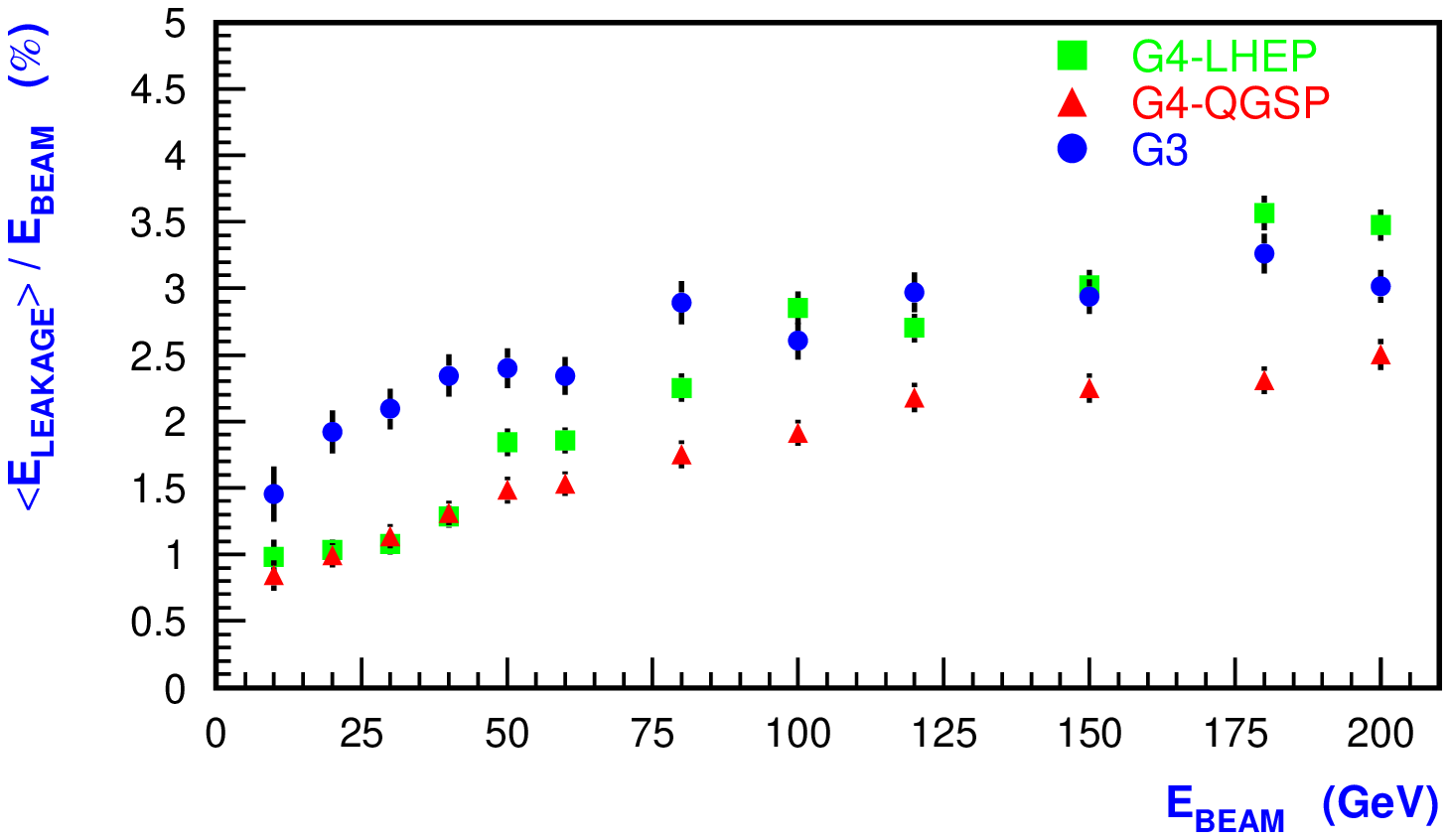,width=1.0\textwidth}}
%    \psfull
  \end{center}
  \caption{Energy dependence for $\pi^-$ of the longitudinal leakage for
  the beam test as obtained from different MC models.}
  \label{nim_4_simulation2}
\end{figure}

\begin{figure}[htb]
  \begin{center}
%    \psdraft   
    \mbox{\epsfig{file=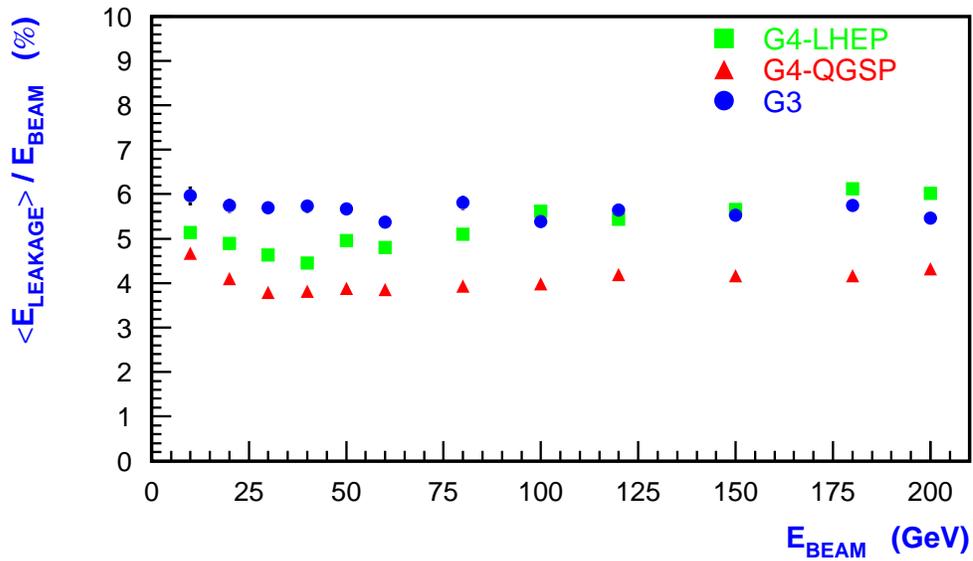,width=1.0\textwidth}}
%    \psfull
  \end{center}
  \caption{Energy dependence for $\pi^-$ of the total leakage for
  the beam test as obtained from different MC models.}
  \label{nim_4_simulation3}
\end{figure}

The response signals from MC simulations and data have to be compared
at a well-defined scale. For the MC the inverse sampling ratio is
used, that is the ratio between the electron beam energy and the total
visible energy.  For the HEC this ratio has been obtained from the HEC
stand alone simulations as $23.3$ ($23.7$) for GEANT~3 (GEANT~4)
respectively.  Simulations of the combined EMEC/HEC beam test runs
have been used to obtain this ratio for the EMEC. The corresponding
values are $11.35$ ($12.05$) for GEANT~3 (GEANT~4) respectively. The
beam test set-up causes some lateral and longitudinal leakage for
hadronic showers. To evaluate the amount of leakage and its influence
on the calorimeter performance, `virtual' leakage detectors, placed
laterally and longitudinally with respect to the EMEC and HEC modules,
have been implemented in the simulations. In these `detectors' the
kinetic energy is summed up for all particles leaving the calorimeter
modules via the related module boundaries.
Figs.~\ref{nim_4_simulation1},~\ref{nim_4_simulation2}
and~\ref{nim_4_simulation3} show the average fraction of energy
leaking with respect to the beam energy for charged pions ($\pi^-$) of
different energies. Shown are the results for the three MC models
considered.  At low energies the lateral leakage dominates, at high
energies the longitudinal leakage increases. The amount of total
energy leakage differs somewhat for the various MC options: GEANT~3
predicts about $6\,$\%, while GEANT~4 QGSP gives $4\,$\% leakage. The
energy dependence of leakage is in both options rather weak. GEANT~4
LHEP is in agreement with GEANT~3 at high energies, but yields
somewhat smaller leakage at low energies.

The simulation of the noise uses real data (see section electronic
noise). For the EMEC the noise has been obtained from the muon data,
excluding the channels hit directly by the muon.  For the HEC the
first time samples are preceding the signal pulse and have been used
to derive the $\sigma$ of the noise distribution.

  \clearpage
  
\section{Electron Results}
%auto-ignore
The main goal of this beam test is the determination of the hadronic
calibration -- constants and procedures -- in the ATLAS region of
$|\eta|\simeq1.8$. The analysis of the pion data is of most relevance
for this task. Nevertheless, to be able to assess the performance of
the EMEC and to obtain the electromagnetic scale, which is the basis
for the hadronic calibration, the analysis of electrons is the initial
step in this project. As already discussed this beam test set-up
geometry is slightly non-projective, in contrast to the ATLAS
detector. Thus the definitive reference for the performance of
electrons in ATLAS is~\cite{r-emec} rather than this paper. As it
turns out the electron results obtained in this beam test are in close
agreement to the previous analysis~\cite{r-emec}.

\subsection{Corrections}
\begin{figure}[htb]
  \begin{center}
%    \psdraft   
    \mbox{\epsfig{file=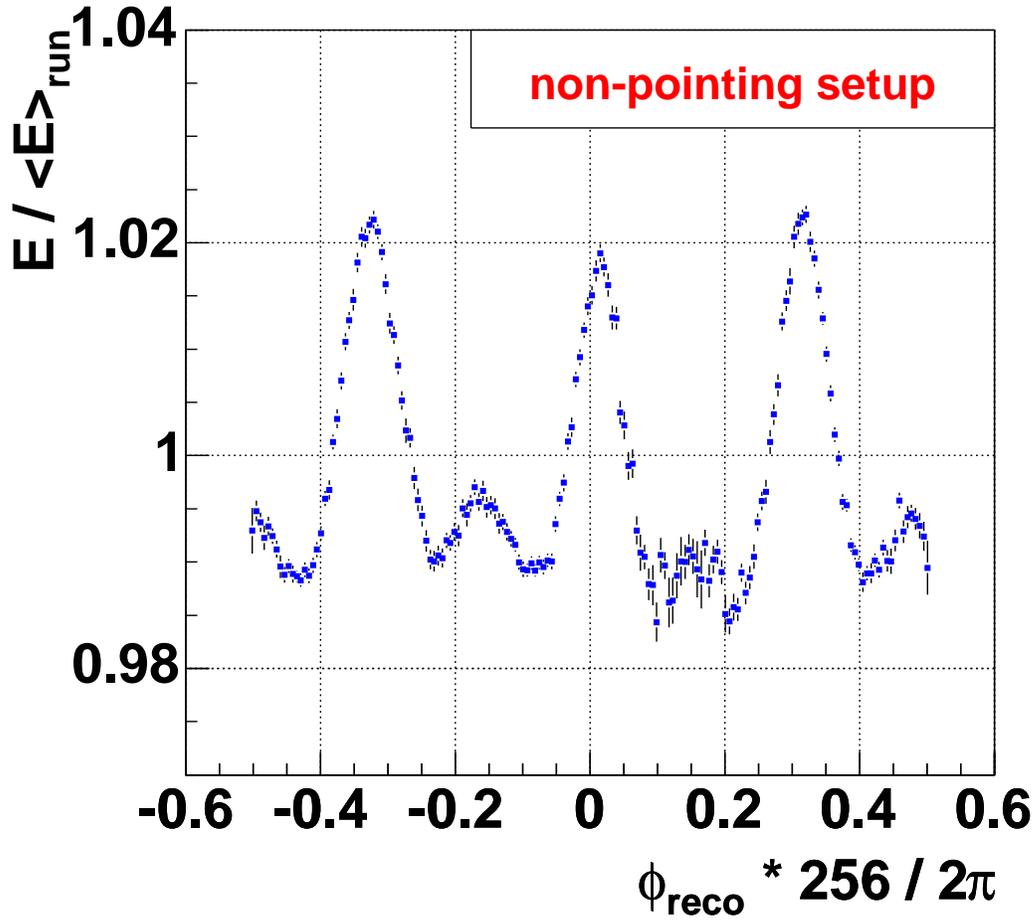,width=1.0\textwidth}}
%    \psfull
  \end{center}
  \caption{Relative response of electrons of $119\,\GeV$ as function
  of $\phi$.  Here $\phi$ is given in units of the absorber structure
  spacing.}
  \label{nim_5_corrections1}
\end{figure}

  %auto-ignore
The end-cap geometry of the EMEC requires small
corrections~\cite{r-emec} for the energy reconstruction of
electrons. The projective HV sectors yield a variation of the signal
response with $\eta$ (corresponding to $R$ in the beam test
geometry). Given the limited $\eta$ range accessible in the beam test
set-up and the non-pointing geometry, where the beam spread covers a
larger $\eta$ range than in the pointing geometry, this variation on
average is smaller in comparison to the azimuthal $\phi$ variation and
has been ignored.  However, electric field and sampling fraction
non-uniformities as well as the non-pointing geometry, cause a
significant $\phi$ variation of the signal response.
Fig.~\ref{nim_5_corrections1} shows the relative response of
$119\,\GeV$ electrons as function of $\phi$ in cell units. Here $E_{run}$ is
the average energy obtained for the corresponding run.
A variation at the level of $\sim\pm1.5\,$\% can be seen.  A
numerical correction defined bin by bin has been used for the data
correction.

  \subsection{Energy Resolution}
\begin{figure}[htb]
  \begin{center}
%    \psdraft   
    \mbox{\epsfig{file=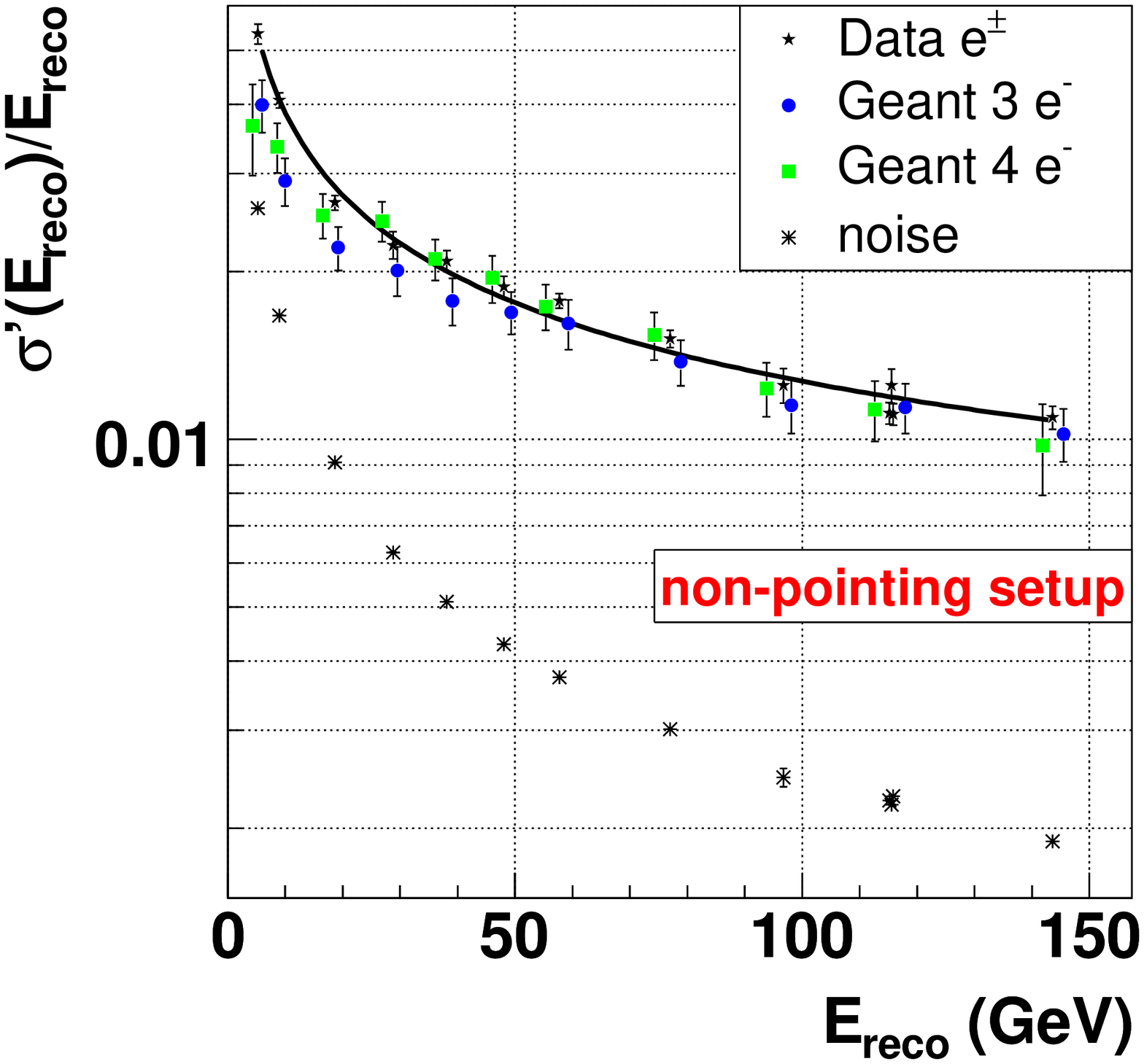,width=1.0\textwidth}}
%    \psfull
  \end{center}
  \caption{Energy dependence of the energy resolution of electrons for
  the impact point~I. Shown are the noise subtracted data as well as
  the various MC predictions. Also shown is the noise contribution at
  the various energies. The line shows the result of the fit to the
  data.}
  \label{nim_5_eresol1}
\end{figure}

  %auto-ignore
The energy resolution has been studied using the cluster algorithm
with the corrections mentioned above. As an example
Fig.~\ref{nim_5_eresol1} shows the energy resolution for impact
point~I.
The data have not been corrected for energy depositions outside the
reconstructed cluster. The data are shown with the noise subtracted.
The noise contribution is shown as well. A fit to the data (solid
line) with $\frac{\sigma(E)}{E} = \frac{a}{\sqrt{E}} \oplus b$
typically yields a sampling term $a=(12.1\pm0.2)\,$\%$\sqrt{\GeV}$ and a
constant term $b=(0.4\pm0.1)\,$\%. The GEANT~3 and GEANT~4 simulations are in
good agreement and yield for GEANT~3 a sampling term of
$(9.3\pm0.6)\,$\%$\sqrt{\GeV}$ and a constant term of 
$(0.8\pm0.1)\,$\% while GEANT~4 yields a sampling term of 
$(10.6\pm0.7)\,$\%$\sqrt{\GeV}$ and a constant term of $(0.7\pm0.2)\,$\%
Further tuning is underway for ATLAS to improve the fair agreement with
the data. For example the amplitude of the $\phi$ modulation is twice as big
in GEANT~4 as in GEANT~3 and the data. Due to the beam spread this variation
is corrected as for the data (see section Corrections).
This causes some systematic
uncertainty in the MC prediction due to the exact beam impact position
and beam spread.  Since the impact angle for the combined beam test is
different from the geometry of the pointing EMEC stand-alone beam
test~\cite{r-emec}, the energy resolution is slightly worse compared
to the sampling term of $\approx 11\,$\%$\sqrt{\GeV}$ and the constant
term of $\approx 0.4\,$\% obtained in the stand-alone EMEC analysis.

\begin{figure}[htb]
  \begin{center}
%    \psdraft   
    \mbox{\epsfig{file=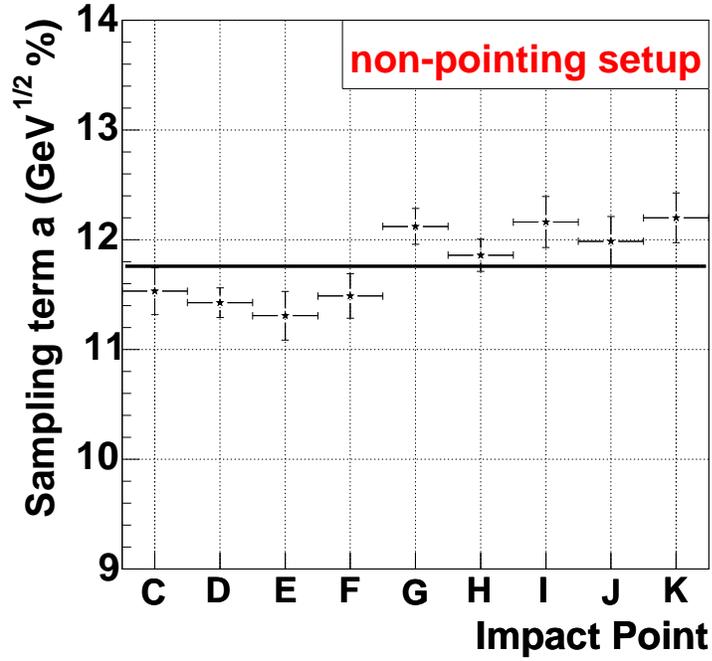,width=0.7\textwidth}}
%    \psfull
  \end{center}
  \caption{Energy resolution for electrons: variation of the sampling
  term with the impact point. The left and right set of points
  correspond to different $\eta$ regions.}
  \label{nim_5_eresol2}
\end{figure}

\begin{figure}[htb]
  \begin{center}
%    \psdraft   
    \mbox{\epsfig{file=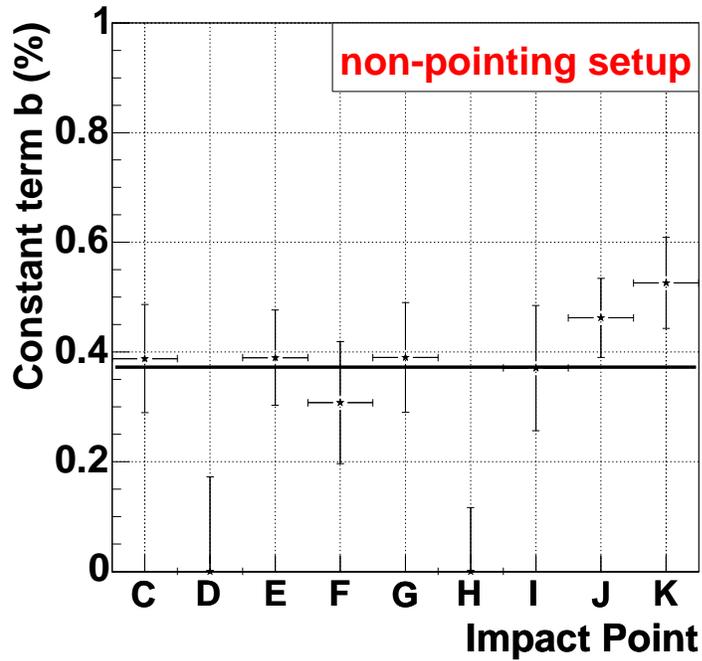,width=0.7\textwidth}}
%    \psfull
  \end{center}
  \caption{Energy resolution for electrons: variation of the constant
  term with the impact point. The left and right set of points
  correspond to different $\eta$ regions.}
  \label{nim_5_eresol3}
\end{figure}

The energy resolution has been studied for various impact points.
Fig.~\ref{nim_5_eresol2} shows the sampling term for various impact
points and Fig.~\ref{nim_5_eresol3} the constant term. Because of the
residual weak correlation between the constant term and the sampling
term some care has to be taken when comparing the various points.
Nevertheless, the differences (left and right set of points) seem to
indicate some $\eta$-dependence. Partially this is related to the
$\eta$-variation of the sampling fraction and also to the weak
$\eta$-dependence of the electric field, which has not been taken into
account.

  \subsection{Linearity and Electromagnetic Scale}
\begin{figure}[htb]
  \begin{center}
%    \psdraft   
    \mbox{\epsfig{file=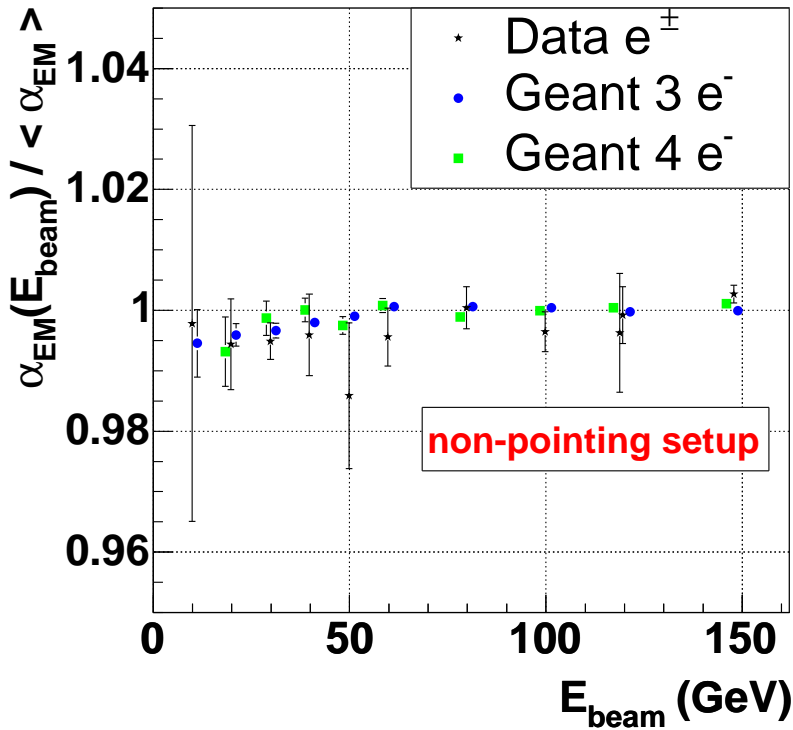,width=0.8\textwidth}}
%    \psfull
  \end{center}
  \caption{Linearity of the electron response: shown is the relative
  variation of the electromagnetic scale $\alpha_{\rm em}$ for
  different energies. The data are compared with MC predictions.}
  \label{nim_5_linearity1}
\end{figure}

\begin{figure}[htb]
  \begin{center}
%    \psdraft   
    \mbox{\epsfig{file=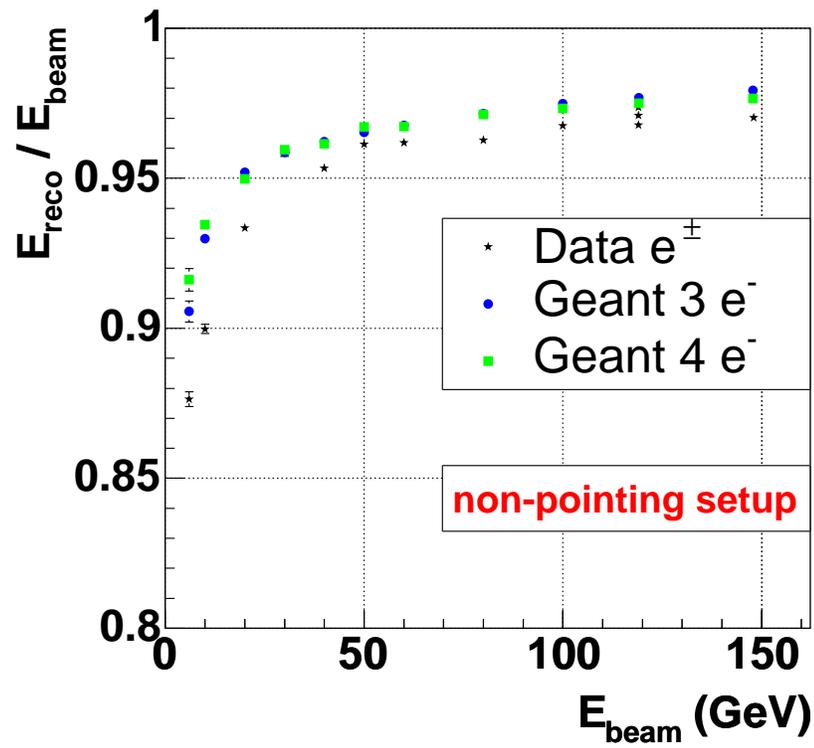,width=0.8\textwidth}}
%    \psfull
  \end{center}
  \caption{Fraction of energy leakage outside the reconstructed
  electron cluster for different energies. The data are compared with
  MC predictions.}
  \label{nim_5_linearity2}
\end{figure}

  %auto-ignore
 To obtain the electromagnetic scale one has to correct for the energy
 deposited outside the reconstructed cluster. For electrons the
 leakage beyond the detector boundaries is known from
 the MC to be negligible. The energy lost outside the reconstructed cluster can easily
 be obtained both for data and MC. For the energy leaking into the
 HEC, the electromagnetic scale of the HEC1 measured in previous HEC
 stand-alone beam tests can be used.  For a given impact point
 Fig.~\ref{nim_5_linearity1} shows the relative variation of the
 electromagnetic scale $\alpha_{\rm em}$ with energy.  In the energy
 range considered, the variation is typically within
 $\pm0.5\,$\%. Both MC models, GEANT~3 as well as GEANT~4, show a
 similarly good linearity of the electromagnetic scale $\alpha_{\rm
 em}$. The data yield an average value of $\alpha_{\rm
 em}=(0.430\pm0.001)\,\MeV/\nA$. Because of the symmetric noise cuts
 any bias in the determination of this electromagnetic scale due to
 noise can be neglected.  Given the uncertainty in the signal shape
 reconstruction ($\approx 1\,$\%) and the uncertainty due to $\eta$
 dependent corrections ($\approx 1\,$\%), which are harder to obtain
 due to the non-pointing geometry and have been neglected, we
 attribute an overall systematic error of $2\,$\% to the
 electromagnetic scale $\alpha_{\rm em}$, corresponding to
 $\pm0.009\,\MeV/\nA$. The sampling ratio for electrons obtained from
 simulations depends on the model and energy and range cuts used -- see
  the Monte Carlo simulation section.
The fraction of the energy leakage outside the electron cluster is
shown in Fig.~\ref{nim_5_linearity2}.
Again the data are compared with MC predictions. Typically the
leakage outside the cluster is between $2\,$\% and $4\,$\%, except for
low energies.  Both MC models show a somewhat smaller leakage,
especially at low energies.  This might be due to small effects of
dead material in front of the EMEC or some cross-talk effects in the
EMEC, which are not fully described in the MC.

 \clearpage
 
% \subsection{Position Resolution}
%  \input{nim_5_position}  
  \subsection{Variation of the Response and Resolution with Material in front of
  the EMEC.}
\begin{figure}[htb]
  \begin{center}  
    \mbox{\epsfig{file=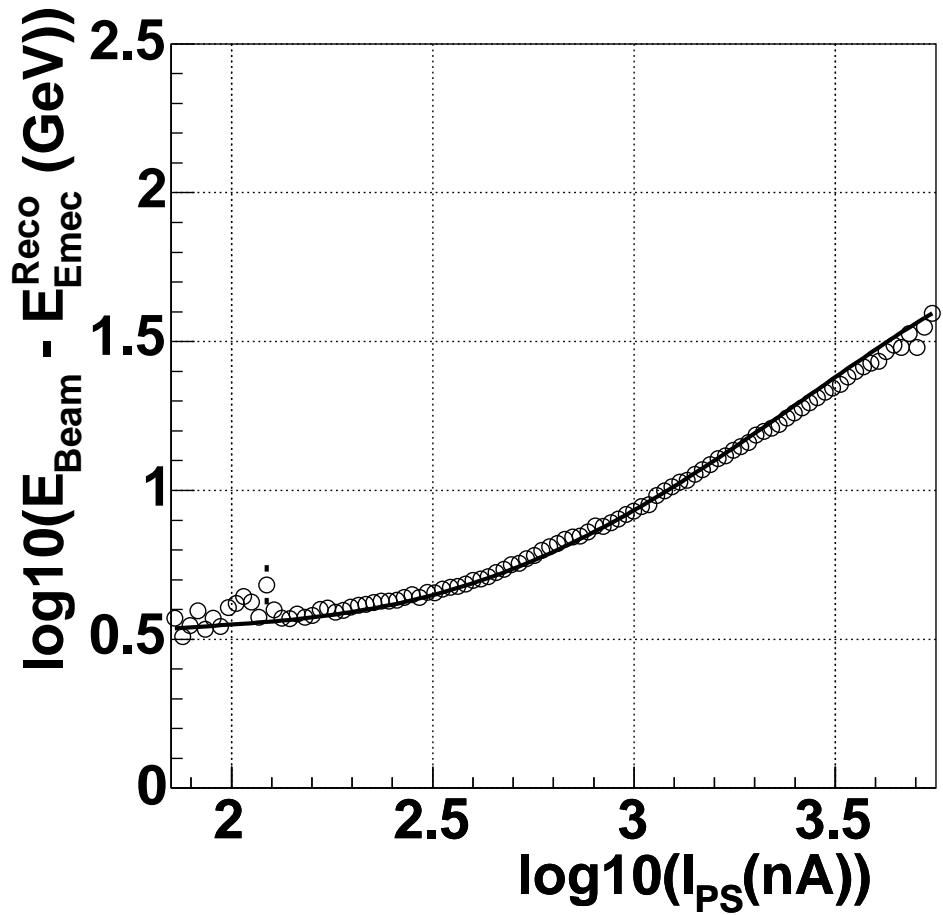,width=1.0\textwidth}}
  \end{center}
 \caption{Dependence of the missing EMEC energy on the presampler
  signal for electrons of $10$, $20$, $119$ and $148\,\GeV$ with
  $2.73\,X_0$ in front of the cryostat.  The solid line shows the
  result of the fit to the data.}
  \label{nim_5_dead_new1}
\end{figure}

\begin{figure}[htb]
  \begin{center}  
    \mbox{\epsfig{file=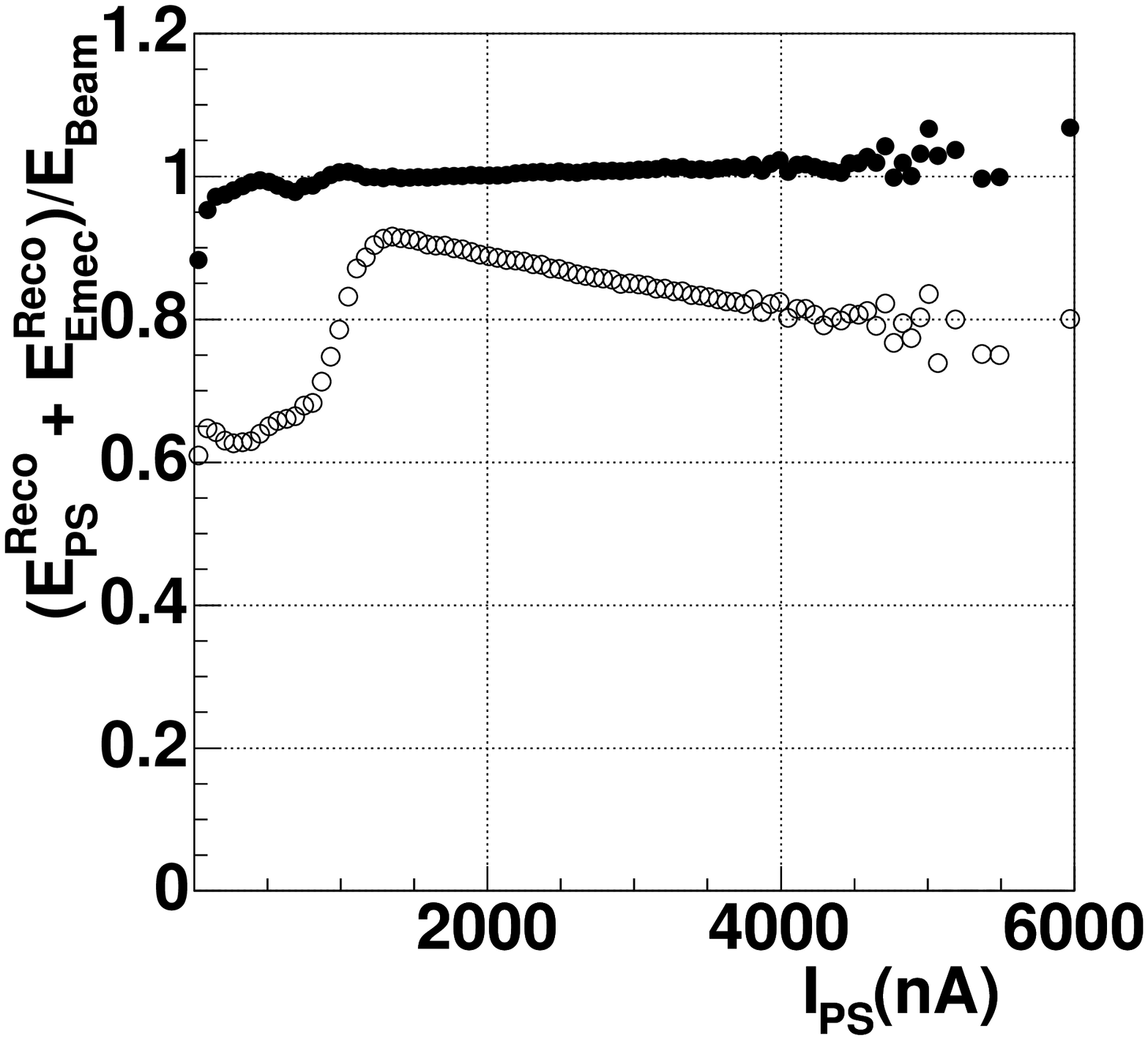,width=1.0\textwidth}}
  \end{center}
  \caption{Electrons with additional $2.73\,X_0$ dead material in
  front of the cryostat: Ratio of the reconstructed energy over
  the beam energy for electrons of $10$, $20$, $119$ and $148\,\GeV$
  before (open points) and after (solid points) the correction as
  a function of presampler signal.}
  \label{nim_5_dead_new2}
\end{figure}

\begin{figure}[htb]
  \begin{center}  
    \mbox{\epsfig{file=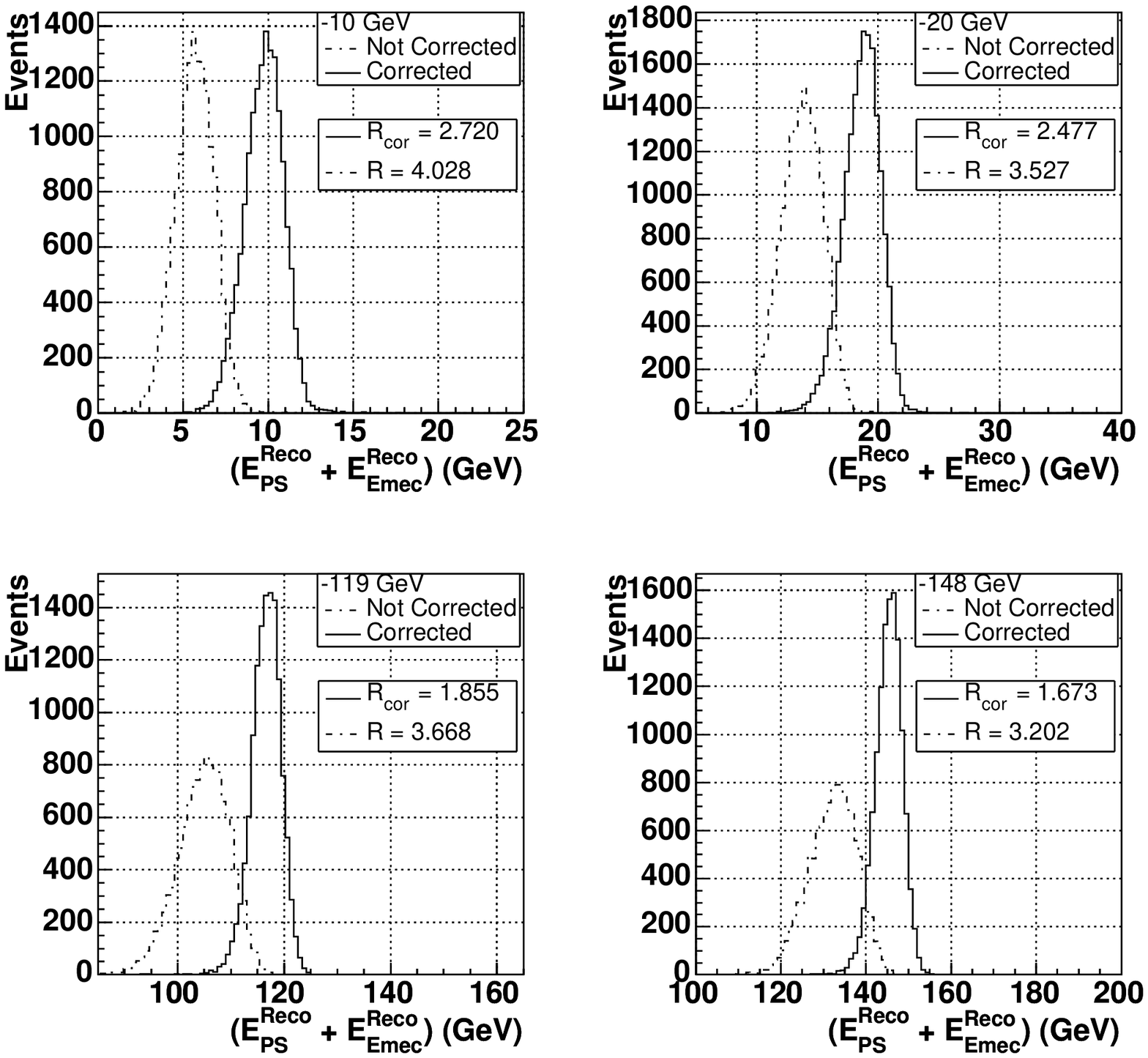,width=1.0\textwidth}}
  \end{center}
  \caption{Response of electrons with additional $2.73\,X_0$ dead
  material in front of the cryostat for energies of $10$, $20$, $119$
  and $148\,\GeV$ before (dashed histogram) and after correction
  (solid histogram). Also given is the corresponding ratio
  $R=[\sigma/E]_{2.73\,X_0}\ /[\sigma/E]_{0\,X_0}$ of the energy
  resolution relative to the resolution without any extra dead
  material in front of the cryostat.}
  \label{nim_5_dead_new3}
\end{figure}

  %auto-ignore
In ATLAS the amount of dead material in front of the calorimeter
varies with $\eta$. The response in the presampler will be used to
correct for the energy loss of electrons in this dead
material. Therefore in this beam test electron data have been taken
varying the amount of dead material in front of the cryostat. From
this data presampler corrections to the measured electron energy in
EMEC can be studied.  The main goal of the correction is to achieve a
good linearity of the electron signal, but in parallel a substantial
improvement of the resolution can be
obtained. Fig.~\ref{nim_5_dead_new1} shows the dependence of the
missing energy on the presampler signal for electrons of $10$, $20$,
$119$ and $148\,\GeV$ with $2.73\,X_0$ in front of the cryostat. A
4$^{\rm th}$ order polynomial fit (solid line) to the data yields a
good description of the dependence for a wide range of presampler
signals and for all energies considered.
Fig.~\ref{nim_5_dead_new2} shows the ratio of the reconstructed energy
relative to the beam energy as function of presampler signal for the
data before (open points) and after (solid points) the correction.
Again, the data for all energies studied are shown. For a wide range
of presampler signals and for all energies considered the linearity of
the electron signal is recovered.
Finally Fig.~\ref{nim_5_dead_new3} shows the response of electrons of
$10$, $20$, $119$ and $148\,\GeV$ before (dashed histogram) and after
(solid histogram) the presampler corrections to the EMEC energy.  Also
given for each energy is the ratio of energy resolution relative to
the resolution without any extra dead material in front of the
cryostat: $R=[\sigma/E]_{2.73\,X_0}\ /[\sigma/E]_{0\,X_0}$.  The
dramatic improvement of the resolution is clearly visible: it is close
to a factor of two, in particular at higher energies.

It should be mentioned that an energy dependent correction could
slightly improve the results. These studies, as well as the dependence
on the amount of extra dead material in front, are still ongoing as
part of the ATLAS energy calibration studies. However it was the goal
of this analysis to restrict the corrections to a rather general and
robust approach which can be further refined for ATLAS.

 \clearpage
 
 \section{Pion Results} 
 \subsection{Response at the Electromagnetic Scale}
\begin{figure}[htb]
  \begin{center}
%    \psdraft   
    \mbox{\epsfig{file=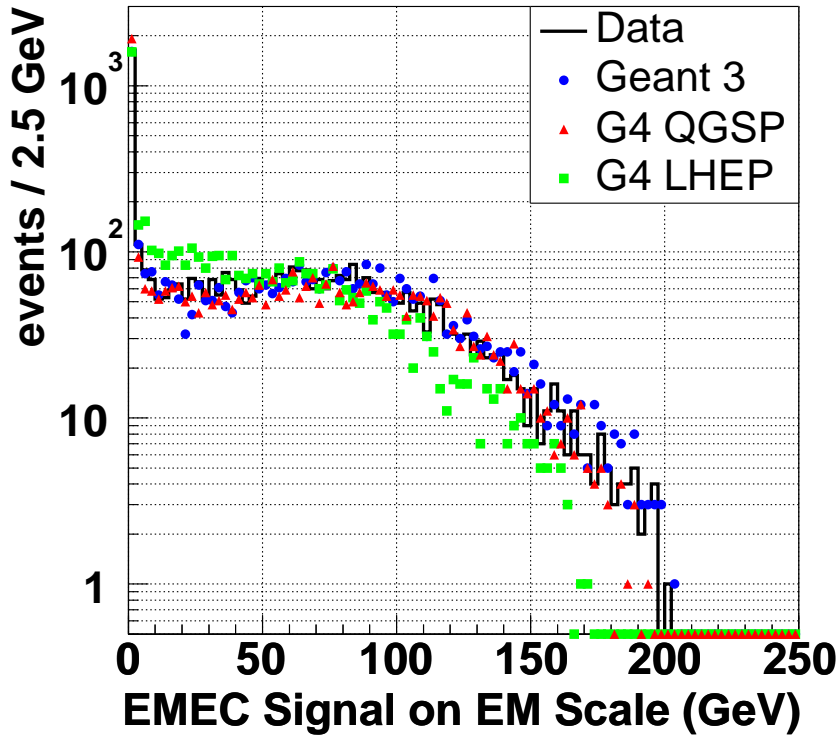,width=0.8\textwidth}}
%    \psfull
  \end{center}
  \caption{Response to $200\,\GeV$ pions in the EMEC on the
  electromagnetic scale. The data are compared to MC predictions.}
  \label{nim_6_response1}
\end{figure}

\begin{figure}[htb]
  \begin{center}
%    \psdraft   
    \mbox{\epsfig{file=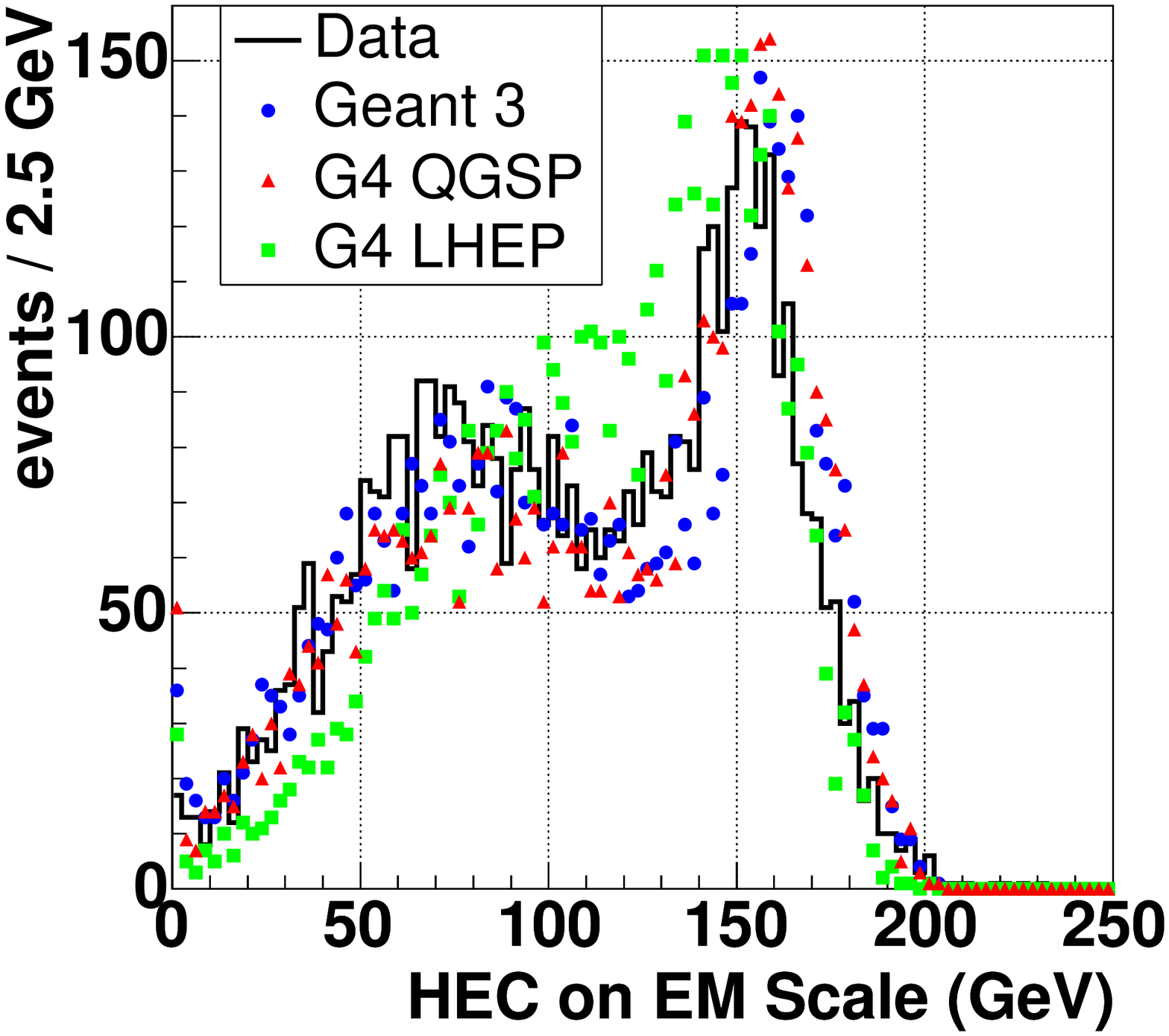,width=0.8\textwidth}}
%    \psfull
  \end{center}
  \caption{Response to $200\,\GeV$ pions in the HEC on the
  electromagnetic scale. The data are compared to MC predictions.}
  \label{nim_6_response2}
\end{figure}

\begin{figure}[htb]
  \begin{center}
%    \psdraft   
    \mbox{\epsfig{file=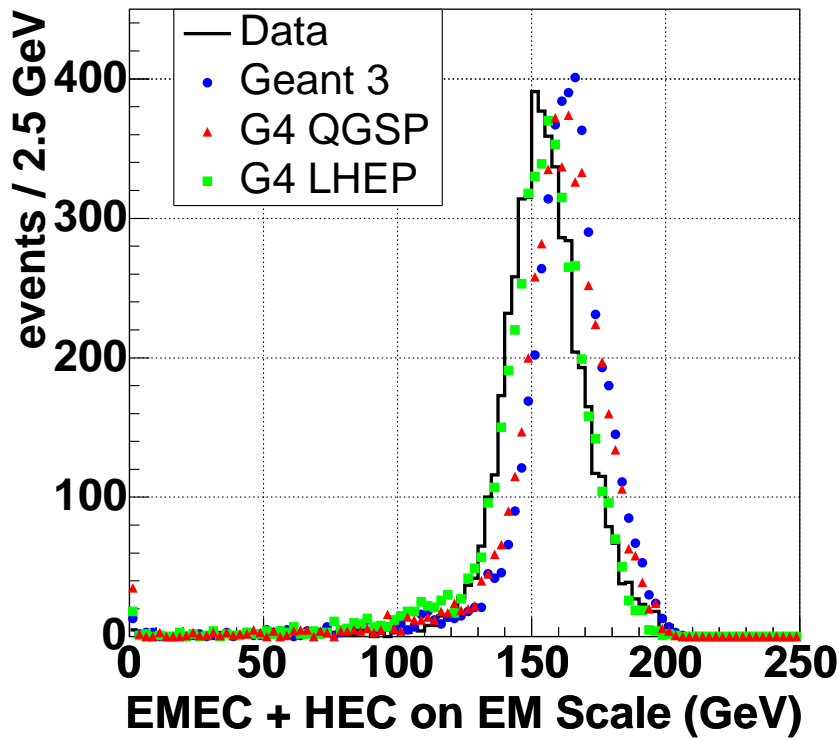,width=0.8\textwidth}}
%    \psfull
  \end{center}
  \caption{Total response to $200\,\GeV$ pions in the EMEC and HEC on
  the electromagnetic scale. The data are compared to MC predictions.}
  \label{nim_6_response3}
\end{figure}

  %auto-ignore
The energy reconstruction of pions is based on the electromagnetic
scale. For the EMEC value $\alpha_{\rm em}^{\rm EMEC}=0.430\,\MeV/\nA$
(see section electrons) has been used, whereas for the HEC the results
from electron data taken in the previous stand-alone beam test
runs~\cite{r-hec} have been used. With respect to~\cite{r-hec} a
correction for the calibration pulse shape had to be applied, finally
yielding a value of $\alpha_{\rm em}^{\rm HEC}=3.27\,\MeV/\nA$. The
statistical (systematic) error of $\alpha_{\rm em}^{\rm HEC}$ is
typically $1\,$\% ($1\,$\%). As an example,
Figs.~\ref{nim_6_response1},~\ref{nim_6_response2}
and~\ref{nim_6_response3} show the energy response in the EMEC, HEC
and the total response respectively to $200\,\GeV$ pions at the impact
point~J. The data are compared with the MC predictions.  Both, GEANT~3
and GEANT~4~QGSP describe the EMEC and HEC data reasonably well, even
though there are some deviations visible. In contrast, the
GEANT~4~LHEP simulation deviates substantially from the data both, for
the EMEC and the HEC.  However the GEANT~4~LHEP simulation yields the
best agreement with data for the total signal.

 \subsection{Energy Reconstruction using the Cluster Weighting Approach}
  %auto-ignore
In the ATLAS calorimeter the response to hadrons differs significantly
from that for electrons of the same energy.  Therefore, for hadronic
energy a calibration coefficient has to be applied to the signals
determined on the corresponding electromagnetic scale. The goal is to
get an equal response for the electromagnetic as well as the pure
hadronic component of a hadronic shower. If this goal were to be
achieved fluctuations in the reconstructed hadronic shower energy
could be minimized and thus the energy resolution substantially
improved.  The technique of signal weighting has been successfully
used in previous experiments (see~\cite{r-H1weight1,r-H1weight2} and
the references therein). It exploits the fact that local energy
deposits of high density are mainly due to electromagnetic
interactions while for hadronic interactions the corresponding density
is substantially lower. Thus for a segmented calorimeter the energy
deposited in individual read-out cells can be on a statistical basis
identified to be of electromagnetic or hadronic origin.  For the
hadronic calibration of the ATLAS calorimeter a similar approach is
envisaged~\cite{r-hadcal}. Methods and algorithms are still under
development. The goal is to use weighting constants and weighting
functions, based on the individual cluster energy, to reconstruct
optimally the related energy. In ATLAS these weighting constants have
to be derived from MC simulations, which to validate the method have
to describe the beam test data well.  The crucial MC information in
this tuning process is the knowledge of the total and visible energy
deposition in each read-out cell.  Thus the weights could be directly
derived from the MC simulation and applied to beam test data.  At
present this MC information is not yet available. Hence a weighting
approach based on individual read-out cell energy deposits as
envisaged cannot yet be achieved. Therefore a different approach has
been applied.

\begin{figure}[htb]
  \begin{center}
%    \psdraft   
    \mbox{\epsfig{file=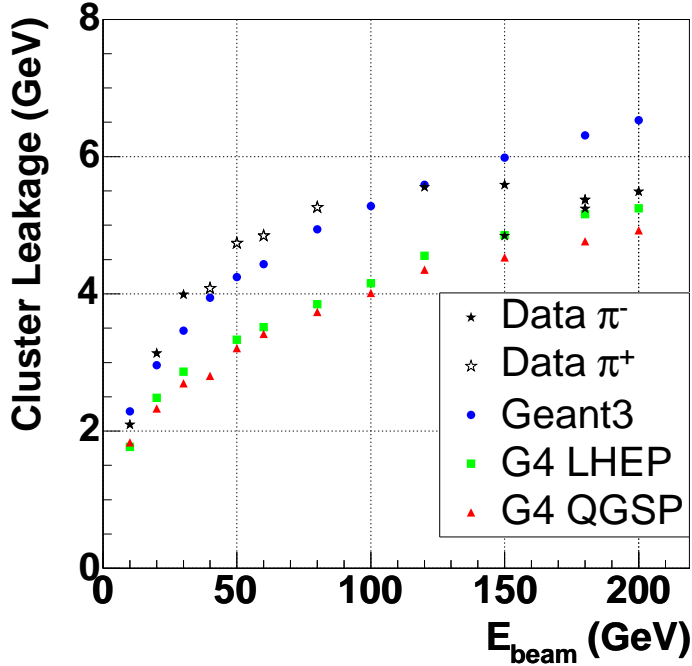,width=0.7\textwidth}}
%    \psfull
  \end{center}
  \caption{Energy dependence of the average energy leakage outside the
   reconstructed cluster for pions. Shown are the data and the 
   different MC expectations.}
  \label{nim_6_clu_weighting1}
\end{figure}

In the EMEC and the HEC the volume of the related clusters can be used
 to obtain the 'EMEC cluster energy density' and the 'HEC cluster
 energy density'.  Here the cluster volume is the sum of all volumes
 of the individual cells contained in the cluster.  With this
 information a correction to the electromagnetic scale can be derived
 using the information of the total energy in the EMEC and HEC on the
 beam energy scale. To obtain an estimate of the `true' energy
 deposited, the energy leakage has to be subtracted from the beam
 energy.  Energy may leak for two reasons:
\begin{itemize}
\item   Energy deposited in the detector, but outside of the reconstructed
 cluster;
\item   Energy leaking outside the detector set-up.
\end{itemize}
The leakage outside the cluster has been obtained for each event by
adding the energies (electromagnetic scale) of all readout cells not
used in the cluster reconstruction. On average the noise
contributions, being symmetric,
cancel. Fig.~\ref{nim_6_clu_weighting1} shows the energy dependence of
this 'cluster leakage'. At low energy it rises up to $5\,\GeV$ and
remains approximately at this level for energies above $60\,\GeV$. The
GEANT~4 MC prediction shows a somewhat smaller leakage, the GEANT~3
prediction is closer to the data but differs in shape.

\begin{figure}[htb]
  \begin{center}
%    \psdraft   
    \mbox{\epsfig{file=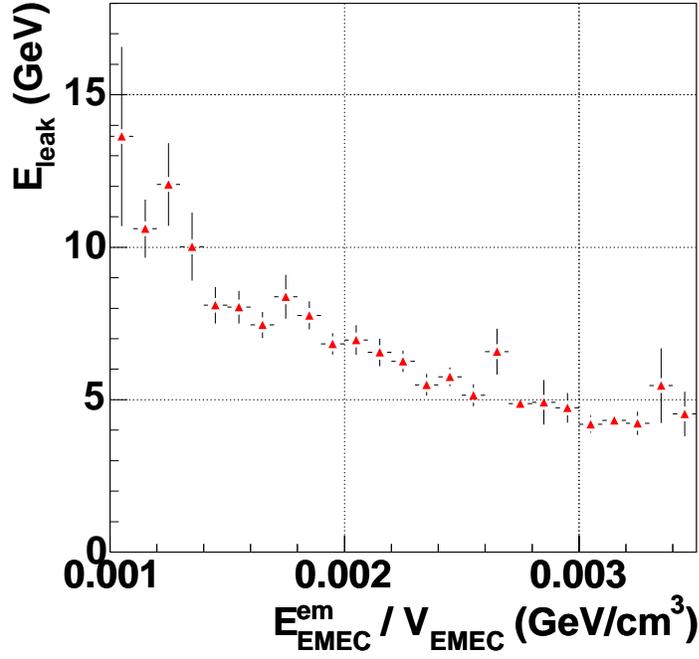,width=0.7\textwidth}}
%    \psfull
  \end{center}
  \caption{Correlation between the cluster energy density $E_{\rm
  EMEC}/V_{\rm EMEC}$ and the total energy leakage $E_{\rm leak}$ for
  the EMEC (at $200\,\GeV$).}
  \label{nim_6_clu_weighting1a}
\end{figure}

\begin{figure}[htb]
  \begin{center}
%    \psdraft   
    \mbox{\epsfig{file=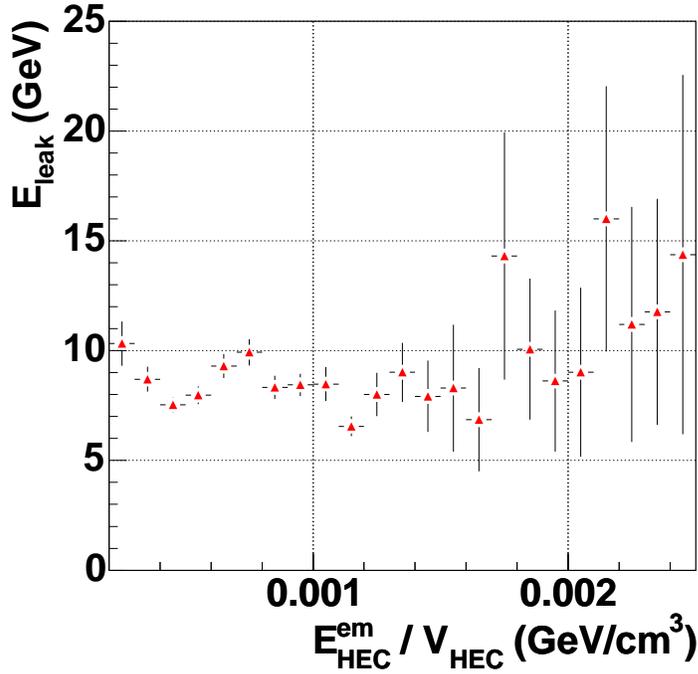,width=0.7\textwidth}}
%    \psfull
  \end{center}
  \caption{Correlation between the cluster energy density $E_{\rm
  HEC}/V_{\rm HEC}$ and the total energy leakage $E_{\rm leak}$ for
  the HEC (at $200\,\GeV$).}
  \label{nim_6_clu_weighting1b}
\end{figure}

To estimate the energy leakage outside the detector for each event,
the MC prediction (GEANT~4~QGSP) has been used. Here the predicted
correlation between the energy density (EMEC and HEC) and the energy
leakage has been used.  Figs.~\ref{nim_6_clu_weighting1a}
and~\ref{nim_6_clu_weighting1b} show this correlation for 200 GeV
pions at a particular impact point as an example.  In particular for
the EMEC the correlation is rather pronounced: whenever the energy
density is low, leakage is getting large, separating more
electromagnetic from hadronic interactions.
This total leakage is on average (see section on Monte Carlo
 simulation) at the level of $4\,$\%.

\begin{figure}[htb]
  \begin{center}
%    \psdraft   
    \mbox{\epsfig{file=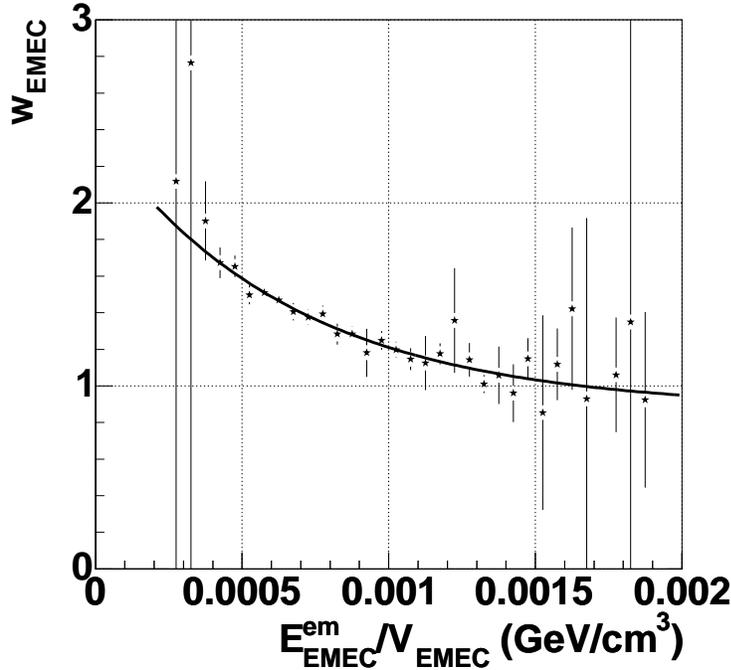,width=0.7\textwidth}}
%    \psfull
  \end{center}
  \caption{Ratio ('hadronic weight') of the true energy over the energy
   measured on the electromagnetic
  scale as function of the cluster energy density
  for pions of $30\,\GeV$ fully contained in the EMEC. The line shows
  the result of the fit (see text).}
  \label{nim_6_clu_weighting2}
\end{figure}

To demonstrate this procedure $30\,\GeV$ pions which have deposited
all their energy in the EMEC have been
selected. Fig.~\ref{nim_6_clu_weighting2} shows the ratio $w_{\rm
EMEC}$ of the `true' deposited energy over the measured energy
on the electromagnetic scale as a function of energy density in the
EMEC.
The data can be well described using the parameterisation:
\begin{equation}\label{eq:Ci}
w_{\rm EMEC}= C_1 \exp\left(-C_2 E^{\rm em}/V\right) + C_3  
\end{equation}

\begin{figure}[htb]
  \begin{center}
%    \psdraft   
    \mbox{\epsfig{file=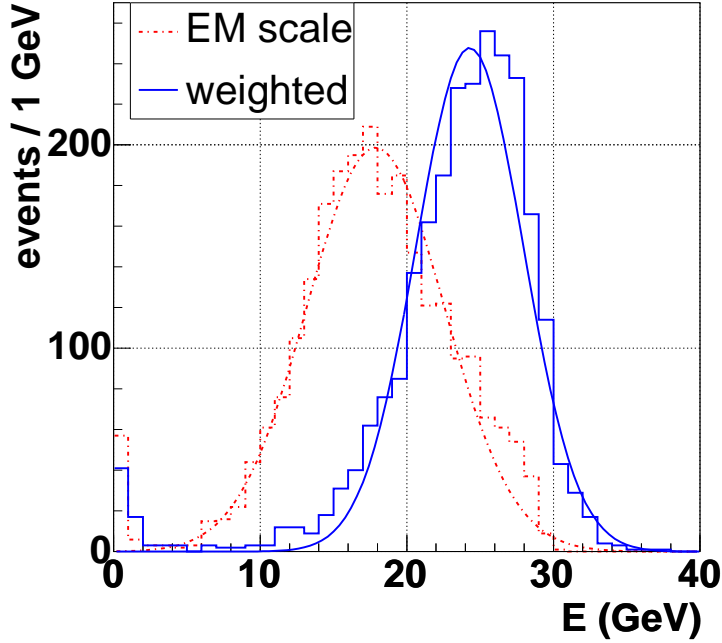,width=0.7\textwidth}}
%    \psfull
  \end{center}
  \caption{Comparison of energy distributions before (broken histogram
  and broken line) and after weighting (solid histogram and solid
  line) for $30\,\GeV$ pions fully contained in the EMEC. The energy
  resolution is improved from $\sigma/E=26.2\,\%$ to
  $\sigma/E=15.1\,\%$.}
  \label{nim_6_clu_weighting3}
\end{figure}

The reconstructed energy
$E^{\rm reco}$ becomes
\begin{equation}
   E^{\rm reco}= w_{\rm EMEC} E^{\rm em}  
\end{equation}
where $E^{\rm em}$ refers to the EMEC cluster energy in the em scale 
and $V$ to the corresponding cluster volume.  The fit describes the
data well. The weighting
parameters $C_i$ from the fit have been applied to the data.
Fig.~\ref{nim_6_clu_weighting3} shows the energy resolution thus
obtained: weighting improves the energy resolution from
$\sigma/E=26.2\,$\% to $\sigma/E=15.1\,$\%. The average measured
energy is less than $30\,\GeV$ because some energy is deposited
outside the cluster reconstructed ($E_{\rm leak}$).
 
For higher energies both subdetectors, the EMEC and the HEC, have to
be considered. Here a $\chi^2$-fit has been used to determine the
parameters $C_i$:
$$\chi^2=\sum_{\rm events} \frac{{\left[E_{\rm beam}-E_{\rm leak}- 
           E_{\rm HEC}^{\rm reco}(C_1^{\rm H},C_2^{\rm H},C_3^{\rm H})- 
           E_{\rm EMEC}^{\rm reco}(C_1^{\rm EM},C_2^{\rm EM},C_3^{\rm EM})\right]}^2}
	   {{(\sigma_{\rm noise}^{\rm reco})}^2 + {(\sigma_{\rm noise}^{\rm leak})}^2},$$
where $\sigma_{\rm noise}^{\rm leak}$ is the noise contribution to the
determined leakage and $\sigma_{\rm noise}^{\rm reco}$ is the
integrated noise in the related cluster. The noise contribution to the
determination of the energy deposits beyond the reconstructed cluster,
$\sigma_{\rm noise}^{\rm leak}$, is given by the quadratic sum of the
noise values on the electromagnetic scale of all calorimeter read-out
cells, which are not included in the clusters. This value has been
found to be rather independent of the cluster size and of the beam
energy; therefore it is set to the average value of $5\,\GeV$.  The
$\chi^2$ has been defined based on noise contributions only,
neglecting shower fluctuations. To a large extent they drop out
because of the correlation between $E_{\rm HEC}^{\rm reco}$ and
$E_{\rm EMEC}^{\rm reco}$. The residual error due to the ansatz used
to achieve an electron to pion compensation on an event by event basis
is non-gaussian and we attribute it to the systematic error of the
method used. It has been neglected in the $\chi^2$ definition
above. But the $\chi^2$ per d.o.f. achieved is typically $3.5$,
indicating a reasonable $\chi^2$ definition.

\begin{figure}[htb]
  \begin{center}
%    \psdraft   
    \mbox{\epsfig{file=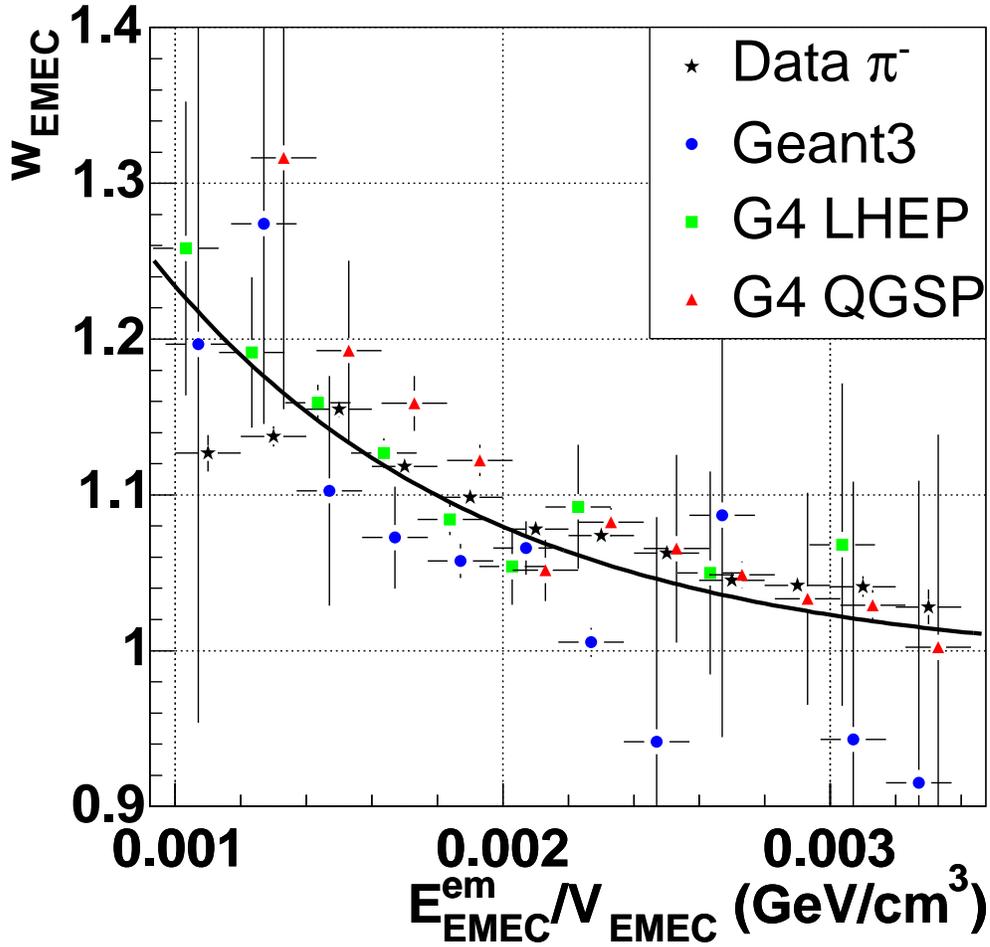,width=1.0\textwidth}}
%    \psfull
  \end{center}
  \caption{EMEC: Ratio ('hadronic weight') of the true energy over the
  energy measured on the
  electromagnetic scale as function of the cluster
  energy density for pions of $200\,\GeV$. The line shows the result
  of the fit to the data (see text).  The data are compared with
  MC predictions.}
  \label{nim_6_clu_weightinga}
\end{figure}

\begin{figure}[htb]
  \begin{center}
%    \psdraft   
    \mbox{\epsfig{file=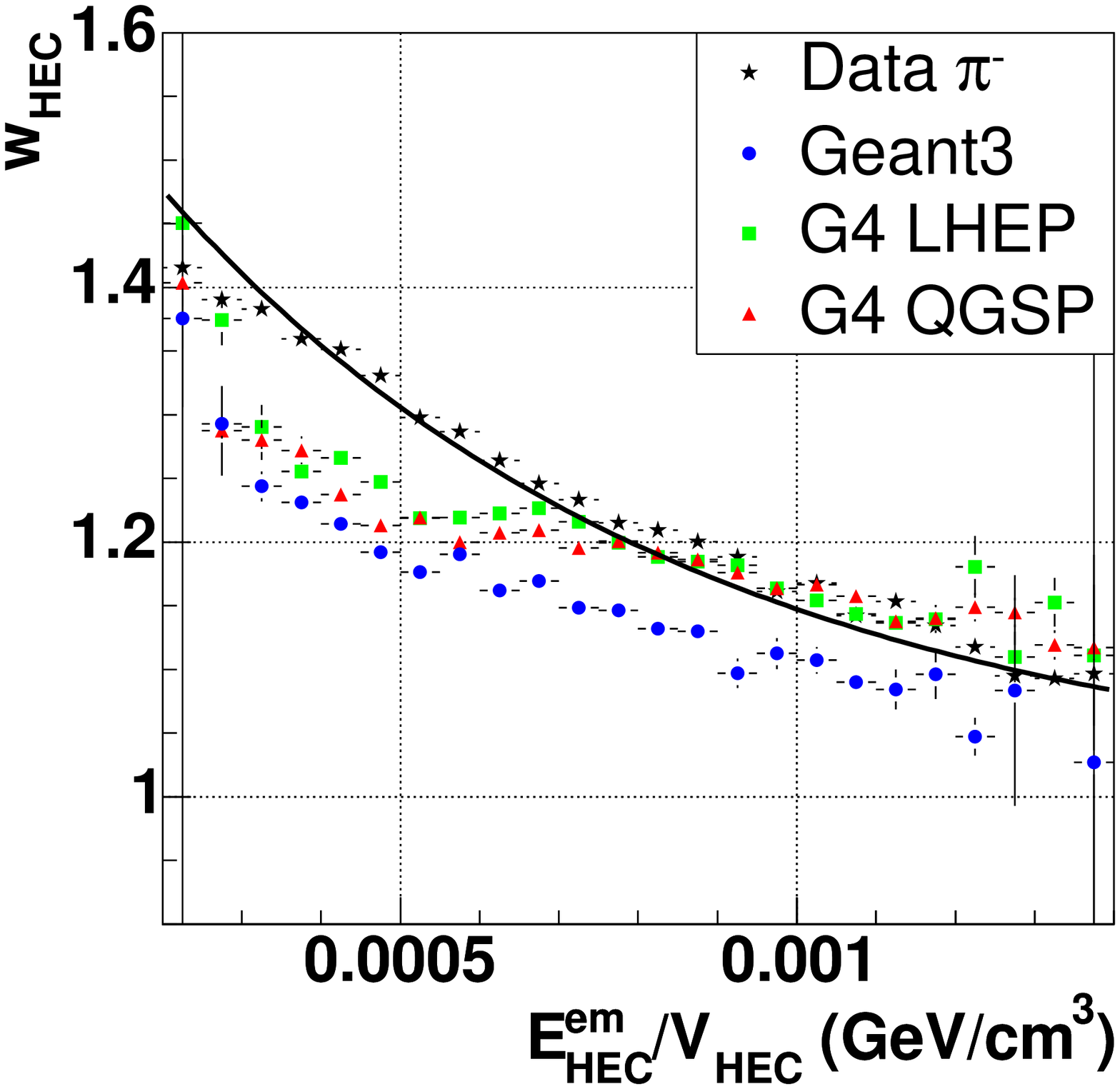,width=1.0\textwidth}}
%    \psfull
  \end{center}
  \caption{HEC: Ratio ('hadronic weight') of the true energy over the
  energy measured on the 
  electromagnetic scale as function of the cluster
  energy density for pions of $200\,\GeV$. The line shows the result
  of the fit to the data (see text). The data are compared with MC
  predictions.}
  \label{nim_6_clu_weightingb}
\end{figure}

Figs.~\ref{nim_6_clu_weightinga} and~\ref{nim_6_clu_weightingb} show
the results for pions of $200\,\GeV$. Shown is again the ratio of the
true energy relative to the measured energy on the electromagnetic
scale for both, the EMEC ($w_{\rm EMEC}$) and the HEC ($w_{\rm HEC}$)
compared with the three MC models.

The line shows the result of the fit to the data using the
parameterisation given above \ref{eq:Ci}. The corresponding fits to
the MC data have been also performed.  Because of the rather strong
correlation between $C_1$ and $C_2$, $C_2$ has been set to a constant
value ($C_2=1000\,\cm^3/\GeV$ for EMEC and $C_2=1500\,\cm^3/\GeV$ for
HEC).  It turned out that leaving $C_2$ as free parameter hardly
improved the $\chi^2$ of the fit.  In the EMEC the energy densities
are well described by both GEANT~4 options, in contrast to GEANT~3; in
the HEC the GEANT~4 models are close to the data, but far from being
optimal. The two GEANT~4 models QGSP and LHEP describe the energy
densities equally well, but GEANT~3 is substantially worse. The
parameterisation mentioned above describes the EMEC as well as the HEC
weights for the individual data sets over a large fraction of energy
densities rather well.

\begin{figure}[htb]
  \begin{center}
%    \psdraft   
    \mbox{\epsfig{file=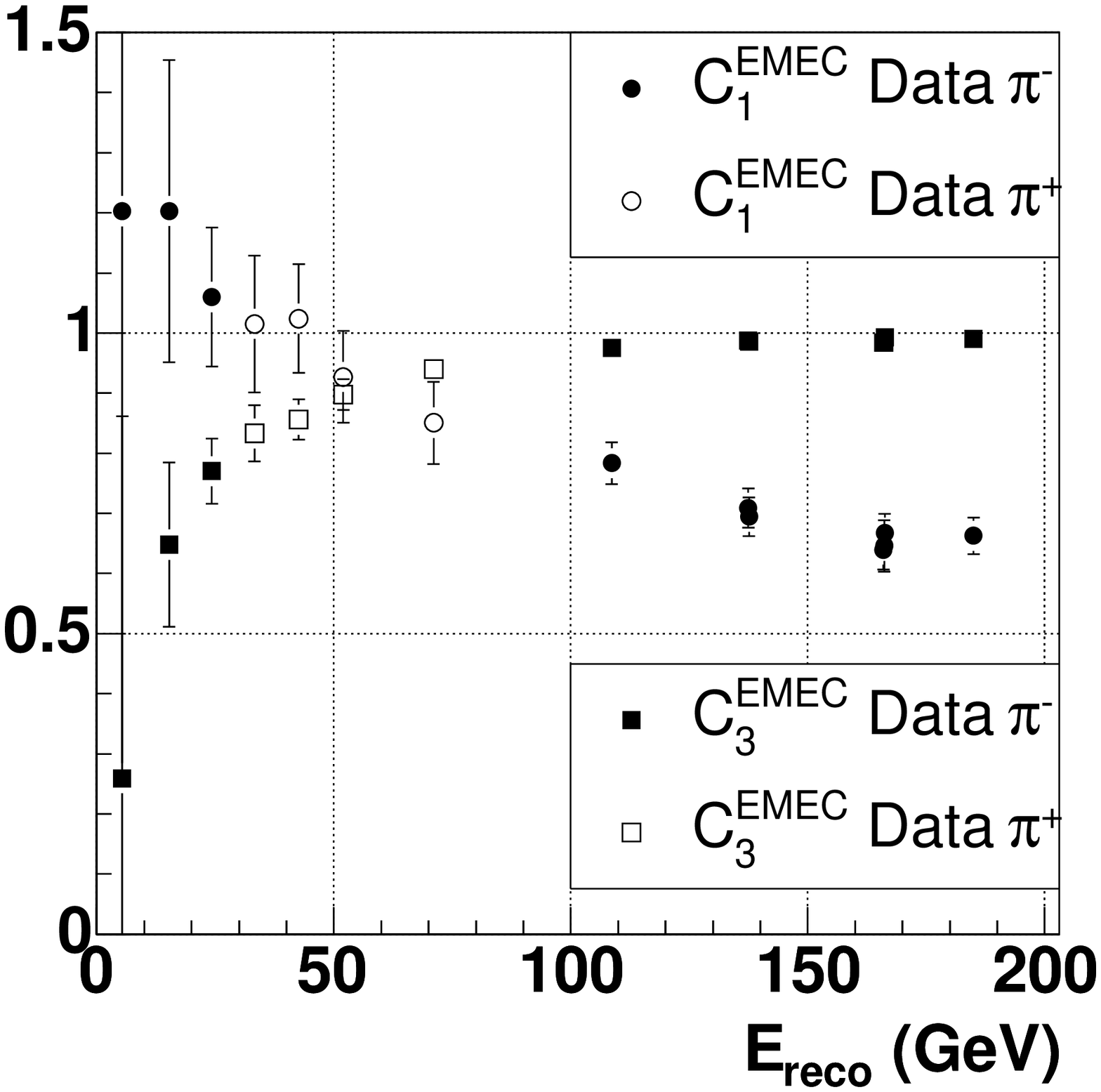,width=0.7\textwidth}}
%    \psfull
  \end{center}
  \caption{Energy dependence of the weighting parameter $C_1^{\rm
  EMEC}$ and $C_3^{\rm EMEC}$ for the data as function of energy.}
  \label{nim_6_clu_weighting10a}
\end{figure}

\begin{figure}[htb]
  \begin{center}
%    \psdraft   
    \mbox{\epsfig{file=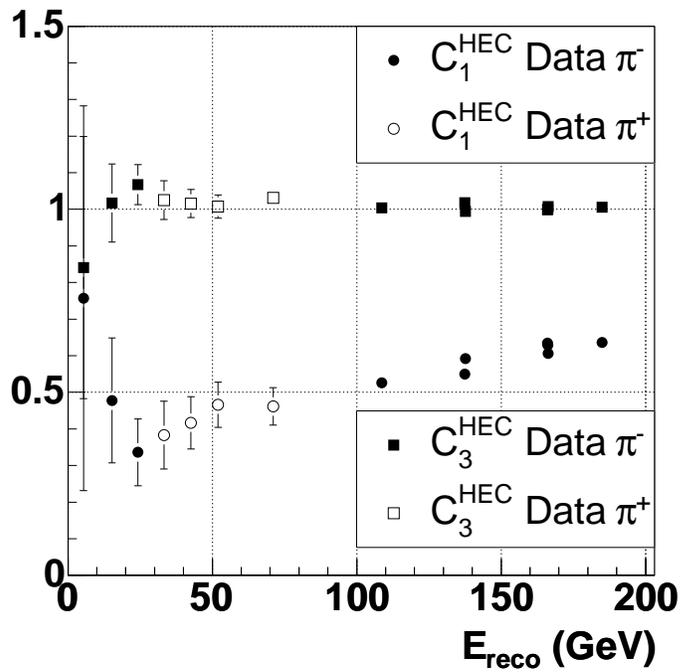,width=0.7\textwidth}}
%    \psfull
  \end{center}
  \caption{Energy dependence of the weighting parameters $C_1^{\rm
  HEC}$ and $C_3^{\rm HEC}$for the data as function of energy.}
  \label{nim_6_clu_weighting12a}
\end{figure}

The energy variation of the fitted parameters $C_1$ and $C_3$ for the
EMEC and the HEC are shown in Figs.~\ref{nim_6_clu_weighting10a}
and~\ref{nim_6_clu_weighting12a}.

\clearpage

The energy dependence of all parameters shows no dramatic variation,
so a parameterization of the energy dependence could easily be
implemented.  As expected $C_3$ approaches unity as given by the
electromagnetic scale. Deviations from 1 are due to residual
correlations with $C_1$, systematic errors in the determination of the
electromagnetic scale or due to systematic errors in signal
reconstruction, calibration and HV-correction.  As seen in the data
these contributions are rather small, both for the EMEC and the HEC.
In ATLAS the in situ determination of $C_3$ with single particles can
yield a powerful crosscheck not only of the electromagnetic scale of
the EMEC, which is well defined from analysing e.g. ${\rm
Z}\rightarrow{\rm e}^+{\rm e}^-$ decays, but also the more difficult
to determine electromagnetic scale of the HEC.
 
\begin{figure}[htb]
  \begin{center}
%    \psdraft   
    \mbox{\epsfig{file=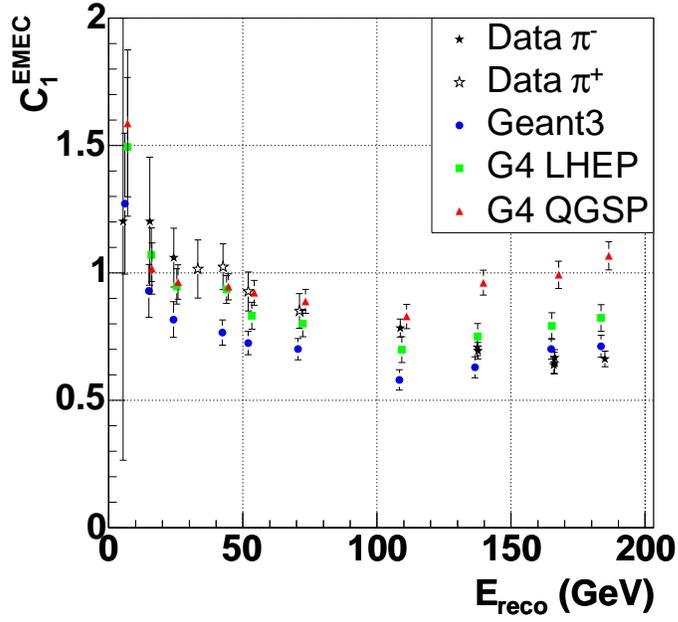,width=0.65\textwidth}}
%    \psfull
  \end{center}
  \caption{Energy dependence of the weighting parameter $C_1^{\rm
  EMEC}$ for the data and the different MC predictions. To optimally
  reveal any differences between data and MC, $C_3^{\rm EMEC}$ has
  been set in the MC to the corresponding value obtained for the
  data.}
  \label{nim_6_clu_weighting14}
\end{figure}

\begin{figure}[htb]
  \begin{center}
%    \psdraft   
    \mbox{\epsfig{file=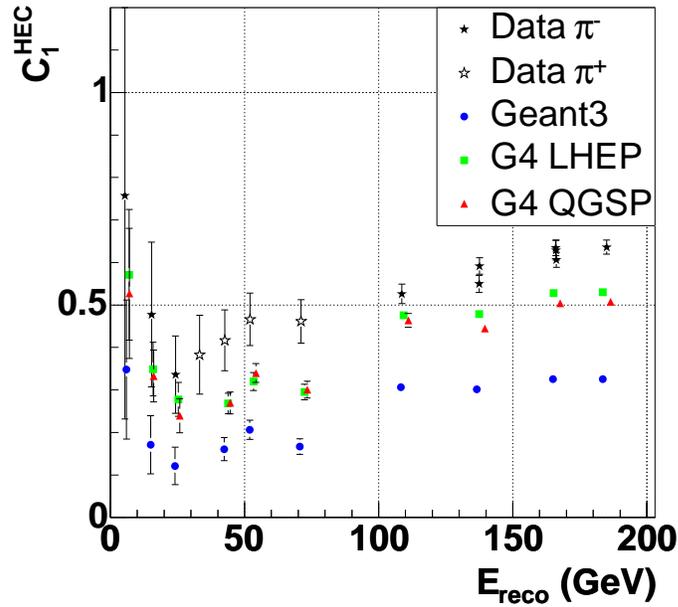,width=0.65\textwidth}}
%    \psful
  \end{center}
  \caption{Energy dependence of the weighting parameter $C_1^{\rm
  HEC}$ for the data and the different MC predictions. To optimally
  reveal any differences between data and MC, $C_3^{\rm HEC}$ has been
  set in the MC to the corresponding value obtained for the data.}
  \label{nim_6_clu_weighting15}
\end{figure}

Some care has to be taken when comparing the $C_i$'s of the data with
MC predictions. The correlations between the individual parameters can
result in rather different energy dependencies, yielding nevertheless
a similar energy dependence of the weights. Therefore for the MC not
only $C_2$ has been kept identical to the data (and fixed, see above),
but also $C_3$ has been set to the value obtained from the fit to the
data. In consequence, $C_1$ will reflect fully any deviation of the MC
from the data, up to a very weak residual correlation of $C_1^{\rm
EMEC}$ and $C_1^{\rm HEC}$.  This approach has been chosen only to
reveal clearly any differences between data and the various MC models.
Figs.~\ref{nim_6_clu_weighting14} and~\ref{nim_6_clu_weighting15} show
the results for $C_1^{\rm EMEC}$ and $C_1^{\rm HEC}$. Only $\pi^-$
interactions have been simulated in the MC. All MC models follow the
general trend of the data, but the differences are significant. The
GEANT~3 predictions deviate strongly from the HEC data and for the
EMEC data none of the MC models can be preferred.

  \clearpage
  
  \subsection{Energy Resolution using the Cluster Weighting Approach}
\begin{figure}[htb]
  \begin{center}
%    \psdraft   
    \mbox{\epsfig{file=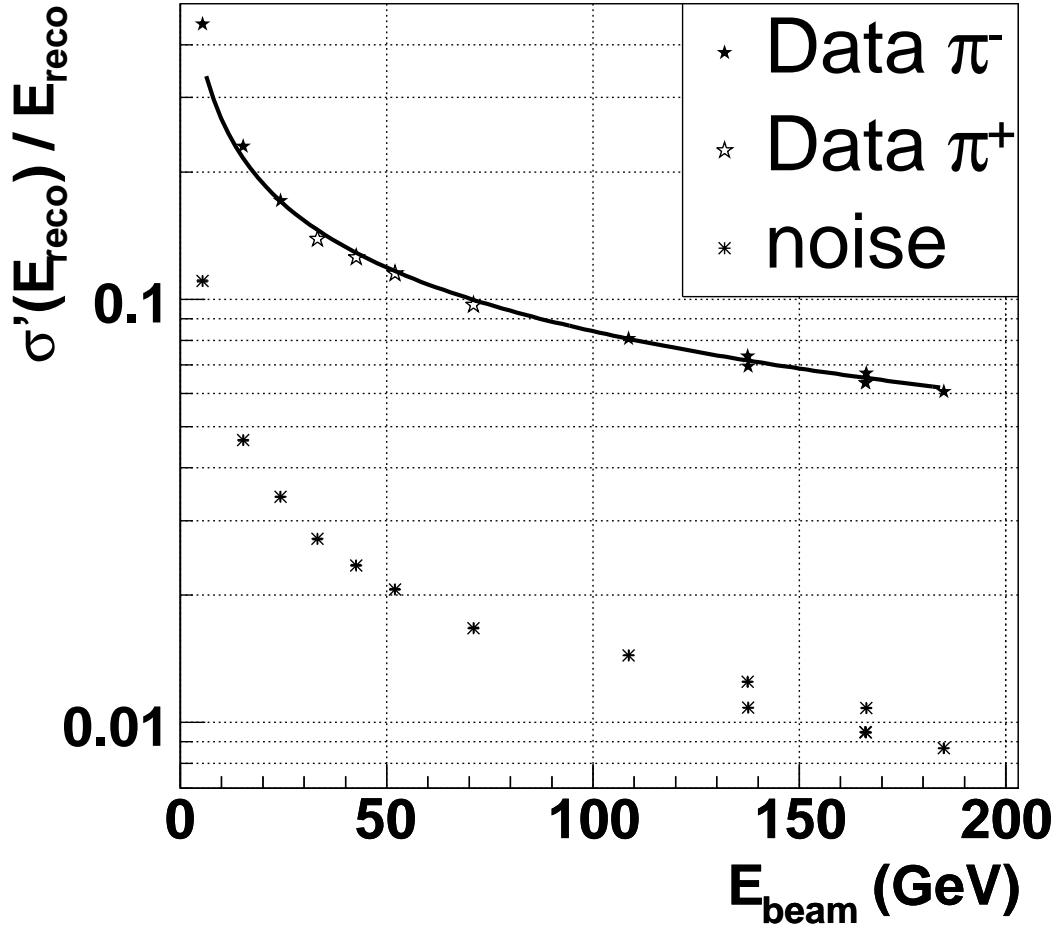,width=1.0\textwidth}}
%    \psfull
  \end{center}

  \caption{Energy dependence of the energy resolution for 
   pion data using the cluster weighting approach. The noise
  has been subtracted, but it is shown explicitly. The line shows the
  result of the fit to the data.}
  \label{nim_6_clu_resolution1}
\end{figure}

\begin{figure}[htb]
  \begin{center}
%    \psdraft   
    \mbox{\epsfig{file=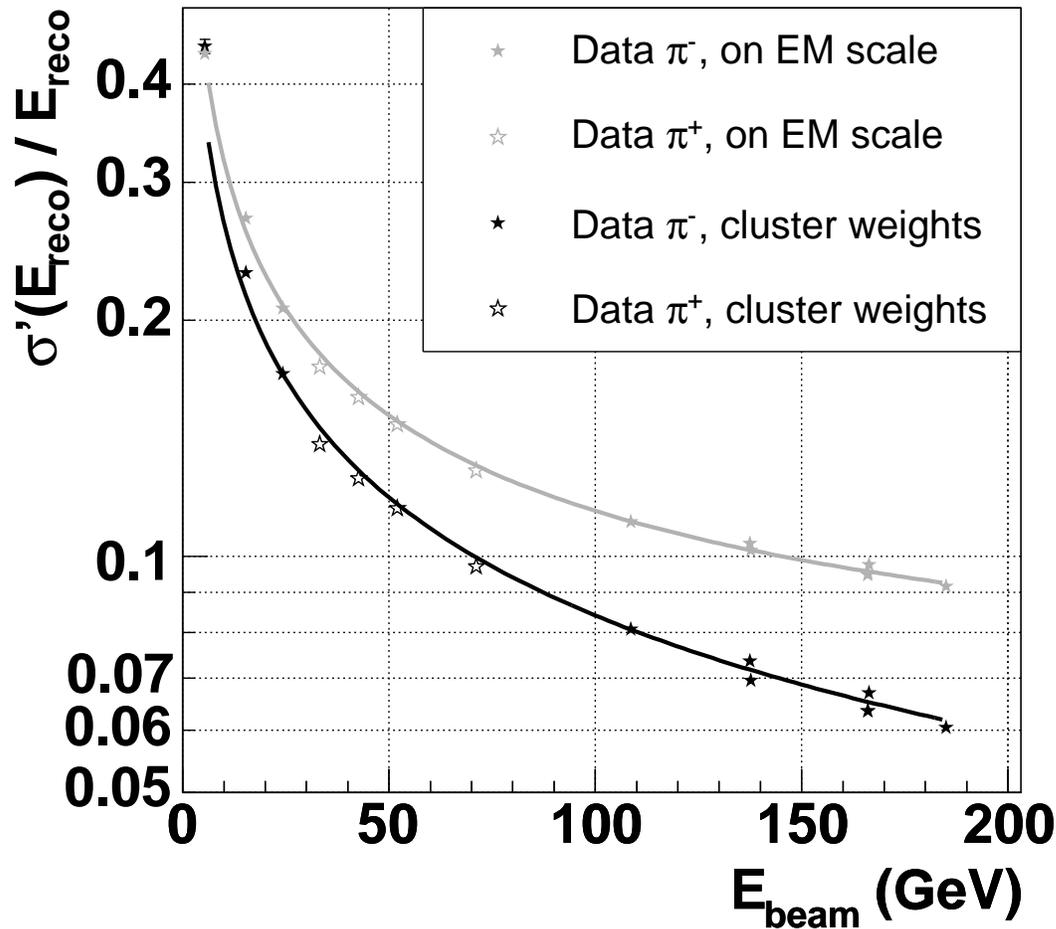,width=1.0\textwidth}}
%    \psfull
  \end{center}
  \caption{Energy dependence of the energy resolution for $\pi$ 
   data using the cluster weighting approach and the
  corresponding results using the electromagnetic scale only. The
  contribution due to the electronic noise has been subtracted. The
  lines show the results of the fits to the data, discussed in the text.}
  \label{nim_6_clu_resolution1a}
\end{figure}

  %auto-ignore
The energy dependence of the energy resolution has been obtained using
the weighting approach discussed above. The results are shown in
Fig.~\ref{nim_6_clu_resolution1} for $\pi$ data after
noise subtraction. The contribution due to electronic noise is shown
explicitly.
A fit to the data with $\frac{\sigma(E)}{E}=\frac{a}{\sqrt{E}}\oplus
b$ yields for the sampling term $(84.1\pm0.3)\,$\%$\sqrt{\GeV}$  and
for the constant term zero within errors. Clearly the energy range
available is not big enough to avoid any correlation between the
sampling and constant term. Nevertheless, the reduction of the
constant term, after the correction for leakage, gives some indication
of the effectiveness of the weighting approach in achieving a good
level of compensation.  The direct comparison with the resolution
obtained on the electromagnetic scale is shown in
Fig.~\ref{nim_6_clu_resolution1a}. The improvement when using the
cluster weighting approach is clearly visible. The lines show the
results of the fits to the data.

\begin{figure}[htb]
  \begin{center}
%    \psdraft   
    \mbox{\epsfig{file=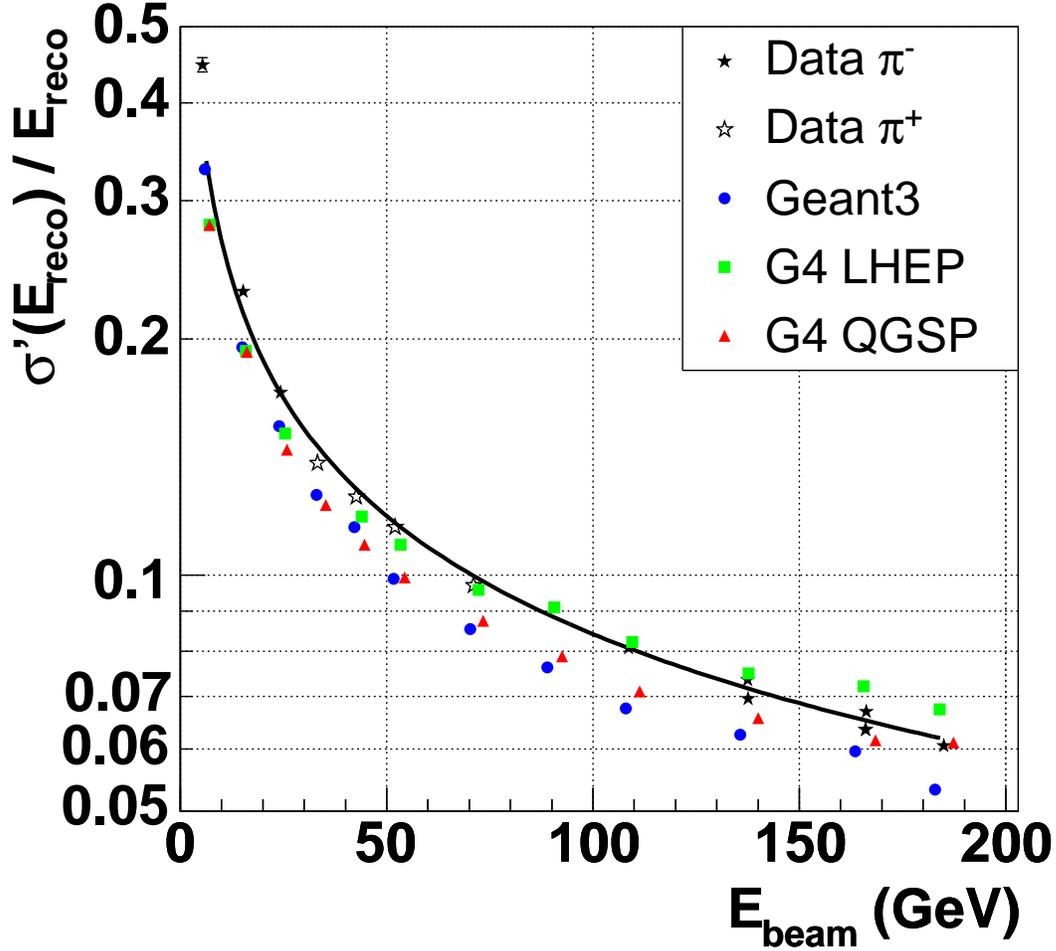,width=1.0\textwidth}}
%    \psfull
  \end{center}
  \caption{Energy dependence of the energy resolution for $\pi$ 
  data using the cluster weighting approach in comparison to
  the different MC predictions. The contribution due to the electronic
  noise has been subtracted. The line shows the result of the fit to
  the data.}
  \label{nim_6_clu_resolution2}
\end{figure}

Finally the data have been compared with MC predictions.
Fig.~\ref{nim_6_clu_resolution2} shows the energy dependence of the
energy resolution, where the line shows the result of the fit to the
data.
The GEANT 3 simulation predicts a sampling term of
$(73.3\pm0.5)\,$\%$\sqrt{\GeV}$ and a vanishing constant term within
errors. The GEANT 4 LHEP and QGSP simulations predict a sampling term
of $(74.0\pm0.5)\,$\%$\sqrt{\GeV}$ and $(72.3\pm0.9)\,$\%$\sqrt{\GeV}$,
respectively.  In general, the GEANT~4 models are closer to the data,
but neither QGSP nor LHEP give an optimal description. The different
energy dependence of the GEANT~4 predictions is also reflected in
non-vanishing constant terms: $(4.1\pm0.1)\,$\% for LHEP and
$(2.5\pm0.3)\,$\% for QGSP. It has been verified that this does not
result from the method of fixing the parameters $C_2$ and $C_3$ to the
data values: the results for both, the constant term and the sampling
term, are almost unchanged when all parameters $C_i$ are obtained from
the fit to the MC data.

  \subsection{${\rm e}/\pi$-Ratio}
\begin{figure}[htb]
  \begin{center}
%    \psdraft   
    \mbox{\epsfig{file=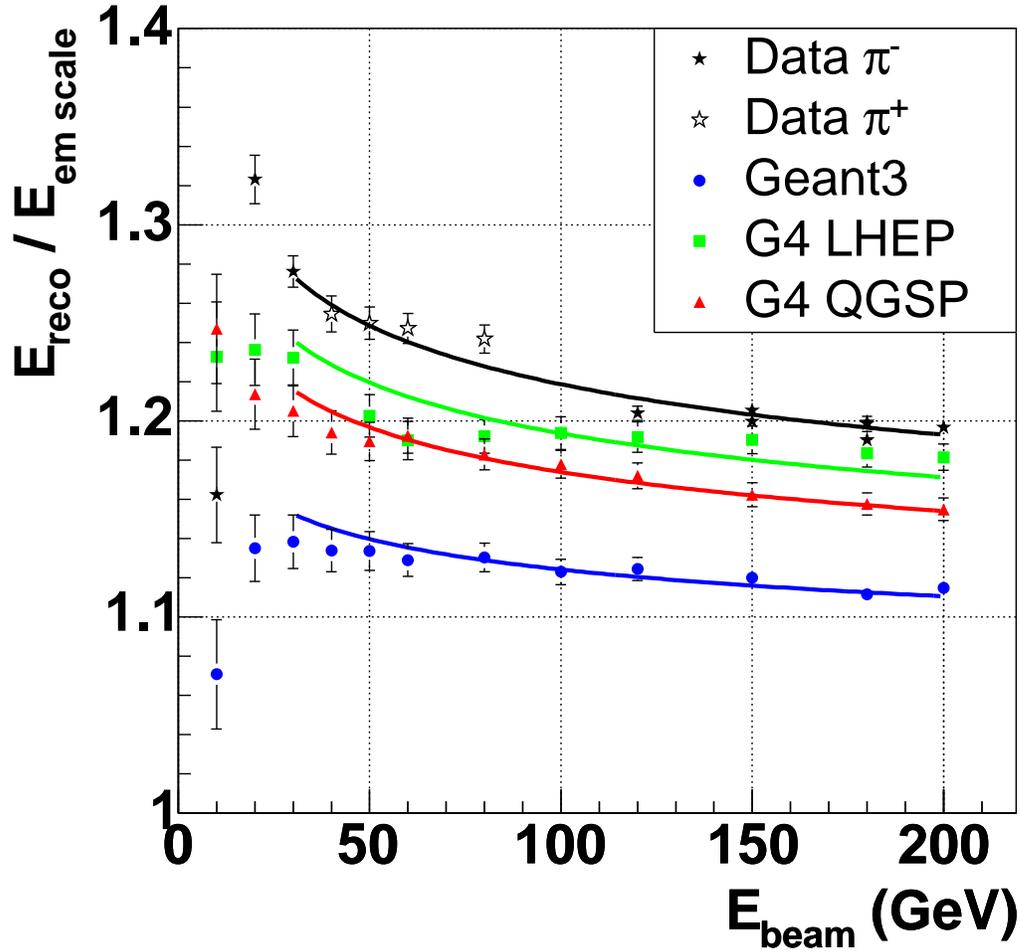,width=1.0\textwidth}}
%    \psfull
  \end{center}
  \caption{${\rm e}/\pi$-ratio as obtained from the cluster weighting
  function.  Shown is the energy dependence for the data as well as
  for the different MC models. The lines show the
  results of fits to the energy dependence.}
  \label{nim_6_etopi1}
\end{figure}

  %auto-ignore
Using the weighting scheme an 'effective' ${\rm e}/\pi$-ratio, the
ratio between the electron and pion response, for the combined set-up
EMEC/HEC can be extracted. Fig.~\ref{nim_6_etopi1} shows this ratio
for the data and for the different MC models.
The energy dependence is in all cases rather similar. However
the MC predictions are substantially below the data,
with GEANT~3 being especially low. From the energy dependence of the
electromagnetic component of the hadronic shower an ${\rm e}/{\rm
h}$-ratio can be obtained~\cite{r-hec,r-eh-Wigmans,r-eh-Groom}.  The
results are using the parameterization~\cite{r-eh-Groom} ${\rm e}/{\rm
h}=1.56\pm0.01$ for the data and ${\rm e}/{\rm h}=1.28\pm0.01$ for GEANT~3,
${\rm e}/{\rm h}=1.48\pm0.01$ for GEANT~4~LHEP and ${\rm e}/{\rm
h}=1.42\pm0.01$ for GEANT~4~QGSP.  The errors are purely
statistical. With the energy dependence being rather similar in all
cases, the 'intrinsic~${\rm e}/{\rm h}$-ratio' reflects closely the
discrepancy seen already in the ${\rm e}/\pi$-ratio. However the
'intrinsic~${\rm e}/{\rm h}$-ratio' for the combined beam test set-up
has no direct interpretation for this composite calorimeter in
contrast to the previous HEC stand alone set-up.

  \subsection{Energy Reconstruction using the Read-out Cell Weighting Approach}
\begin{figure}[htb]
  \begin{center}
%    \psdraft   
    \mbox{\epsfig{file=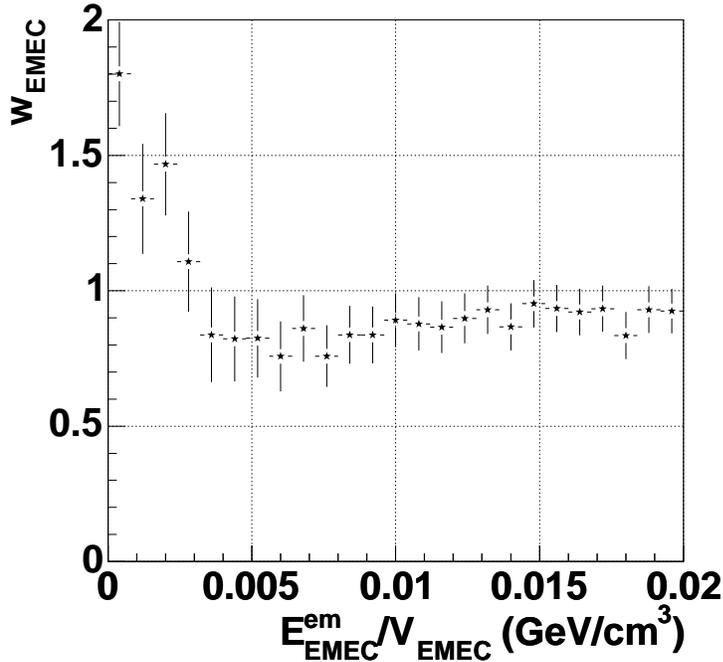,width=0.7\textwidth}}
%    \psfull
  \end{center}
  \caption{Read-out cell weighting approach: weights obtained for the
  EMEC for pions of $200\,\GeV$.}
  \label{nim_6_cell_weighting1}
\end{figure}

\begin{figure}[htb]
  \begin{center}
%    \psdraft   
    \mbox{\epsfig{file=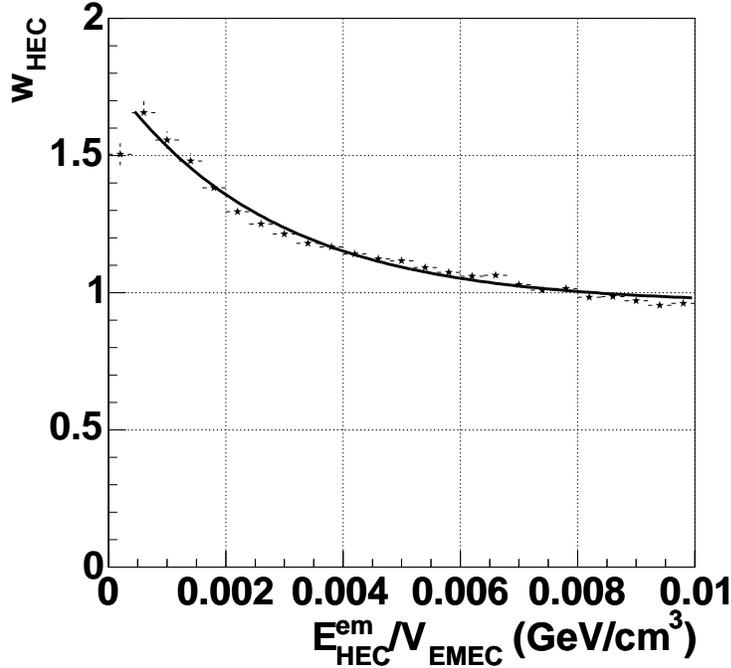,width=0.7\textwidth}}
%    \psfull
  \end{center}
  \caption{Read-out cell weighting approach: weights obtained for the
  HEC for pions of $200\,\GeV$. The line shows the result of the fit.}
  \label{nim_6_cell_weighting2}
\end{figure}

  %auto-ignore
The energy weighting approach on the cluster level described above
yields acceptable results for the case of single particles in beam
tests. However in ATLAS the hadronic calibration requires the
calibration of the calorimeter based on jet interactions. Here the
heavy overlap of many particles dramatically reduces the power of
distinction between hadronic and electromagnetic energy deposits based
on cluster energy density criteria. Therefore for ATLAS a weighting
approach based on the read-out cell level is
envisaged~\cite{r-hadcal}.  The optimal procedure would be to obtain
these weights from MC simulation, where on the read-out cell level the
energy deposits in active and passive material are known. Lacking the
MC instruments (presently in development) an approach using the data
only has been tried. The basic idea is to obtain weights for
individual bins of energy densities $\rho_i$ in the EMEC and HEC by
minimizing:
$$\chi^2=\sum_{\rm events}{\frac{\left[E_{\rm beam}-E_{\rm leak}- 
           E_{\rm HEC}^{\rm reco}- 
           E_{\rm EMEC}^{\rm reco}\right]^2}
	   {\left(\sigma_{\rm noise}^{\rm em}\right)^2 
	+ \left(\sigma_{\rm noise}^{\rm leak}\right)^2}},$$
with
$$E_{\rm EMEC/HEC}^{\rm reco} = \sum_{i=1}^{N}{\left[w_{\rm EMEC/HEC}^{i} 
	\!\!\!\!\!\!\sum_{{\rm cells~with}~\rho_i \leq \rho_{\rm cell} 
	\leq \rho_{i+1}}{\!\!\!\!\!E_{\rm cell}^{\rm em}}\right]}.$$
%$$ E_{EMEC/HEC}^{reco} = \sum_{cell~densities} [w_i \star E_i^{sum}] $$
%$$ E_i^{sum} = \sum_{cells~with~density~i} E_{cell}^{EM scale}, 
%      ~~~~~~~\{\frac{E}{V}\}_i < \frac{E_{cell}^{EM scale}}{V_{cell}} \le 
%      \{\frac{E}{V}\}_{i+1} $$
An important aspect is to correct for the energy leakage prior to
minimizing the $\chi^2$, because the energy leakage can affect
$\chi^2$ more than any difference in the electron to pion response of
the calorimeter. The number of parameters determined in the fit was
typically $N=26$ for EMEC and $N=26$ for the HEC (25 bins in each
energy weighting histogram, plus one overflow bin).
Figs.~\ref{nim_6_cell_weighting1} and~\ref{nim_6_cell_weighting2} show
the results obtained for the weights $w_i$ for the EMEC and HEC for
pions of $200\,\GeV$.

Again, a parameterization of the type 
$$w\left(\frac{E_{\rm cell}^{\rm em}}{V_{\rm cell}}\right) = C_1 \exp \left(-C_2 \frac{E_{\rm cell}^{\rm em}}{V_{\rm cell}} \right) + C_3$$ 
yields at least for the HEC a good parameterization (see
Fig.~\ref{nim_6_cell_weighting2}).

\begin{figure}[htb]
  \begin{center}
%    \psdraft   
    \mbox{\epsfig{file=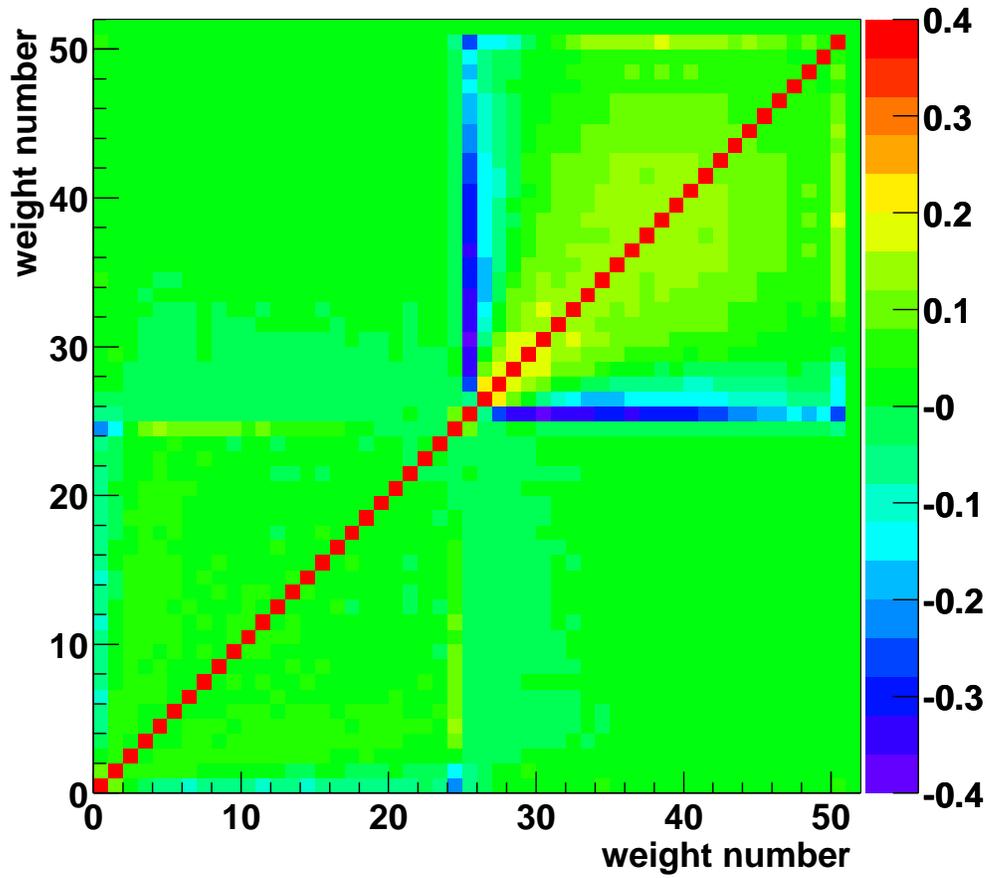,width=1.0\textwidth}}
%    \psfull
  \end{center}
  \caption{Correlation coefficients between the individual weights for
  the $2$ sets of $26$ weights ($26$ weights for the EMEC, $26$
  weights for the HEC) obtained from the fit for pions of
  $200\,\GeV$. The axes run over weight numbers ($0$-$25$ for the EMEC and
  $26$-$51$ for the HEC).}
  \label{nim_6_cell_weighting3}
\end{figure}

\begin{figure}[htb]
  \begin{center}
%    \psdraft   
    \mbox{\epsfig{file=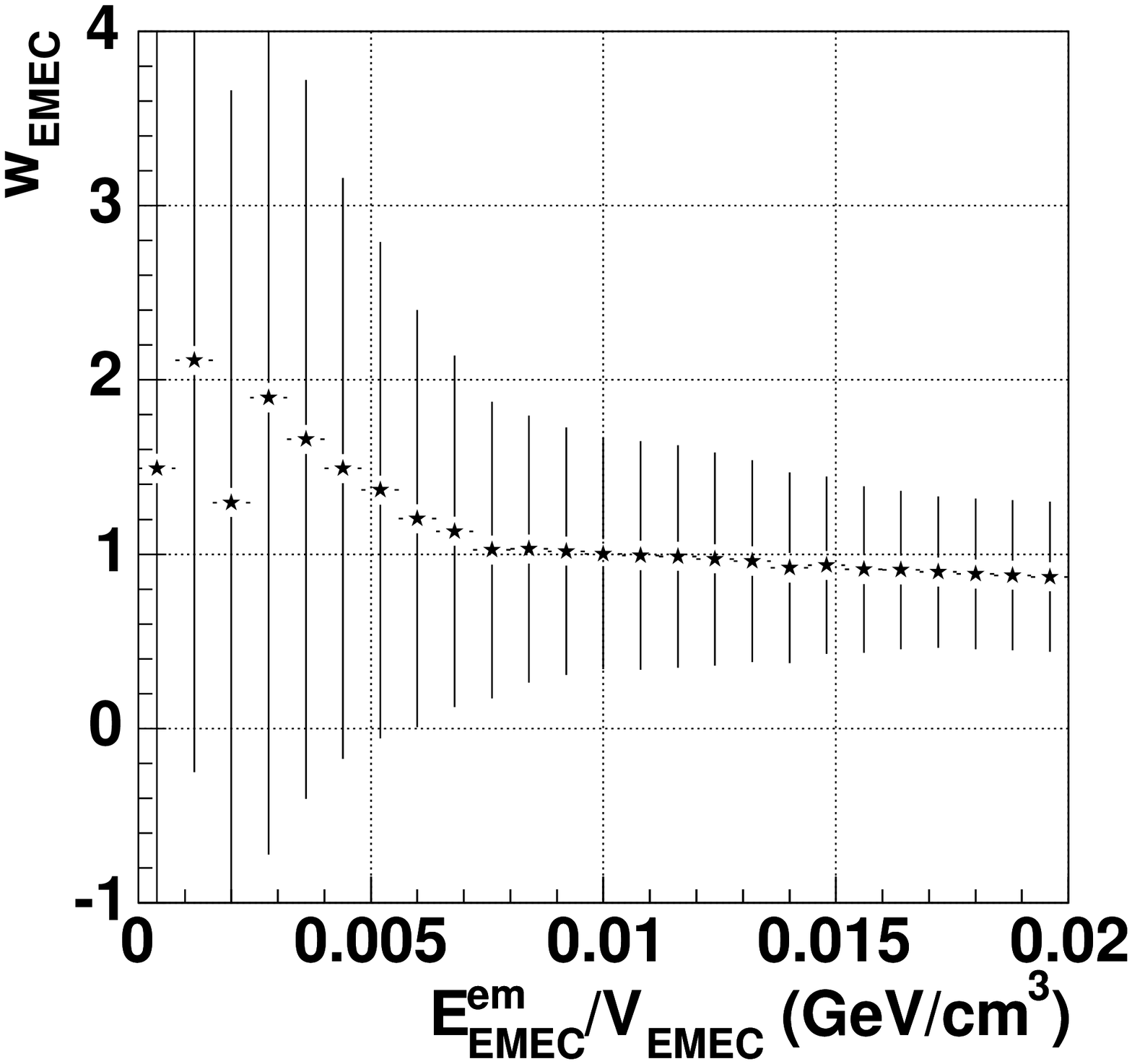,width=0.7\textwidth}}
%    \psfull
  \end{center}
  \caption{Weights obtained from the read-out cell weighting approach
   applied to electrons of $148\,\GeV$. Shown is the dependence on the
   energy density in the EMEC.}
  \label{nim_6_cell_weighting4}
\end{figure}

\begin{figure}[htb]
  \begin{center}
%    \psdraft   
    \mbox{\epsfig{file=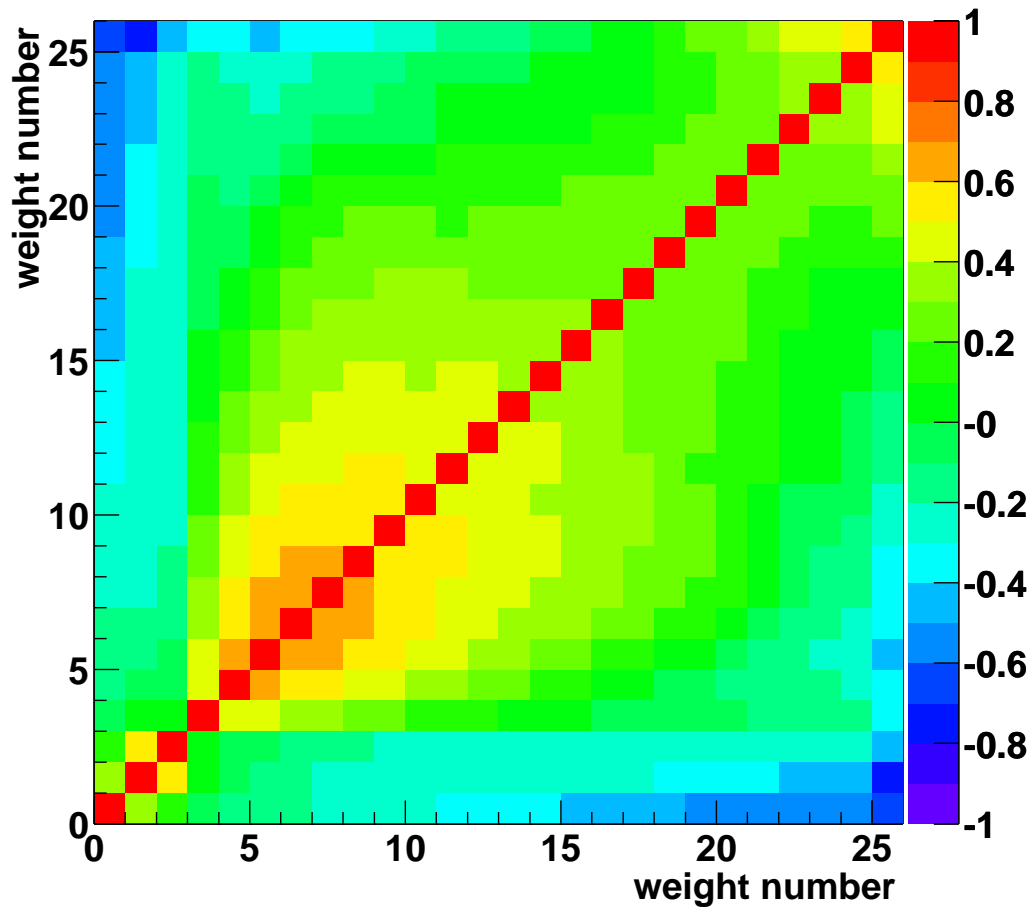,width=1.0\textwidth}}
%    \psfull
  \end{center}
  \caption{Read-out cell weighting approach applied to electrons of
  $148\,\GeV$.  Shown are the correlation coefficients between the
  EMEC weights obtained.  The axes run over weight numbers ($0$-$25$ for
  the EMEC).}
  \label{nim_6_cell_weighting6}
\end{figure}

The method to extract the weights directly from the data rather than
using the MC has a drawback: the weights obtained from the fit are to
some extent correlated and reflect more than the pure e to $\pi$
compensation.  Fig.~\ref{nim_6_cell_weighting3} shows the correlation
coefficients between the individual weights for the $2$ sets of
$26$ weights obtained from the fit for pions of $200\,\GeV$.
The two sets of EMEC and HEC weights are uncorrelated. But
there is a sizeable anti-correlation between the weights for
low densities and all other energy densities for the individual EMEC
and HEC sets of weights. The sensitivity of this method to even
very small imperfections is a severe limitation. This can be
demonstrated nicely by applying the method to
electrons. Fig.~\ref{nim_6_cell_weighting4} shows the weights obtained
in this case for electrons of $148\,\GeV$.
\begin{figure}[htb]
  \begin{center}
%    \psdraft   
    \mbox{\epsfig{file=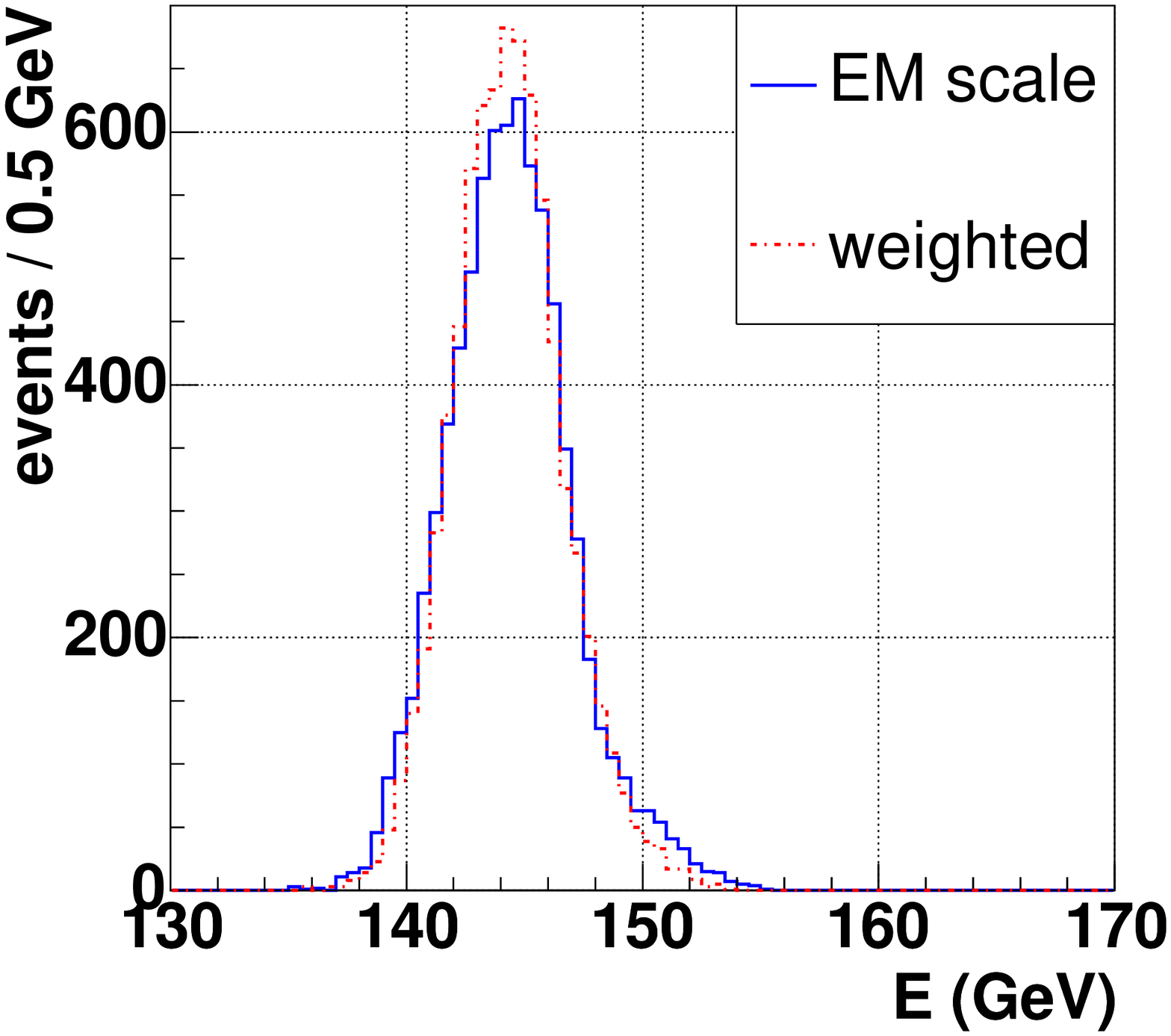,width=0.7\textwidth}}
%    \psfull
  \end{center}
  \caption{Read-out cell weighting approach applied to electrons of
  $148\,\GeV$.  Shown is the energy distribution using the standard
  electromagnetic scale (solid histogram) and the result using the
  read-out cell weights (dashed histogram) determined for electrons.}
  \label{nim_6_cell_weighting5}
\end{figure}
Again, the weights at low energy densities deviate from the nominal
electromagnetic scale, and are anti-correlated with the corresponding
weights at high energy densities.  Fig.~\ref{nim_6_cell_weighting5}
shows the energy distribution for electrons of $148\,\GeV$ using the
nominal electromagnetic scale (solid histogram) and the cell weights
(dashed histogram). The actual energy distribution is hardly changed
at all. The correlation coefficients for the cell weights are shown
in Fig.~\ref{nim_6_cell_weighting6}.

  \clearpage 
    
   \subsection{Energy Resolution using the Read-out Cell Weighting Approach}
\begin{figure}[htb]
  \begin{center}
%    \psdraft   
    \mbox{\epsfig{file=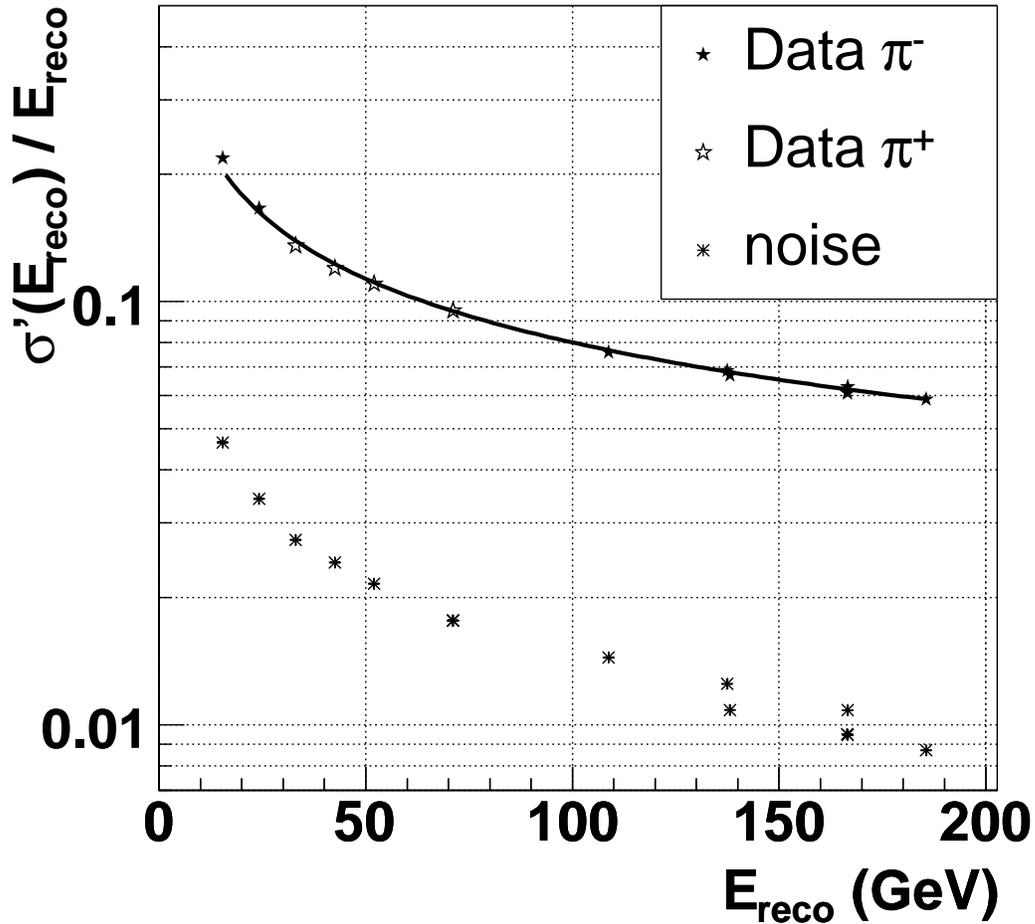,width=1.0\textwidth}}
%    \psfull
  \end{center}
  \caption{Read-out cell weighting approach: Energy dependence of the
   energy resolution for pions. The electronic
   noise has been subtracted, but is shown explicitly.}
  \label{nim_6_cell_resolution1}
\end{figure}

  %auto-ignore
The energy resolution for pions has been determined using the read-out
cell weighting approach. Fig.~\ref{nim_6_cell_resolution1} shows the
energy dependence of the energy resolution obtained. The electronic
noise contribution has been again subtracted, but is shown explicitly.
Using the parameterisation
$\frac{\sigma(E)}{E}=\frac{a}{\sqrt{E}}\oplus b$ a sampling term of
$(80.1\pm0.3)\,$\%$\sqrt{\GeV}$  is obtained.  The constant
term is compatible with being zero. This result is very close to
the result obtained using the cluster weighting approach.

  \subsection{Impact of Hadronic Weights on Energy of Electromagnetic Clusters}
\begin{figure}[htb]
  \begin{center}
%    \psdraft   
    \mbox{\epsfig{file=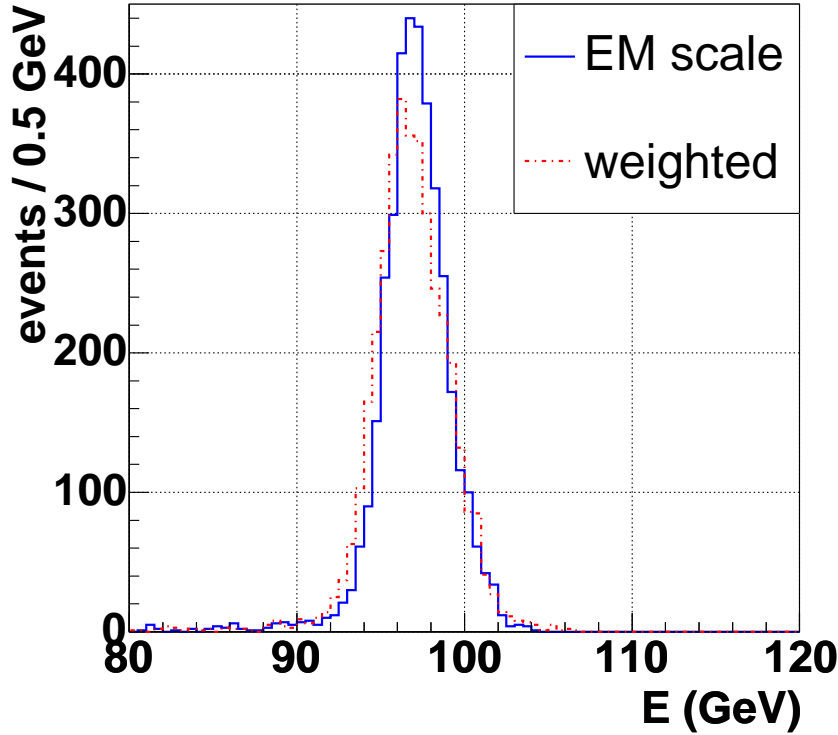,width=0.8\textwidth}}
%    \psfull
  \end{center}
  \caption{Energy distribution of electrons of $100\,\GeV$ using the
  cluster weighting approach (dashed histogram) with pion weights in
  comparison with the energy reconstruction using the electromagnetic
  scale (solid histogram).}
  \label{nim_6_weights_electron1}
\end{figure}

\begin{figure}[htb]
  \begin{center}
%    \psdraft   
    \mbox{\epsfig{file=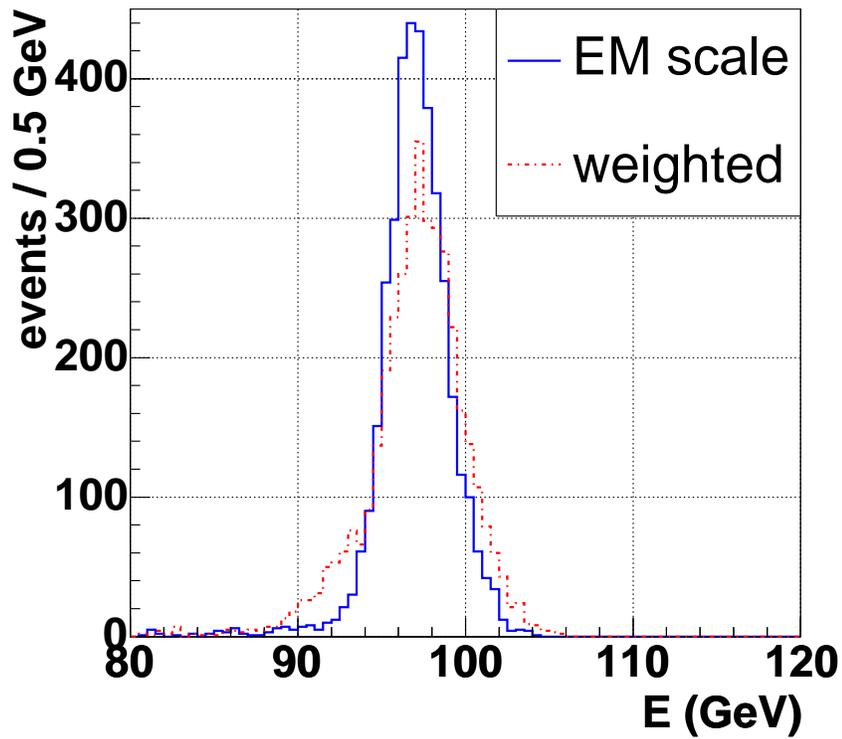,width=0.8\textwidth}}
%    \psfull
  \end{center}
  \caption{Energy distribution of electrons of $100\,\GeV$ using the
  read-out cell weighting approach (dashed histogram) with pion
  weights in comparison with the energy reconstruction using the
  electromagnetic scale (solid histogram).}
  \label{nim_6_weights_electron2}
\end{figure}

  %auto-ignore
In the weighting approach for the hadronic calibration it has to be
guaranteed that even clusters with almost pure electromagnetic energy
are reconstructed at the correct energy scale. To test this, electrons
have been reconstructed using the hadronic weights.
Figs.~\ref{nim_6_weights_electron1} and~\ref{nim_6_weights_electron2}
show the energy distributions for electrons of $100\,\GeV$ using the
cluster weighting and read-out cell weighting approach (weights
defined from pions, dashed histogram) respectively compared to the
normal (electromagnetic scale, solid histogram) reconstruction.
In both cases the mean energy is obtained correctly, but the
resolution is deteriorated. The requirement, not to distort the
electromagnetic energy scale, is important for any weighting approach
to be applied to clusters in the ATLAS calorimeter. Therefore any
algorithm developed to compensate for the different ${\rm e}/\pi$
response in the ATLAS calorimeter has to fulfill this requirement.

%  \clearpage 
  
%  \section{Muon Results}
%  \input{nim_7_muon}

%ls
  \clearpage
\section{Conclusions}
  %auto-ignore
%
% Conclusions
% 
Results of calibration runs with electrons and pions in the ATLAS
end-cap calorimeter corresponding to the region in pseudorapidity
around $|\eta|=1.6$-$1.8$ in ATLAS have been presented. The
electromagnetic calibration constants have been determined for both
subdetectors, the EMEC as well as the HEC. 
First steps of the ATLAS hadronic calibration strategy have been
tested. The signal weighting approach, as used in previous
experiments, improves the pion energy resolution substantially: using
the ansatz $\frac{\sigma (E)}{E}=\frac{a}{\sqrt{E}}\oplus b$ a
sampling term of typically $a\approx80\,$\%$\sqrt{\GeV}$ can be achieved
and a vanishing constant term $b$.  For ATLAS the hadronic calibration
has to deal with jets rather than single particles. Therefore the
transfer of weighting constants from the beam test to ATLAS is only
possible when using MC simulation. One of the important steps in this
procedure is to validate the MC via comparison of the simulation
results with pion data. This has been initiated in detail using
GEANT~3 and GEANT~4 simulations.  In general GEANT~4 yields the better
description of the data, but not yet at the level required. Whereas
the agreement of the GEANT~4 prediction with data for the response at
the electromagnetic scale is rather good, GEANT~4 fails to describe
the details of hadronic shower fluctuations at the level required to
apply weighting techniques. Here further improvements of GEANT~4
parameters and processes are required.

\section*{Acknowledgements}
  %auto-ignore
%
% Acknowledgements
%
The support of the CERN staff operating the SPS and the H6 beam line
is gratefully acknowledged.  We thank the ATLAS-LAr cryogenics
operations team for their invaluable help.

This project has been carried out in the framework of the INTAS
project CERN99-0278, we thank them for the support received.  Further this
work has been supported by the Bundesministerium f\"ur Bildung,
Wissenschaft, Forschung und Technologie, Germany, under contract
numbers $05~{\rm HA}~8{\rm EX}1~6$, $05~{\rm HA}~8{\rm UMA}~8$ and
$05~{\rm HA}~8{\rm PXA}~2$, by the Natural Science and Engineering
Research Council of Canada and by the Slovak funding agency VEGA under
contract number 2/2098/22. We thank all funding agencies for financial
support.

%
%auto-ignore
%
% References
%

%
\end{document}